\makeatletter
\let\latexkernel@label\label
\makeatother

\documentclass[aps,prl,twocolumn,superscriptaddress,amsmath,amssymb,10pt,nofootinbib]{revtex4-2}
\makeatletter
\let\label\latexkernel@label
\makeatother

\usepackage[utf8]{inputenc}
\usepackage[english]{babel}
\usepackage[T1]{fontenc}
\usepackage{graphicx}
\usepackage{dcolumn}
\usepackage{bm}
\usepackage{amsmath,amsfonts,amssymb,latexsym,amsthm,bbm}
\usepackage{tikz}
\usepackage{physics}
\usetikzlibrary{positioning, arrows.meta, calc}
\usepackage{xcolor}
\usepackage{comment}
\usepackage{cancel}
\usepackage{titletoc}
\usepackage{titlesec}
\usepackage[colorlinks]{hyperref}
\hypersetup{linkcolor={black}, urlcolor={blue}, citecolor={blue}}
\usepackage[resetlabels]{multibib}
\usepackage[caption=false]{subfig}

\newcites{app}{References}

\theoremstyle{plain}

\newtheorem*{post*}{Postulate}
\newtheorem{definition}{Definition}

\newcommand{\M}{\tilde{M}_{\rm lin}(\psi)}
\newcommand{\oT}{\Tr(\Omega_2 \psi^{\otimes 8})}
\newcommand{\oF}{\Tr(\Omega_4 \psi^{\otimes 8})}
\newcommand{\TrO}[1]{\operatorname{Tr}\!\left(\Omega_{#1}\,\psi^{\otimes 8}\right)}

\newcommand{\hi}{\mathcal{H}}

\newcommand{\ot}{\otimes}
\newcommand{\sswap}{{\rm T}}

\newcommand{\id}{\mathbbm{1}}
\newcommand{\sym}{\text{sym}}

\newcommand{\pur}{\operatorname{Pur}}
\newcommand{\psit}{\psi^{\otimes 4}}
\newcommand{\rhot}{\rho^{\otimes 4}}
\newcommand{\exv}{\mathbb{E}}
\newcommand{\exvpsi}{\exv_U}
\newcommand{\lin}{\operatorname{lin}}
\newcommand{\stp}{\operatorname{SP}}
\newcommand{\pauli}{\mathbb{P}}
\newcommand{\ttt}{\sswap}
\newcommand{\mlin}{{\Tilde{M}_{\lin}}}
\newcommand{\mtwo}{{\Tilde{M}_{2}}}
\newcommand{\zz}{\mathbb{Z}_2}
\newcommand{\mbar}{\Bar{M}}

\newcommand{\db}{D_B\rho}
\newcommand{\exvrmps}[1]{\exv_{\rho \sim #1}}
\newcommand{\nnf}{\mathcal{N}_4}
\newcommand{\nnt}{\mathcal{N}_2}
\newcommand{\pstab}{{\rm PSTAB}}
\newcommand{\stabzero}{{\rm STAB_0}}
\newcommand{\free}{\mathcal{F}}
\newcommand{\freeop}{\mathcal{O}}
\newcommand{\exvc}{\exv_C}

\titleformat{\paragraph}
{\normalfont\normalsize\bfseries}{\theparagraph}{1em}{}
\titlespacing*{\paragraph}
{0pt}{3.25ex plus 1ex minus .2ex}{1.5ex plus .2ex}

\makeatletter
\AtBeginDocument{\let\auto@bib\@empty}
\makeatother

\begin{document}

	\title{Stabilizer entropy is  trustworthy  for mixed states}
	
\author{Gianluca Esposito}
\affiliation{Scuola Superiore Meridionale, Via Mezzocannone, 4, Napoli, 80134, Italy.}
\affiliation{Istituto Nazionale di Fisica Nucleare (INFN) Sezione di Napoli}

\author{Michele Viscardi}
\affiliation{Dipartimento di Fisica “E.R. Caianiello”, Università di Salerno, Via Giovanni Paolo II, 132, I-84084 Fisciano (SA), Italy}

\author{Alioscia Hamma}
\affiliation{Scuola Superiore Meridionale, Via Mezzocannone, 4, Napoli, 80134, Italy.}
\affiliation{Istituto Nazionale di Fisica Nucleare (INFN) Sezione di Napoli}
\affiliation{Università\'a degli Studi di Napoli Federico II , Dipartimento di Fisica Ettore Pancini}
	\date{\today}
	
	\begin{abstract}

Quantifying non-stabilizerness in mixed states is provably intractable, as any strict monotone requires superexponential time. We propose a linear Stabilizer Entropy that acts as a proper non-stabilizerness monotone with overwhelming probability when restricted to non-adaptive Clifford channels acting on flat mixed stabilizer states. Analytical and numerical results for Haar-random states, Clifford orbits, and random matrix product states show that monotonicity violation probabilities are lower than $\exp-\eta N$. We also prove the validity of Stabilizer Entropy in specific many-body systems undergoing partial measurements, where the amount of resource never increases for each measurement outcome as well as when averaged over outcome probabilities. Given the hardness of strict alternatives, Stabilizer Entropy emerges as a practical and theoretically justified resource measure.
	\end{abstract}
	
	\maketitle
	\begin{figure*}
	    \centering
	    \includegraphics[width=0.9\linewidth]{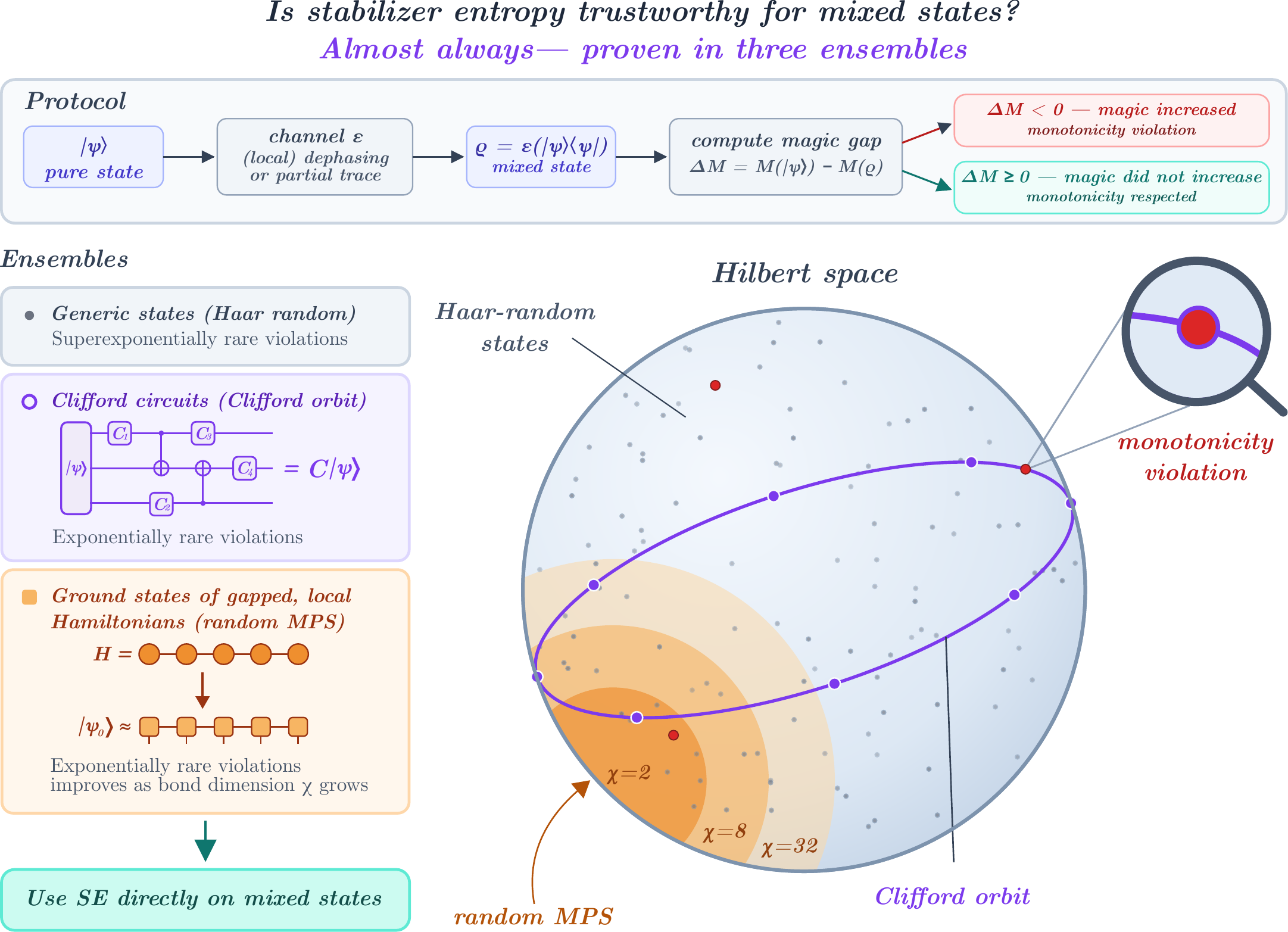}
	    \caption{Stabilizer entropy provides a computationally efficient proxy for exact measures, with provably rare failure across Haar-random, Clifford, and random MPS ensembles, and allows direct evaluation on mixed states where an exact monotone is computationally hard.}
	    \label{fig:graphical_abstract}
	\end{figure*}
    
	\textit{Introduction}--- In recent years, non-stabilizerness has established itself as a necessary ingredient, together with entanglement, to achieve quantum advantage and to uncover the true nature of quantum systems, from exposing the breaking of efficient classical simulation algorithms \cite{gottesman1998HeisenbergRepresentationQuantuma,aaronson2004ImprovedSimulationStabilizer,gu2024magic} to signaling onset of chaotic behavior both in quantum circuits \cite{leone2021quantum,Magni2025quantumcomplexity} and in many-body systems \cite{odavic2022ComplexityFrustrationNew,Jasser_Odavic_Hamma_2025,Odavic_Viscardi_Hamma_2025,tirrito_anticoncentration,turkeshi_magic_many_body} . Many non-stabilizerness monotones have emerged in the literature \cite{veitch2014ResourceTheoryStabilizer,leone2022stabilizer, howard2017ApplicationResourceTheory,haug2023ScalableMeasuresMagic}, with the aim of tracking the behavior of resource during a quantum computation or a physical process, and to obtain bounds on the extractable resource from a given quantum state. However, a recent work proves computational hardness of quantifying non-stabilizerness in the realm of mixed states~\cite{Leone_Eisert_Oliviero_2026_unbearable}, stating that any strict non-stabilizerness monotone for mixed states lies in the complexity class $\rm QP^2$, meaning that its cost scales superexponentially in the number of qubits. In this scenario, a practical non-stabilizerness measure seems unfeasible if applied to realistic physical applications, both experimental and theoretical, where many physical processes are modeled through maps that do not preserve state purity. This situation is akin as the quantum separability problem for entanglement in mixed states, which is also computationally hard \cite{separability_NP,gurvits_NP_sep}: in that field the hurdle has been partially overcome by the creation of entanglement witnesses~\cite{peres_separability,Eisert_comparison_entanglement,vidal_computable_entanglement,plenio_log_negativity,zyczkowski_volume}, which sacrifice faithfulness to obtain efficient computability, and more generic frameworks for resource witnesses are also being pursued \cite{tang2026witnessexpansionunifiedframework, Bermejo_Braccia_Mele_Diaz_Deneris_Larocca_Cerezo_2025}.

    A different, more general question that can be asked for mixed states is whether they admit a purification that is a stabilizer state, in other words, what is the \textit{non-stabilizerness of purification} of a mixed state. It is well-known that generic convex combinations of pure stabilizer states, referred to as the stabilizer polytope, do not admit a stabilizer purification despite having no distillable non-stabilizer resources. The only mixtures of stabilizer states that do admit a stabilizer purification are those whose Pauli tomography weights are flat, and we dub this set as $\stabzero$ with an explicit definition in the following section. 
    
    The computational problem of quantifying non-stabilizerness of purification has not (to our knowledge) been fully explored, but it is likely to be hard, akin to it counterpart in entanglement theory, namely entanglement of purification~\cite{Terhal_Horodecki_Leung_DiVincenzo_2002_ent_pur,Nguyen_Devakul_Halbasch_Zaletel_Swingle_2018_ent_pur,Leone_Rizzo_Eisert_Jerbi_2025}.
    The unresolved nature of this question led us to investigate whether it exists a measure of non-stabilizerness of purification that is efficiently computable. Stabilizer Entropy (SE) \cite{leone2022stabilizer} appeared as the perfect candidate: computable \cite{haug2023QuantifyingnonstabilizernessMatrix, huang2026fastexactapproachstabilizer}, faithful and experimentally accessible \cite{oliviero2022MeasuringMagicQuantum,ahmad2025experimentaldemonstrationnonlocalmagic}. For pure states, as expected, SE coincides with distillable non-stabilizerness, but its naive extension to mixed states, despite being faithful on the set of $\stabzero$, leads to violations of monotonicity under stabilizer operations, as correctly pointed out in \cite{Haug2023stabilizerentropies}, and a proper mixed-state extension of SE, presented in \cite{Leone_Bittel_2024}, leads again to a computationally hard quantity. 
    
	In this paper, we present computable extensions of Stabilizer Entropy to mixed states, and provide analytical and numerical guarantees on its monotonicity under free operations of non-stabilizerness resource theory, namely partial trace and non-post-selected measurement of the system in the computational basis. We tested the validity of our claims in three relevant state ensembles: Haar random states, states with fixed non-stabilizerness, and random Matrix Product States with fixed bond dimension $\chi$. Across the state ensembles, the message is clear: violation of monotonicity of linear Stabilizer Entropy are exponentially rare, i.e. linear SE behaves as a true mixed-state non-stabilizerness monotone with overwhelming probability.  We introduce the formalism of the resource theory of non-adaptive stabilizer computation, and our definition of linear Stabilizer Entropy for mixed states. We also define the concept of \textit{$\eta-$resource proxy} as a resource quantifier whose probability of monotonicity violations decays as $\exp(-\eta N)$. We then compute the monotonicity violation probability for SE in Haar random states, in the Clifford orbit of fixed-non-stabilizerness states, and in random MPS with a fixed bond dimension. 
    
    Moreover, we numerically analyze the efficacy of logarithmic stabilizer entropy in the ground space of the XY model. We ascertain that, upon partial measurement of the system in the computational basis, SE averaged over outcome probabilities never increases with respect to its values before measurement. While this is to be expected for linear SE~\cite{Leone_Bittel_2024}, it holds for its logarithmic mixed state~\cite{leone2022stabilizer} counterpart as well across the XY model phase diagram.
   
	 This work paves the way for the use of Stabilizer Entropy to reliably and efficiently quantify non-stabilizer resources in realistic physical settings, from many-body systems where access to probe non-stabilizerness properties of subsystems is needed \cite{frau_2024_magic_entanglement,viscardi_2026_interplay, Iannotti2025entanglement, tirrito2023quantifying, Tarabunga_2025,iannotti2026nonlocalmagicresourcesfermionic}, to noisy quantum information processes involving measurements.

	% ---------------------------------------------------------------
	\textit{Resource proxies and linear stabilizer entropy}---Quantum Resource Theories (QRT) are a useful conceptual tool for quantifying cost of quantum operations~\cite{chitambar2019QuantumResourceTheories,gour2024resources}: the main idea consists in categorizing quantum states and operations as either ``free'' or ``costly'', motivated on operational constraints or the laws of physics: the set of free states and operations uniquely characterize the QRT. Along with a resource theory, a resource monotone is also needed, namely a function $f:\mathcal{D}(\mathcal{H})\rightarrow\mathbb{R}^+$ that quantifies the amount of the resource contained in a specific quantum state. Such a function needs to be \textit{i)} faithful, namely $f(\rho)=0 \iff \rho$ is a free state, and \textit{ii)} monotone under free operations, namely $f(\mathcal{E}(\rho))\leq f(\rho)$ whenever $\mathcal{E}$ is a free operation in the resource theory.
	In this work we relax the condition of monotonicity and define the concept of $\eta$-resource proxy in the context of qubit systems:
	\begin{definition}[Resource proxy]\label{proxy_def}
		Given a $N-$qubit system Hilbert space $\hi=\mathbb{C}^{2\ot N}$, a quantum resource theory defined by a set of free states $\mathcal{F}$ and a set of free operations $\mathcal{O}$, and another set of states $\mathcal{S}$, a function $f:\mathcal{D}(\hi)\rightarrow \mathbb{R}^+$ is called a $\boldsymbol{\eta}-$\textbf{resource proxy} in the ensemble $\mathcal{S}$ if i) it is faithful; ii) for every free operation $\mathcal{E}\in \mathcal{O}$ and $\rho \in \mathcal{S}$, then there exists $\eta>0$ such that \begin{equation}{\rm Pr}\Big[f(\rho)-f[\mathcal{E}(\rho)]<0\Big]\leq \exp(-\eta N) \end{equation}
	\end{definition}
    \noindent
	Notice that $\mathcal{S}$ does not need to coincide with the set of free states. In other words, $f$ is a resource proxy in a state ensemble if it behaves as a resource monotone with overwhelming probability when states are randomly picked from said ensemble. The higher the value of $\eta$, the better proxy $f$ is: values of $\eta=O(1)$ imply that monotonicity violations are exponentially vanishing, and higher scalings of $\eta$ imply even rarer violations. Notice that the definition of resource proxy is more strict than the definition of resource witnesses (regularly employed in mixed-state entanglement theory \cite{peres_separability,Eisert_comparison_entanglement,vidal_computable_entanglement,plenio_log_negativity,zyczkowski_volume}), which does not require faithfulness.
	
	The stabilizer formalism resource theory~\cite{veitch2014ResourceTheoryStabilizer} in qubit systems, in its non-adaptive form, i.e. not including operations conditioned on the outcome of a measurement, allows the set denoted here $\rm STAB_0$ as free states. The set comprises the density matrices that allow the following decomposition:
	\begin{equation}
	\stabzero\!\ni\!\rho=\!d^{-1}\sum_{P\in G}\phi_P P\,,G\!\subset \!\pauli_N\,{\rm abelian}\,,\phi_P=\pm 1\,,
	\end{equation}	 
	where $\pauli_N$ denotes the set of $N-$qubit Pauli strings modulo phases. Notice that pure stabilizer states are contained in this set. The operations that preserve this set are: i) Clifford unitaries; ii) partial trace; iii) tensor product with $\stabzero$ states; iv) measuring in the computational basis. We dub these operations and concatenations of them \textit{non-adaptive stabilizer operations} \cite{Yashin_2025}.

	Motivated by the proven computational hardness of computing a true non-stabilizerness monotone for mixed states, in this work we introduce the linear version of mixed-state Stabilizer Entropy~\cite{leone2022stabilizer}:
	\begin{definition}[Linear Stabilizer entropy]
		Given a $N$-qubit state $\rho$, with $d=\mathrm{dim}(\hi)=2^N$, we define 
		\begin{equation}
			\mlin(\rho):=\pur(\rho)-\stp(\rho)
		\end{equation}
		with $\pur(\rho)=\Tr(\rho^2)$ and $\stp(\rho)=d\Tr(Q\rhot)\,,Q=d^{-2}\sum_{P\in\pauli_N}P^{\otimes 4}$.
	\end{definition}
	In the rest of the paper we will show that $\mlin$ is a resource proxy for the non-adaptive stabilizer formalism resource theory. To do that, we first show that $\mlin$ is faithful on the free set, i.e. $\mlin(\rho)=0\iff \rho \in \stabzero$. The $\Leftarrow$ implication is straightforward since for states in $\stabzero$ $\pur(\rho)=\stp(\rho)=|G|/d$, whereas we refer to \cite{leone2022stabilizer} for a proof of the other direction of the implication. Moreover $\mlin$ is also invariant under Clifford operations, for the same reasons why Stabilizer R\'enyi entropy is invariant under Cliffords~\cite{leone2022stabilizer}. $\mlin$ is also non increasing under the appending of free states, as $\mlin(\psi\ot \rho)=\pur(\psi)\pur(\rho)-\stp(\psi)\stp(\rho)=|G_\rho|/d\mlin(\psi)\leq \mlin(\psi)$. The rest of the paper will be devoted to quantitatively assess that $\mlin$ almost never violates monotonicity under the remaining two free maps of this resource theory, namely partial trace and complete dephasing onto the computational basis $D_B(\cdot)$. To do this, we define the linear SE gap $\mbar^\mathcal{E} (\rho):=\mlin(\rho)-\mlin(\mathcal{E}(\rho))$ of a state upon application of the map $\mathcal{E}$, and then we compute its average over the previously mentioned state ensembles, namely $\mbar^{\mathcal{E}}_\mathcal{S}:=\exv_{\rho \sim \mathcal{S}}\mbar^\mathcal{E} (\rho)$ with $\mathcal{E}\in \{\Tr_A(\cdot)\,,D_B(\cdot)\}\}$. A positive gap means that $\mlin$ is non-increasing under the map $\mathcal{E}$ on average: we then use typicality arguments, using Chebyshev inequality or L\'evy lemma, to obtain analytical and numerical bounds on the probability of monotonicity violations.
%---------------------------------------------------------------

	\begin{figure}[h!]
	    \centering
	    \includegraphics[width=0.8\linewidth]{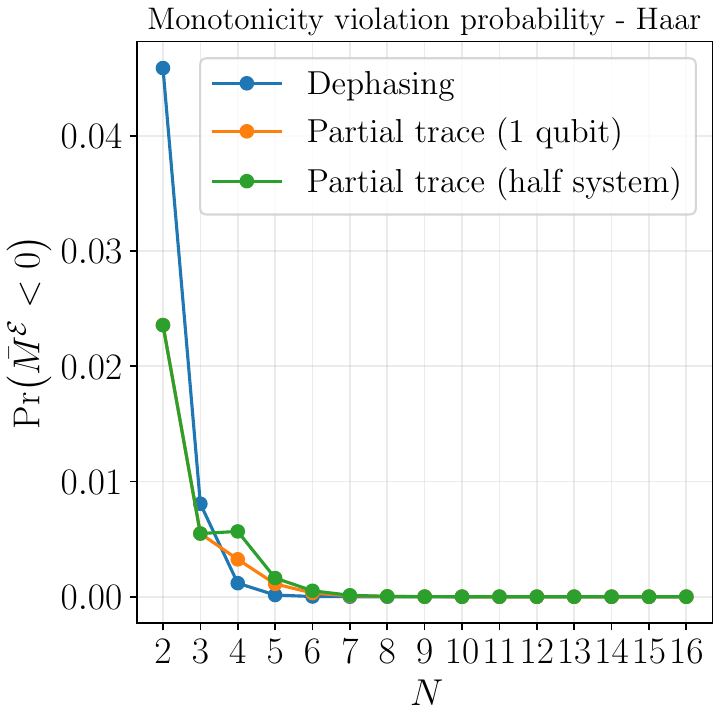}
	    \caption{Behavior of the analytical violation probability bound over the Haar ensemble under dephasing, partial trace of one qubit and partial trace of half the system.}
	    \label{fig:probviol_haar}
	\end{figure}
	\textit{SE as a proxy in Haar random states}--- Starting from an initially pure $N-$qubit state $\rho$, we define $\rho_U:=U\rho U^{\dag}$ with $U$ being a random Haar unitary operator, and compute the average $\mbar^{\mathcal{E}}_{\rm Haar}:=\exvpsi[ \mlin(\rho_U)-\mlin(\mathcal{E}(\rho_U))]=O(1)>0$ for both partial trace and dephasing using standard Haar average techniques~\cite{mele2024introduction}. Applying L\'evy's lemma we prove that violation of monotonicity are super-exponentially rare, namely:
	\begin{equation}
		{\rm Pr}(\mbar^{\mathcal{E}}\leq 0)\leq O(\exp(-d))\,,\mathcal{E}\in \{\Tr_A(\cdot),D_B(\cdot)\}
	\end{equation}
	
	This means that $\mlin $ is a $\eta=O(d)-$proxy for mixed-state non-stabilizerness for generic quantum states, as one can also see in Fig. \ref{fig:probviol_haar}. However, typicality results in the random Haar ensemble do not hold great significance for realistic frameworks, as either physically relevant or computationally useful states are far from being Haar random states. Thus we narrowed our focus on more restricted ensemble of states.

	% --------------------------------------------------------------

	\textit{SE as a proxy in the Clifford orbit}--- Again, starting from a pure $N-$qubit state $\rho$, we define $\rho_C:=C\rho C^\dag$ with $C$ being a Clifford unitary drawn at random with uniform probability and compute $\mbar^{\mathcal{E}}_{\rm Clifford}:=\exv_C [\mlin(\rho_C)-\mlin(\mathcal{E}(\rho_C))]=O(\mlin(\rho))>0$ for both partial trace and dephasing, using Clifford averages techniques found in~\cite{leone2021QuantumChaosQuantum,Bittel_Eisert_Leone_Mele_Oliviero_2025,magni2026anticoncentration}. In this case, computing the variance of this random variable over the Clifford group and using Chebyshev inequality we show that
		\begin{equation}
		{\rm Pr}(\mbar^{\mathcal{E}}\leq 0)\leq O(\exp(-N))\,,\mathcal{E}\in \{\Tr_A(\cdot),D_B(\cdot)\}
	\end{equation}
	 meaning that $\mlin$ is an $\eta=O(1)-$ proxy for mixed state non-stabilizerness in this ensemble. Although typicality is weaker than the Haar case due to the much smaller ensemble given by the finite Clifford group, monotonicity violations are still exponentially rare.

	\begin{figure}[h!]
	    \centering
	    \includegraphics[width=0.8\linewidth]{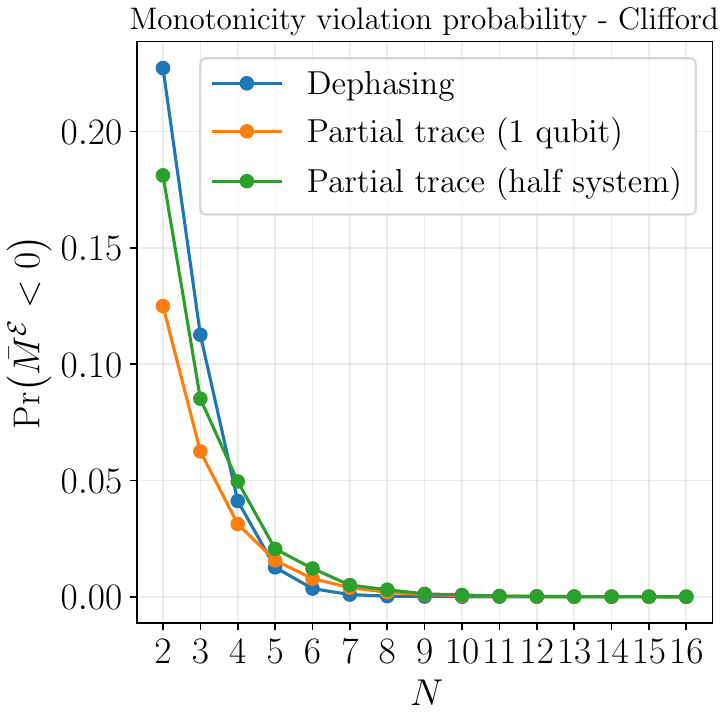}
        \label{fig:probviol_clif}
	    \caption{Behavior of the analytical violation probability bound over the Clifford orbit under dephasing, partial trace of one qubit and partial trace of half the system.}
	\end{figure}
	% 

	%-----------------------------------------------------------------------------------------------

\begin{figure}[h!]
	    \centering
        \includegraphics[width=0.8\linewidth]{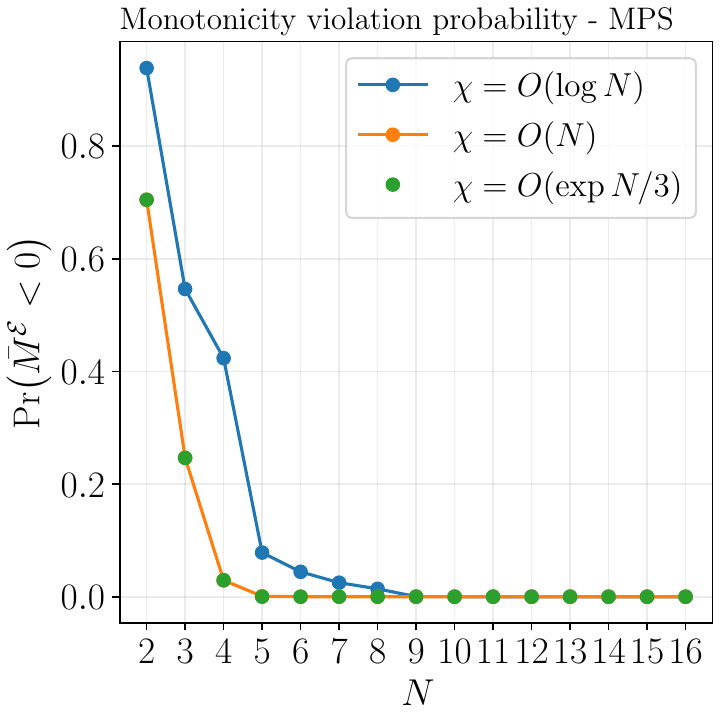}
	    \caption{Behavior of the analytical violation probability bound over the random MPS ensemble under both dephasing and partial trace (the bound is identical for the two), with different scalings of the bond dimension with the number of qubits.}
		\label{fig:probviol_mps}
	\end{figure}

	\textit{SE as a proxy in random Matrix Product States}--- We now move to another important ensemble of states, ubiquitous in quantum physics, namely Matrix Product States (MPS) \cite{Perez-Garcia_Verstraete_Wolf_Cirac_2007}. MPS and their 2D extension, PEPS, are shown to be an invaluable tool for efficiently representing ground states of local (gapped) Hamiltonians \cite{Verstraete_Cirac_2006,Hastings_2006} by encoding all relevant physical properties of such states in low-rank tensors, as well as capturing physical symmetries of quantum states, contained in its tensor representation, through their entanglement structure, \cite{Cirac_Perez-Garcia_Schuch_Verstraete_2021}.
	In addition to their conceptual importance, MPS and, in general, tensor networks methods are also at the core of various efficient numerical simulations of many-body quantum states \cite{Schollwoeck_2011} through RG flows, able to efficiently compute expectation values of local observables and correlation functions, as well as Stabilizer Entropy~\cite{haug2023QuantifyingnonstabilizernessMatrix, tarabunga2024nonstabilizerness,perfectPauliSamplingMPS,huang2026fastexactapproachstabilizer, xiao2026exponentiallyacceleratedsamplingpauli,sierant2026computingquantummagicstate}. %\textbf{INSERT OTHER CIT}.  
	Given the importance of this class of states, we show that $\mlin$ is a good non-stabilizerness proxy in the ensemble of random MPS. A generic with periodic boundary conditions (PBC) MPS is a state in the following form: $\ket\psi=\sum_{i_1,\dots,i_N}\Tr[A^{i_1}[1]\dots A^{i_N}[N] ]\ket{i_1,\dots,i_N}$, where $A^{i_k}$ are $\chi\times \chi$ matrices: $\chi$ is the \textit{bond dimension} of the state and the determining factor of this representation. A given qubit MPS with bond dimension $\chi$ needs $O(2N\chi^2)$ parameters to be completely characterized, much less than a $2^N-$vector representing a generic qubit state. To obtain the ensemble of random MPS with a certain bond dimension value $\chi$, we have drawn and developed the techniques of MPS sequential generation~\cite{Schoen_Hammerer_Wolf_Cirac_Solano_2007}, namely seeing each $A^{i_k}$ tensor as the result of the unitary interaction of a virtual ancillary $\chi-$dimensional qubit with the physical one, with the ancilla subsequently traced out: drawing the interaction unitaries from the Haar measure one obtains the ensemble of random MPS with bond dimension $\chi$. More details on these techniques are found in~\cite{Schoen_Hammerer_Wolf_Cirac_Solano_2007,pz_typicality_mps, Garnerone_Oliveira_Haas_Zanardi_2010, Chen_Garcia_Bu_Jaffe_2024,Lami_Haug_Nardis_2024_CAMPS,turkeshi2026lecturenotesreplicatensor,suppl}. We compute the average linear SE gap over the random MPS ensemble for both partial trace and dephasing, and, again, applying L\'evy's lemma, we found that for both maps monotonicity violations are exponentially rare:
	
	\begin{equation}
		\begin{split}
			{\rm Pr}(\mbar(\psi)^{\mathcal{E}}&\leq 0)\leq 2 \exp [-\frac{(2N-1)\chi^2-1 }{4096\pi}]\,,\\&\mathcal{E}\in \{\Tr_A(\cdot),D_B(\cdot)\}
		\end{split}
	\end{equation}
	thus making $\mlin$ a $\eta=O(\chi^2)-$non-stabilizerness proxy in this ensemble. Even with constant $\chi$, monotonicity violations are exponentially unlikely, with increasing typicality as $\chi$ increases. Fig~\ref{fig:probviol_mps} shows the behavior of the monotonicity violation probability with different scalings of the bond dimension with the number of qubits. 

\begin{figure*}[t!]
        \centering
        \includegraphics[width=0.9\linewidth]{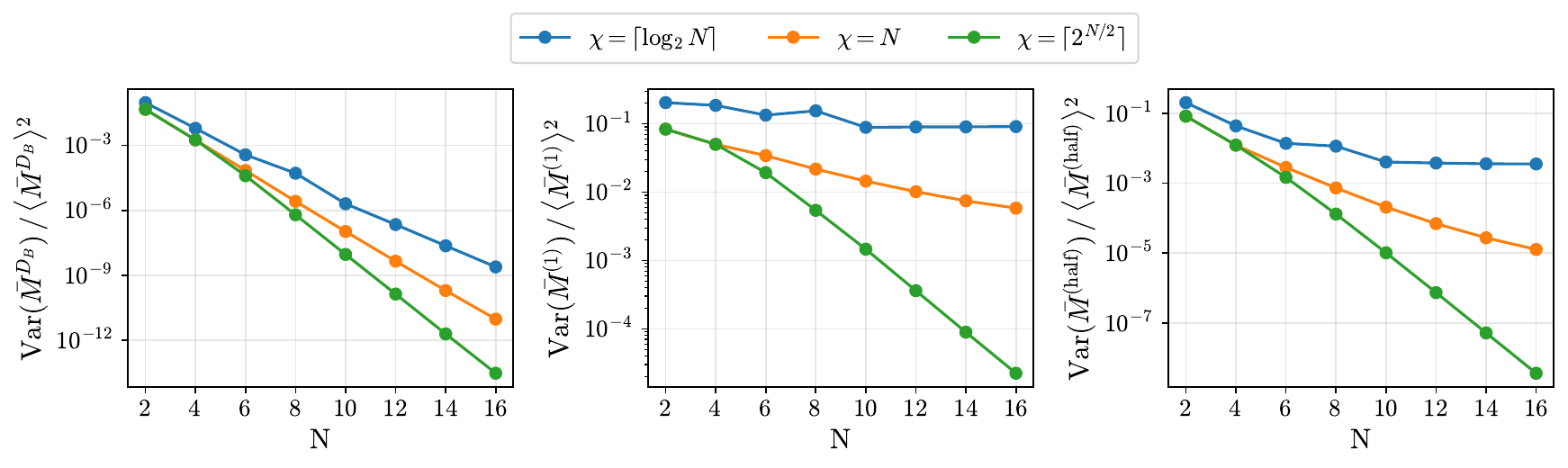}
        \caption{Chebyshev upper bound on the violation probability for $\mlin$ as
        a function of $N$ for three $\chi(N)$ scalings. Left: complete dephasing
        $D_B$. Center: partial trace of the first qubit. Right: partial
        trace of the first half of the system.}
        \label{fig:cheb_UB_Mlin_scaling}
    \end{figure*}

    Having established the behavior of non-stabilizerness gaps for MPS with PBC, we now turn to the open boundary conditions (OBC) setting. OBC are the natural choice in tensor-network algorithms such as DMRG~\cite{Schollwoeck_2011}, and the different topology of such states may, in principle, affect the positivity and the scaling of the non-stabilizerness gap. Since the analytical bound derived for PBC does not directly extend to the OBC MPS ensemble, we numerically evaluate $\mlin$ and the logarithmic mixed-state SE~\cite{leone2022stabilizer} $\mtwo(\psi):=M_2(\psi)-S_2(\psi)$ with $S_2(\psi):=-\log(\pur(\psi))$ being the 2-R\'enyi entropy of the state. We ran numerical simulations for such state ensemble across several system sizes, bond dimension scalings, and choices of quantum channels, inspecting whether both non-stabilizerness measures gaps remain positive.
    For each pair $(N,\chi)$, we sample $3.5\times 10^4$ random OBC MPS~\footnote{Random OBC MPS at fixed bond dimension are generated using the ITensor Julia package~\cite{ITensor}.} and evaluate the non-stabilizerness gap~\footnote{We compute purities and stabilizer purities of states before and after dephasing or partial trace by extending the methods of \cite{huang2026fastexactapproachstabilizer}. In particular, we leverage the Walsh-Hadamard transform to evaluate $\mlin$ exactly, as further detailed in Supplemental Material. } with respect to each of the three channels of interest: complete dephasing in computational basis, partial trace of the first qubit and partial trace of the first half of the system.
    No monotonicity violations were observed in any sample, precluding a frequentist estimate of the violation probability. We therefore apply Chebyshev's inequality to derive an upper bound.
    Figure~\ref{fig:cheb_UB_Mlin_scaling} shows that the upper bound decreases in both $\chi$ and $N$ across all settings and channels. The behavior, however, differs across channels. For the dephasing channel, the upper bound decreases exponentially in the system size $N$ for all considered scalings of the bond dimension. For the partial trace channel, exponential decay in $N$ is retained under an exponential scaling of the bond dimension, while the remaining scalings yield a slower decrease. However, the absence of observed monotonicity violations suggests that the upper bounds may be loose.

\begin{figure*}[t!]
    \centering
    \includegraphics[width=0.9\linewidth]{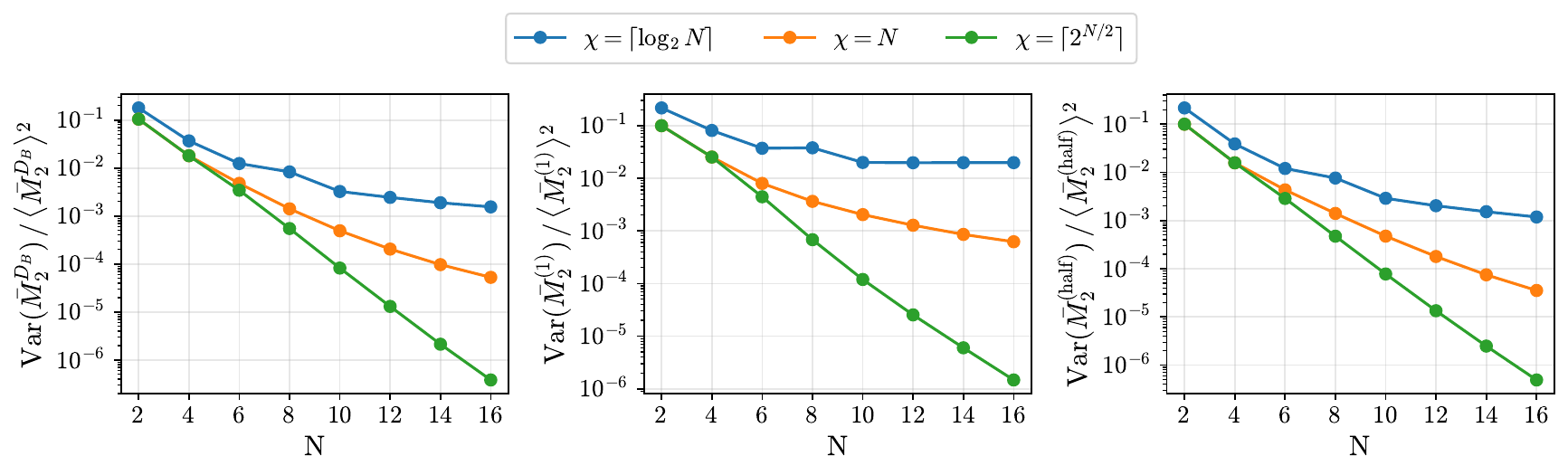}
    \caption{Chebyshev upper bound on the violation probability for $\mtwo$ as
        a function of $N$ for three $\chi(N)$ scalings. Left: complete dephasing
        $D_B$. Center: partial trace of the first qubit. Right: partial
        trace of the first half of the system.}
    \label{fig:cheb_UB_M2_scaling}
\end{figure*}

    The same qualitative behavior is observed for $\mtwo$ (see Fig.~\ref{fig:cheb_UB_M2_scaling}), with the only difference that a slower decay also appears under complete dephasing.

    \textit{SE as a proxy in the XY model}--- Having established that the non-stabilizerness monotonicity violations are extremely rare in random MPS, we now ask if the same behavior survives in ground states of a concrete many-body Hamiltonian. This constitutes a considerably more stringent test: rather than sampling a kinematical ensemble of tensor-network states, we restrict our attention to an ensemble of states - the family of the ground states of a model - which evades the typicality arguments underlying our analytical bounds.
    
    To this end, we consider a paradigmatic example of a one-dimensional spin system with a non-trivial phase diagram: the spin-$1/2$ quantum XY chain~\cite{Lieb1961,Niemeijer1967}. The model Hamiltonian is the following:
    \begin{equation}
    H_{\rm XY}=
     J \sum_j \Big[ (1 + \gamma) S^x_j S^x_{j+1} + (1 - \gamma) S^y_j S^y_{j+1} \Big] - h \sum_j S^z_j
    ,
    \end{equation}
    where $J$ is the exchange interaction term, $\gamma$ is an anisotropy parameter and $h$ is the transverse field intensity.
    We take all parameters to be dimensionless and fix the energy scale by setting $J = 1$.
    In the following discussion we restrict our attention to positive values for both $h$ and $\gamma$, since the model is symmetric for $\pi/2$-rotations about the $z$-axis, which sends $\gamma$ to $-\gamma$, and reflections across the $x-y$ plane, which maps $h$ to $-h$.

    The quantum XY model is exactly solvable and can be mapped to free fermions~\cite{Sachdev_2011,Franchini2017,Mbeng_2024}. The model exhibits two quantum phase transitions, occurring at the isotropic line $\gamma = 0$ (with $|h| \leq 1$) and the critical magnetic field $|h| = 1$. The model hosts an ordered phase for $|h|<1$, in which the $\mathbb{Z}_2$ parity symmetry of the Hamiltonian is spontaneously broken. For $|h|>1$, the system enters a disordered phase where the symmetry is restored and the degeneracy is lifted.
    We study the phase diagram of this model through the lens of the $M_2-$quantified non-stabilizerness  gap with respect to partial post-selected measurements in the computational basis. The reason to compute $M_2$ gaps in post-selected measurements is that it is already known that the pure-state $M_{\rm lin}$ is a strong monotone under stabilizer protocols \cite{Leone_Bittel_2024}, namely that for every stabilizer protocol $\mathcal{E}$ with $\mathcal{E}(\psi)=\sum_i p_i \ketbra{\phi_i}$, then $M_{\lin}(\psi)\geq \sum_i p_i M_{\lin}(\ketbra{\phi_i})$. However, this property is not generally shared by the logarithmic stabilizer entropy $M_2$, so it is of interest to verify whether this feature is at least shared in specific physical systems and highly non-generic but realistic stabilizer channels such as post-selected measurements.
    In detail, we consider the quantum XY model for $N=16$ spins. For each point $(h, \gamma)$, with $h,\gamma \in[0,2]$, we compute the ground state in both parity sectors and perform complete post-selected measurements on a subsystem $B$ of four qubits. Denoting by $\rho_0^{(p)}$ the ground state in the parity sector $p\in\{e,o\}$, with $e$ corresponding to even and $o$ to odd parity, the outcome-averaged post-measurement non-stabilizerness gap is defined as:
    \begin{equation}
        \begin{split}
            \Delta_B \mtwo^{(p)}
            &:=
            \mtwo(\rho_0^{(p)})
            -
            \sum_{\mathbf b}
            p_{\mathbf b}\,
            \mtwo(\rho_{\mathbf b}^{(p)})
            %\\
            %&\equiv
            %M_2^{\rm full}
            %-
            %\overline{M_2^{\rm aft.\ meas.\ }B},
        \end{split}
        \label{def:avg_post_meas_magic_gap}
    \end{equation}
where $\rho_{\mathbf b}^{(p)}$ denotes the normalized post-selected state associated with outcome $\mathbf b$ and $p_{\mathbf b}$ its probability~\footnote{
In the definition in Eq.~\ref{def:avg_post_meas_magic_gap} we adopt the SE $M_2$ rather than $M_{\rm lin}$, since it is not bounded by 1 and therefore retains higher numerical resolution in the many-body regime.}.

\begin{figure}[t]
\centering
\makebox[\columnwidth][l]{
  \subfloat[Even parity sector.]{
    \includegraphics[width=0.958\columnwidth]{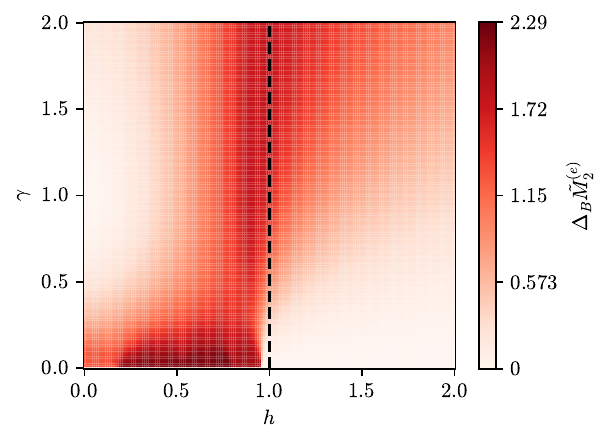}
    \label{fig:xy_gap_even}}}\\[1ex]
\makebox[\columnwidth][l]{
  \subfloat[Odd parity sector.]{
    \includegraphics[width=\columnwidth]{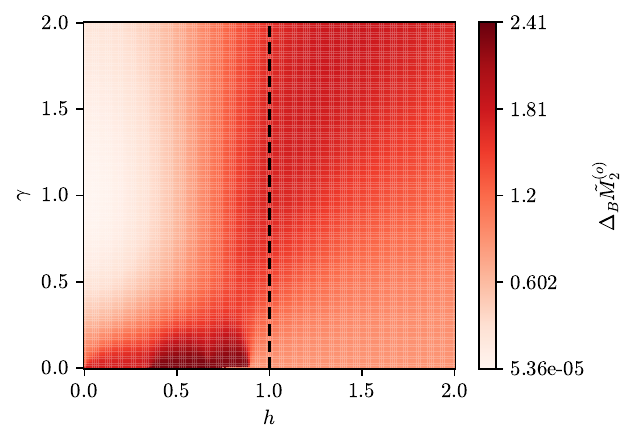}
    \label{fig:xy_gap_odd}}}
\caption{
Phase diagrams of the outcome-averaged post-measurement non-stabilizerness gap
$\Delta_B \mtwo^{(p)}$
for the XY chain with $N=16$, obtained after complete post-selected measurements in computational basis on subsystem $B$ ($N_A=12$, $N_B=4$). The dashed vertical line marks the Ising critical line at $h=1$, while the lower boundary $\gamma=0$ corresponds to the isotropic limit. The gap remains positive throughout the explored phase diagram in both parity sectors, meaning that the average stabilizer entropy of the post-measurement states never exceeds the stabilizer entropy of the ground state.}
\label{fig:xy_gap}
\end{figure}

The numerical results are summarized in Fig.~\ref{fig:xy_gap}. The most significant observation is that the outcome-averaged post-measurement non-stabilizerness gap remains positive throughout the entire explored phase diagram, implying that partial measurements in computational basis lower the stabilizer entropy on average.

The two parity sectors display a highly consistent overall picture. In both cases the positivity of $\Delta_B \mtwo$ is preserved over the entire parameter space. 

Also, both sectors' ground states develop a visible oscillatory behavior as the isotropic limit is approached. Such behavior is consistent with the well-known finite-size parity structure of the XY model, where parity-ground-states energy crossings occur near the isotropic line~\cite{DePasquale_Facchi_09}.

Taken together, these results extend our findings on the positivity of the SE gap for random MPS in two directions. First, we show that the gap stays positive for an ensemble of states (the ground states of an integrable model) that are far from being random. Second, we show it for a different operation, projective measurement of a subsystem in the computational basis, complementing the previous results for the partial-trace and dephasing channels.

	% ---------------------------------------------------------------
	\textit{Discussion}--- In this work, we introduced the concept of resource proxies as quantifiers that exhibit monotonicity with exponentially vanishing failure probability, and we proposed a linear version of stabilizer entropy which we proved to be a genuine proxy for non-stabilizerness in the non-adaptive stabilizer computation resource theory. We have shown that SE behaves as a non-stabilizerness monotone with overwhelming probability not only in the Haar random ensemble, generally used for typicality benchmarks, but also in state ensembles that are more relevant in the many-body and quantum information processing settings, namely the Clifford orbit and random matrix product states. Moreover, we also verified the strong monotonicity of $M_2$ in a paradigmatic many-body model such as the XY-model, with no strong monotonicity violations across the phase diagram. Since having a mixed-state non-stabilizerness monotone is computationally hard \cite{Leone_Eisert_Oliviero_2026_unbearable}, the inception of resource proxies \cite{tang2026witnessexpansionunifiedframework} is a real necessity rather than a matter of convenience. Typicality of rare monotonicity violations paves the way for use of stabilizer entropy in practical settings where measurements \cite{oliviero2022MeasuringMagicQuantum,ahmad2025experimentaldemonstrationnonlocalmagic} and reduced density matrices are analyzed, \cite{frau_2024_magic_entanglement,viscardi_2026_interplay, Iannotti2025entanglement, tirrito2023quantifying, Tarabunga_2025} as well as monitored systems~\cite{fux2024, scocco2026, Tarabunga_Tirrito_2025,PRXQuantum.5.030332,paviglianiti_2025} with strong monotonicity guarantees in physical and computationally significant state ensembles. Moreover, further avenues, such as a more systematic verification of the strong monotonicity of $M_2$ in practical scenarios are to be explored, and more questions arise: can we tighten the analytical violation probability bound to match the numerical evidence? What happens in other state ensembles such as $t-$doped stabilizer states and will SE monotonicity violations transition to Haar values as seen in other probes of chaos \cite{cusumano2026probeschaoscliffordgroup}? 
	Is SE a good non-stabilizerness proxy in 2D many-body state variational ansatze such as PEPS or in Clifford-enhanced MPS~\cite{Lami_Haug_Nardis_2024_CAMPS,qian_2024,huang_2025}?

	\begin{acknowledgments}
		\textit{Acknowledgements.} GE thanks S.F.E. Oliviero, L. Leone and G. Styliaris for useful discussions and comments. 
	\end{acknowledgments}
	
	% ---------------------------------------------------------------
\bibliographystyle{apsrev4-2}
\bibliography{prl_main}

@article{leone2022stabilizer,
  title={Stabilizer {{R}}{\'e}nyi {{E}}ntropy},
  author={Leone, Lorenzo and Oliviero, Salvatore FE and Hamma, Alioscia},
  journal={Physical Review Letters},
  volume={128},
  number={5},
  pages={050402},
  year={2022},
  publisher={APS}
}

@article{fux2024,
  title = {Entanglement -- nonstabilizerness separation in hybrid quantum circuits},
  author = {Fux, Gerald E. and Tirrito, Emanuele and Dalmonte, Marcello and Fazio, Rosario},
  journal = {Phys. Rev. Res.},
  volume = {6},
  issue = {4},
  pages = {L042030},
  numpages = {6},
  year = {2024},
  month = {Oct},
  publisher = {American Physical Society},
  doi = {10.1103/PhysRevResearch.6.L042030},
  url = {https://link.aps.org/doi/10.1103/PhysRevResearch.6.L042030}
}

@article{Yashin_2025,
   title={Characterization of non-adaptive {C}lifford channels},
   volume={24},
   ISSN={1573-1332},
   url={http://dx.doi.org/10.1007/s11128-025-04682-0},
   DOI={10.1007/s11128-025-04682-0},
   number={3},
   journal={Quantum Information Processing},
   publisher={Springer Science and Business Media LLC},
   author={Yashin, Vsevolod I. and Elovenkova, Maria A.},
   year={2025},
   month=mar }

@misc{Leone_Eisert_Oliviero_2026_unbearable, title={The unbearable hardness of deciding about magic}, url={http://arxiv.org/abs/2602.22330}, DOI={10.48550/arXiv.2602.22330}, abstractNote={Identifying the boundary between classical and quantum computation is a central challenge in quantum information. In multi-qubit systems, entanglement and magic are the key resources underlying genuinely quantum behaviour. While entanglement is well understood, magic -- essential for universal quantum computation -- remains relatively poorly characterised. Here we show that determining membership in the stabilizer polytope, which defines the free states of magic-state resource theory, requires super-exponential time $exp( n^2)$ in the number of qubits $n$, even approximately. We reduce the problem to solving a $3$-SAT instance on $n^2$ variables and, by invoking the exponential time hypothesis, the result follows. As a consequence, both quantifying and certifying magic are fundamentally intractable: any magic monotone for general states must be super-exponentially hard to compute, and deciding whether an operator is a valid magic witness is equally difficult. As a corollary, we establish the robustness of magic as computationally optimal among monotones. This barrier extends even to classically simulable regimes: deciding whether a state lies in the convex hull of states generated by a logarithmic number of non-Clifford gates is also super-exponentially hard. Together, these results reveal intrinsic computational limits on assessing classical simulability, distilling pathological magic states, and ultimately probing and exploiting magic as a quantum resource.}, note={arXiv:2602.22330 [quant-ph]}, number={arXiv:2602.22330}, publisher={arXiv}, author={Leone, Lorenzo and Eisert, Jens and Oliviero, Salvatore F. E.}, year={2026}, month=feb }

@article{separability_NP,
	author = {Gharibian, Sevag},
	title = {Strong {NP}-hardness of the quantum separability problem},
	year = {2010},
	issue_date = {March 2010},
	publisher = {Rinton Press, Incorporated},
	address = {Paramus, NJ},
	volume = {10},
	number = {3},
	issn = {1533-7146},
	journal = {Quantum Info. Comput.},
	month = mar,
	pages = {343–360},
	numpages = {18}
}

@article{Tarabunga_2025,
	title={Efficient mutual magic and magic capacity with matrix product states},
	volume={19},
	ISSN={2542-4653},
	url={http://dx.doi.org/10.21468/SciPostPhys.19.4.085},
	DOI={10.21468/scipostphys.19.4.085},
	number={4},
	journal={SciPost Physics},
	publisher={Stichting SciPost},
	author={Tarabunga, Poetri Sonya and Haug, Tobias},
	year={2025},
	month=oct }

@inproceedings{gurvits_NP_sep,
	author = {Gurvits, Leonid},
	title = {Classical deterministic complexity of {E}dmonds' {P}roblem and quantum entanglement},
	year = {2003},
	isbn = {1581136749},
	publisher = {Association for Computing Machinery},
	address = {New York, NY, USA},
	url = {https://doi.org/10.1145/780542.780545},
	doi = {10.1145/780542.780545},
	booktitle = {Proceedings of the Thirty-Fifth Annual ACM Symposium on Theory of Computing},
	pages = {10–19},
	numpages = {10},
	location = {San Diego, CA, USA},
	series = {STOC '03}
}

@article{scocco2026,
  title = {Rise and fall of nonstabilizerness via random measurements},
  author = {Scocco, Annarita and Mok, Wai-Keong and Aolita, Leandro and Collura, Mario and Haug, Tobias},
  journal = {Phys. Rev. Res.},
  volume = {8},
  issue = {1},
  pages = {013217},
  numpages = {18},
  year = {2026},
  month = {Feb},
  publisher = {American Physical Society},
  doi = {10.1103/31sq-k4m3},
  url = {https://link.aps.org/doi/10.1103/31sq-k4m3}
}

@article{leone2021quantum,
  title={Quantum chaos is quantum},
  author={Leone, Lorenzo and Oliviero, Salvatore FE and Zhou, You and Hamma, Alioscia},
  journal={Quantum},
  volume={5},
  pages={453},
  year={2021},
  publisher={Verein zur F{\"o}rderung des Open Access Publizierens in den Quantenwissenschaften}
}

@article{gour2024resources,
  title={Resources of the quantum world},
  author={Gour, Gilad},
  journal={arXiv preprint arXiv:2402.05474},
  year={2024}
}

@article{Cirac_Perez-Garcia_Schuch_Verstraete_2021, title={Matrix Product States and Projected Entangled Pair States: Concepts, Symmetries, and Theorems}, volume={93}, ISSN={0034-6861, 1539-0756}, DOI={10.1103/RevModPhys.93.045003}, abstractNote={The theory of entanglement provides a fundamentally new language for describing interactions and correlations in many body systems. Its vocabulary consists of qubits and entangled pairs, and the syntax is provided by tensor networks. We review how matrix product states and projected entangled pair states describe many-body wavefunctions in terms of local tensors. These tensors express how the entanglement is routed, act as a novel type of non-local order parameter, and we describe how their symmetries are reflections of the global entanglement patterns in the full system. We will discuss how tensor networks enable the construction of real-space renormalization group flows and fixed points, and examine the entanglement structure of states exhibiting topological quantum order. Finally, we provide a summary of the mathematical results of matrix product states and projected entangled pair states, highlighting the fundamental theorem of matrix product vectors and its applications.}, note={arXiv:2011.12127 [quant-ph]}, number={4}, journal={Reviews of Modern Physics}, author={Cirac, Ignacio and Perez-Garcia, David and Schuch, Norbert and Verstraete, Frank}, year={2021}, month=dec, pages={045003} }

@article{mele2024introduction,
  title={Introduction to {H}aar Measure Tools in {Q}uantum {I}nformation: A Beginner's Tutorial},
  author={Mele, Antonio Anna},
  journal={Quantum},
  volume={8},
  pages={1340},
  year={2024},
  publisher={Verein zur F{\"o}rderung des Open Access Publizierens in den Quantenwissenschaften}
}

@article{gu2024magic,
  title={Magic-induced computational separation in entanglement theory},
  author={Gu, Andi and Oliviero, Salvatore FE and Leone, Lorenzo},
  journal={arXiv preprint arXiv:2403.19610},
  year={2024}
}

@misc{ahmad2025experimentaldemonstrationnonlocalmagic,
      title={Experimental demonstration of non-local magic in a superconducting quantum processor}, 
      author={Halima Giovanna Ahmad and Gianluca Esposito and Viviana Stasino and Jovan Odavic and Carlo Cosenza and Alessandro Sarno and Pasquale Mastrovito and Michele Viscardi and Stefano Cusumano and Francesco Tafuri and Davide Massarotti and Alioscia Hamma},
      year={2025},
      eprint={2511.15576},
      archivePrefix={arXiv},
      primaryClass={quant-ph},
      url={https://arxiv.org/abs/2511.15576}, 
}

@Article{viscardi_2026_interplay,
	title={{Interplay of entanglement structures and stabilizer entropy in spin models}},
	author={Michele Viscardi and Marcello Dalmonte and Alioscia Hamma and Emanuele Tirrito},
	journal={SciPost Phys. Core},
	volume={9},
	pages={012},
	year={2026},
	publisher={SciPost},
	doi={10.21468/SciPostPhysCore.9.1.012},
	url={https://scipost.org/10.21468/SciPostPhysCore.9.1.012},
}

@article{aaronson2004ImprovedSimulationStabilizer,
  title = {Improved Simulation of Stabilizer Circuits},
  author = {Aaronson, Scott and Gottesman, Daniel},
  year = {2004},
  month = nov,
  journal = {Physical Review A},
  volume = {70},
  pages = {052328--052328},
  doi = {10.1103/PhysRevA.70.052328}
}

@misc{suppl,
title = {See {S}upplemental {M}aterial for formal proofs, definitions and derivations.},
year=2026
}

@article{Jasser_Odavic_Hamma_2025, title={Stabilizer entropy and entanglement complexity in the Sachdev-Ye-Kitaev model}, volume={112}, DOI={10.1103/rz86-47h3}, abstractNote={The Sachdev-Ye-Kitaev (SYK) model is of paramount importance for the understanding of both strange metals and a microscopic theory of two-dimensional gravity. We study the interplay between stabilizer Rényi entropy and entanglement entropy in both the ground state and highly excited states of the SYK−4+SYK−2 model, interpolating the highly chaotic four-body interactions model with the integrable two-body interactions one. The interplay between these quantities is also assessed through universal statistics of the entanglement spectrum and its antiflatness. We find that SYK-4 is indeed characterized by a complex pattern of both entanglement and nonstabilizer resources, while SYK-2 is nonuniversal and not complex. We discuss the fragility and robustness of these features depending on the interpolation parameter.}, number={17}, journal={Physical Review B}, publisher={American Physical Society}, author={Jasser, Barbara and Odavić, Jovan and Hamma, Alioscia}, year={2025}, month=nov, pages={174204} }

@article{Odavic_Viscardi_Hamma_2025, title={Stabilizer entropy in nonintegrable quantum evolutions}, volume={112}, DOI={10.1103/y9r6-dx7p}, abstractNote={Entanglement and stabilizer entropy are both involved in the onset of complex behavior in quantum many-body systems. Their interplay is at the root of complexity of simulability, scrambling, thermalization, and typicality. In this work, we study the dynamics of entanglement, stabilizer entropy, and the antiflatness of the entanglement spectrum after a quantum quench in a spin chain. We find that free-fermion theories show a gap in the long-time behavior of these resources compared to their random matrix theory value while nonintegrable models saturate it.}, number={10}, journal={Physical Review B}, publisher={American Physical Society}, author={Odavić, J. and Viscardi, M. and Hamma, A.}, year={2025}, pages={104301} }

@article{Verstraete_Cirac_2006, title={Matrix product states represent ground states faithfully}, volume={73}, rights={http://link.aps.org/licenses/aps-default-license}, ISSN={1098-0121, 1550-235X}, DOI={10.1103/PhysRevB.73.094423}, number={9}, journal={Physical Review B}, author={Verstraete, F. and Cirac, J. I.}, year={2006}, month=mar, pages={094423}}

@article{Hastings_2006, title={Solving gapped Hamiltonians locally}, volume={73}, rights={http://link.aps.org/licenses/aps-default-license}, ISSN={1098-0121, 1550-235X}, DOI={10.1103/PhysRevB.73.085115}, number={8}, journal={Physical Review B}, author={Hastings, M. B.}, year={2006}, month=feb, pages={085115}}

@misc{tang2026witnessexpansionunifiedframework,
      title={Witness expansion: A unified framework for analytical and measurable mixed-state resource detection}, 
      author={Yifan Tang and Chengkai Zhu and Yuzhen Zhang and Jens Eisert and Zi-Wen Liu and Ingo Roth and Otfried Gühne and Xin Wang and Zhenhuan Liu},
      year=2026,
      eprint={2606.27105},
      archivePrefix={arXiv},
      primaryClass={quant-ph},
      url={https://arxiv.org/abs/2606.27105}, 
}

@misc{Bermejo_Braccia_Mele_Diaz_Deneris_Larocca_Cerezo_2025, title={Characterizing quantum resourcefulness via group-Fourier decompositions}, url={http://arxiv.org/abs/2506.19696}, DOI={10.48550/arXiv.2506.19696}, note={arXiv:2506.19696 [quant-ph]}, number={arXiv:2506.19696}, publisher={arXiv}, author={Bermejo, Pablo and Braccia, Paolo and Mele, Antonio Anna and Diaz, Nahuel L. and Deneris, Andrew E. and Larocca, Martin and Cerezo, M.}, year=2025, month=jun }

@misc{cusumano2026probeschaoscliffordgroup,
	title={Probes of chaos over the {C}lifford group and approach to {H}aar values}, 
	author={Stefano Cusumano and Gianluca Esposito and Alioscia Hamma},
	year={2026},
	eprint={2603.29695},
	archivePrefix={arXiv},
	primaryClass={quant-ph},
	url={https://arxiv.org/abs/2603.29695}, 
}

@article{Lami_Haug_Nardis_2024_CAMPS,
  title = {Quantum State Designs with {C}lifford-Enhanced Matrix Product States},
  author = {Lami, Guglielmo and Haug, Tobias and De Nardis, Jacopo},
  journal = {PRX Quantum},
  volume = {6},
  issue = {1},
  pages = {010345},
  numpages = {14},
  year = {2025},
  month = {Mar},
  publisher = {American Physical Society},
  doi = {10.1103/PRXQuantum.6.010345},
  url = {https://link.aps.org/doi/10.1103/PRXQuantum.6.010345}
}

@article{Perez-Garcia_Verstraete_Wolf_Cirac_2007,
author = {Perez-Garcia, D. and Verstraete, F. and Wolf, M. M. and Cirac, J. I.},
title = {Matrix product state representations},
year = {2007},
issue_date = {July 2007},
publisher = {Rinton Press, Incorporated},
address = {Paramus, NJ},
volume = {7},
number = {5},
issn = {1533-7146},
journal = {Quantum Info. Comput.},
month = jul,
pages = {401–430},
numpages = {30}
}

@article{Schoen_Hammerer_Wolf_Cirac_Solano_2007, title={Sequential Generation of Matrix-Product States in Cavity QED}, volume={75}, ISSN={1050-2947, 1094-1622}, DOI={10.1103/PhysRevA.75.032311}, abstractNote={We study the sequential generation of entangled photonic and atomic multi-qubit states in the realm of cavity QED. We extend the work of C. Schoen et al. [Phys. Rev. Lett. 95, 110503 (2005)], where it was shown that all states generated in a sequential manner can be classified efficiently in terms of matrix-product states. In particular, we consider two scenarios: photonic multi-qubit states sequentially generated at the cavity output of a single-photon source and atomic multi-qubit states generated by their sequential interaction with the same cavity mode.}, note={arXiv:quant-ph/0612101}, number={3}, journal={Physical Review A}, author={Schoen, C. and Hammerer, K. and Wolf, M. M. and Cirac, J. I. and Solano, E.}, year={2007}, pages={032311} }

@misc{Bittel_Eisert_Leone_Mele_Oliviero_2025, title={A complete theory of the {C}lifford commutant}, url={http://arxiv.org/abs/2504.12263}, DOI={10.48550/arXiv.2504.12263}, abstractNote={The Clifford group plays a central role in quantum information science. It is the building block for many error-correcting schemes and matches the first three moments of the Haar measure over the unitary group -a property that is essential for a broad range of quantum algorithms, with applications in pseudorandomness, learning theory, benchmarking, and entanglement distillation. At the heart of understanding many properties of the Clifford group lies the Clifford commutant: the set of operators that commute with $k$-fold tensor powers of Clifford unitaries. Previous understanding of this commutant has been limited to relatively small values of $k$, constrained by the number of qubits $n$. In this work, we develop a complete theory of the Clifford commutant. Our first result provides an explicit orthogonal basis for the commutant and computes its dimension for arbitrary $n$ and $k$. We also introduce an alternative and easy-to-manipulate basis formed by isotropic sums of Pauli operators. We show that this basis is generated by products of permutations -which generate the unitary group commutant- and at most three other operators. Additionally, we develop a graphical calculus allowing a diagrammatic manipulation of elements of this basis. These results enable a wealth of applications: among others, we characterize all measurable magic measures and identify optimal strategies for stabilizer property testing, whose success probability also offers an operational interpretation to stabilizer entropies. Finally, we show that these results also generalize to multi-qudit systems with prime local dimension.}, note={arXiv:2504.12263 [quant-ph]}, number={arXiv:2504.12263}, publisher={arXiv}, author={Bittel, Lennart and Eisert, Jens and Leone, Lorenzo and Mele, Antonio A. and Oliviero, Salvatore F. E.}, year={2025}, month=apr }

@article{turkeshi_magic_many_body,
	title = {Pauli spectrum and nonstabilizerness of typical quantum many-body states},
	author = {Turkeshi, Xhek and Dymarsky, Anatoly and Sierant, Piotr},
	journal = {Phys. Rev. B},
	volume = {111},
	issue = {5},
	pages = {054301},
	numpages = {12},
	year = {2025},
	month = {Feb},
	publisher = {American Physical Society},
	doi = {10.1103/PhysRevB.111.054301},
	url = {https://link.aps.org/doi/10.1103/PhysRevB.111.054301}
}

@article{tirrito_anticoncentration,
	title = {Anticoncentration and Nonstabilizerness Spreading under Ergodic Quantum Dynamics},
	author = {Tirrito, Emanuele and Turkeshi, Xhek and Sierant, Piotr},
	journal = {Phys. Rev. Lett.},
	volume = {135},
	issue = {22},
	pages = {220401},
	numpages = {9},
	year = {2025},
	month = {Nov},
	publisher = {American Physical Society},
	doi = {10.1103/1jzy-sk9r},
	url = {https://link.aps.org/doi/10.1103/1jzy-sk9r}
}

@article{Magni2025quantumcomplexity,
	doi = {10.22331/q-2025-12-24-1956},
	url = {https://doi.org/10.22331/q-2025-12-24-1956},
	title = {Quantum {C}omplexity and {C}haos in {M}any-{Q}udit {D}oped {C}lifford {C}ircuits},
	author = {Magni, Beatrice and Turkeshi, Xhek},
	journal = {{Quantum}},
	issn = {2521-327X},
	publisher = {{Verein zur F{\"{o}}rderung des Open Access Publizierens in den Quantenwissenschaften}},
	volume = {9},
	pages = {1956},
	month = dec,
	year = {2025}
}

@misc{huang2026fastexactapproachstabilizer,
	title={A fast and exact approach for stabilizer {R}\'enyi entropy via the {XOR-FWHT} algorithm}, 
	author={Xuyang Huang and Han-Ze Li and Ching Hua Lee and Jian-Xin Zhong},
	year={2026},
	eprint={2512.24685},
	archivePrefix={arXiv},
	primaryClass={quant-ph},
	url={https://arxiv.org/abs/2512.24685}, 
}

@article{Iannotti2025entanglement,
	doi = {10.22331/q-2025-07-21-1797},
	url = {https://doi.org/10.22331/q-2025-07-21-1797},
	title = {Entanglement and {S}tabilizer entropies of random bipartite pure quantum states},
	author = {Iannotti, Daniele and Esposito, Gianluca and Campos Venuti, Lorenzo and Hamma, Alioscia},
	journal = {{Quantum}},
	issn = {2521-327X},
	publisher = {{Verein zur F{\"{o}}rderung des Open Access Publizierens in den Quantenwissenschaften}},
	volume = {9},
	pages = {1797},
	month = jul,
	year = {2025}
}

@article{chitambar2019QuantumResourceTheories,
  title = {Quantum Resource Theories},
  author = {Chitambar, Eric and Gour, Gilad},
  year = {2019},
  month = apr,
  journal = {Review of Modern Physics},
  volume = {91},
  pages = {025001--025001},
  doi = {10.1103/RevModPhys.91.025001}
}

@article{paviglianiti_2025,
  title = {Estimating Nonstabilizerness Dynamics Without Simulating It},
  author = {Paviglianiti, Alessio and Lami, Guglielmo and Collura, Mario and Silva, Alessandro},
  journal = {PRX Quantum},
  volume = {6},
  issue = {3},
  pages = {030320},
  numpages = {17},
  year = {2025},
  month = {Aug},
  publisher = {American Physical Society},
  doi = {10.1103/msm2-vmg7},
  url = {https://link.aps.org/doi/10.1103/msm2-vmg7}
}

@article{qian_2024,
  title = {Augmenting Density Matrix Renormalization Group with Clifford Circuits},
  author = {Qian, Xiangjian and Huang, Jiale and Qin, Mingpu},
  journal = {Phys. Rev. Lett.},
  volume = {133},
  issue = {19},
  pages = {190402},
  numpages = {6},
  year = {2024},
  month = {Nov},
  publisher = {American Physical Society},
  doi = {10.1103/PhysRevLett.133.190402},
  url = {https://link.aps.org/doi/10.1103/PhysRevLett.133.190402}
}

@article{huang_2025,
  title = {Clifford circuits augmented matrix product states for fermion systems},
  author = {Huang, Jiale and Qian, Xiangjian and Qin, Mingpu},
  journal = {Phys. Rev. B},
  volume = {112},
  issue = {20},
  pages = {205106},
  numpages = {7},
  year = {2025},
  month = {Nov},
  publisher = {American Physical Society},
  doi = {10.1103/lwwp-6rqk},
  url = {https://link.aps.org/doi/10.1103/lwwp-6rqk}
}

@inproceedings{gottesman1998HeisenbergRepresentationQuantuma,
  title = {The {{Heisenberg}} Representation of Quantum Computers},
  booktitle = {International {{Conference}} on {{Group Theoretic Methods}} in {{Physics}}},
  author = {Gottesman, Daniel},
  year = {1998},
  url = {http://citeseerx.ist.psu.edu/viewdoc/summary?doi=10.1.1.252.9446}
}

@article{haug2023QuantifyingNonstabilizernessMatrix,
  title = {Quantifying Nonstabilizerness of Matrix Product States},
  author = {Haug, Tobias and Piroli, Lorenzo},
  year = {2023},
  month = jan,
  journal = {Phys. Rev. B},
  volume = {107},
  number = {3},
  pages = {035148},
  publisher = {{American Physical Society}},
  doi = {10.1103/PhysRevB.107.035148}
}

@article{haug2023ScalableMeasuresMagic,
  title = {Scalable {{Measures}} of {{Magic Resource}} for {{Quantum Computers}}},
  author = {Haug, Tobias and Kim, M.S.},
  year = {2023},
  month = jan,
  journal = {PRX Quantum},
  volume = {4},
  number = {1},
  pages = {010301},
  publisher = {{American Physical Society}},
  doi = {10.1103/PRXQuantum.4.010301}
}

@article{howard2017ApplicationResourceTheory,
  title = {Application of a {{Resource Theory}} for {{Magic States}} to {{Fault-Tolerant Quantum Computing}}},
  author = {Howard, Mark and Campbell, Earl},
  year = {2017},
  month = mar,
  journal = {Physical Review Letters},
  volume = {118},
  pages = {090501--090501},
  doi = {10.1103/PhysRevLett.118.090501}
}

@article{leone2021QuantumChaosQuantum,
  title = {Quantum {{Chaos}} Is {{Quantum}}},
  author = {Leone, Lorenzo and Oliviero, Salvatore F. E. and Zhou, You and Hamma, Alioscia},
  year = {2021},
  month = may,
  journal = {Quantum},
  volume = {5},
  pages = {453--453},
  doi = {10.22331/q-2021-05-04-453}
}

@misc{odavic2022ComplexityFrustrationNew,
  title = {Complexity of Frustration: A New Source of Non-Local Non-Stabilizerness},
  shorttitle = {Complexity of Frustration},
  author = {Odavi{\'c}, J. and Haug, T. and Torre, G. and Hamma, A. and Franchini, F. and Giampaolo, S. M.},
  year = {2022},
  month = sep,
  number = {arXiv:2209.10541},
  eprint = {2209.10541},
  eprinttype = {arxiv},
  primaryclass = {cond-mat, physics:quant-ph},
  publisher = {{arXiv}},
  doi = {10.48550/arXiv.2209.10541},
  archiveprefix = {arXiv},
  pages = {2209.10541},
  journaltitle = {}
}

@article{oliviero2022MeasuringMagicQuantum,
  title = {Measuring Magic on a Quantum Processor},
  author = {Oliviero, Salvatore F. E. and Leone, Lorenzo and Hamma, Alioscia and Lloyd, Seth},
  year = {2022},
  month = dec,
  journal = {npj Quantum Inf},
  volume = {8},
  number = {1},
  pages = {1--8},
  publisher = {{Nature Publishing Group}},
  issn = {2056-6387},
  doi = {10.1038/s41534-022-00666-5},
  copyright = {2022 The Author(s)},
  langid = {english}
}

@article{peres_separability,
	title = {Separability Criterion for Density Matrices},
	author = {Peres, Asher},
	journal = {Phys. Rev. Lett.},
	volume = {77},
	issue = {8},
	pages = {1413--1415},
	numpages = {0},
	year = {1996},
	month = {Aug},
	publisher = {American Physical Society},
	doi = {10.1103/PhysRevLett.77.1413},
	url = {https://link.aps.org/doi/10.1103/PhysRevLett.77.1413}
}

@article{zyczkowski_volume,
	title = {Volume of the set of separable states},
	author = {\ifmmode \dot{Z}\else \.{Z}\fi{}yczkowski, Karol and Horodecki, Pawe\l{} and Sanpera, Anna and Lewenstein, Maciej},
	journal = {Phys. Rev. A},
	volume = {58},
	issue = {2},
	pages = {883--892},
	numpages = {0},
	year = {1998},
	month = {Aug},
	publisher = {American Physical Society},
	doi = {10.1103/PhysRevA.58.883},
	url = {https://link.aps.org/doi/10.1103/PhysRevA.58.883}
}

@article{Eisert_comparison_entanglement,
author = {Jens Eisert and Martin B. Plenio},
title = {A comparison of entanglement measures},
journal = {Journal of Modern Optics},
volume = {46},
number = {1},
pages = {145--154},
year = {1999},
publisher = {Taylor \& Francis},
doi = {10.1080/09500349908231260},
}

@article{vidal_computable_entanglement,
	title = {Computable measure of entanglement},
	author = {Vidal, G. and Werner, R. F.},
	journal = {Phys. Rev. A},
	volume = {65},
	issue = {3},
	pages = {032314},
	numpages = {11},
	year = {2002},
	month = {Feb},
	publisher = {American Physical Society},
	doi = {10.1103/PhysRevA.65.032314},
	url = {https://link.aps.org/doi/10.1103/PhysRevA.65.032314}
}

@article{plenio_log_negativity,
	title = {Logarithmic Negativity: A Full Entanglement Monotone That is not Convex},
	author = {Plenio, M. B.},
	journal = {Phys. Rev. Lett.},
	volume = {95},
	issue = {9},
	pages = {090503},
	numpages = {4},
	year = {2005},
	month = {Aug},
	publisher = {American Physical Society},
	doi = {10.1103/PhysRevLett.95.090503},
	url = {https://link.aps.org/doi/10.1103/PhysRevLett.95.090503}
}

@article{frau_2024_magic_entanglement,
	title = {Nonstabilizerness versus entanglement in matrix product states},
	author = {Frau, M. and Tarabunga, P. S. and Collura, M. and Dalmonte, M. and Tirrito, E.},
	journal = {Phys. Rev. B},
	volume = {110},
	issue = {4},
	pages = {045101},
	numpages = {13},
	year = {2024},
	month = {Jul},
	publisher = {American Physical Society},
	doi = {10.1103/PhysRevB.110.045101},
	url = {https://link.aps.org/doi/10.1103/PhysRevB.110.045101}
}

@article{veitch2014ResourceTheoryStabilizer,
  title = {The {{Resource Theory}} of {{Stabilizer Quantum Computation}}},
  author = {Veitch, Victor and Mousavian, S. A. Hamed and Gottesman, Daniel and Emerson, Joseph},
  year = {2014},
  month = jan,
  journal = {New Journal of Physics},
  volume = {16},
  number = {1},
  pages = {013009--013009},
  doi = {10.1088/1367-2630/16/1/013009}
}

@article{Chen_Garcia_Bu_Jaffe_2024, title={Magic of Random Matrix Product States}, volume={109}, ISSN={2469-9950, 2469-9969}, DOI={10.1103/PhysRevB.109.174207}, abstractNote={Magic, or nonstabilizerness, characterizes how far away a state is from the stabilizer states, making it an important resource in quantum computing, under the formalism of the Gotteman-Knill theorem. In this paper, we study the magic of the $1$-dimensional Random Matrix Product States (RMPSs) using the $L_{1}$-norm measure. We firstly relate the $L_{1}$-norm to the $L_{4}$-norm. We then employ a unitary $4$-design to map the $L_{4}$-norm to a $24$-component statistical physics model. By evaluating partition functions of the model, we obtain a lower bound on the expectation values of the $L_{1}$-norm. This bound grows exponentially with respect to the qudit number $n$, indicating that the $1$D RMPS is highly magical. Our numerical results confirm that the magic grows exponentially in the qubit case.}, note={arXiv:2211.10350 [quant-ph]}, number={17}, journal={Physical Review B}, author={Chen, Liyuan and Garcia, Roy J. and Bu, Kaifeng and Jaffe, Arthur}, year={2024}, month=may, pages={174207} }

@article{Garnerone_Oliveira_Haas_Zanardi_2010, title={Statistical properties of random matrix product states}, volume={82}, ISSN={1050-2947, 1094-1622}, DOI={10.1103/PhysRevA.82.052312}, note={arXiv:1003.5253 [quant-ph]}, number={5}, journal={Physical Review A}, author={Garnerone, Silvano and Oliveira, Thiago R. de and Haas, Stephan and Zanardi, Paolo}, year={2010}, month=nov, pages={052312} }

@misc{pz_typicality_mps, title={Typicality in random matrix product states}, url={https://arxiv.org/abs/0908.3877v2}, DOI={10.1103/PhysRevA.81.032336}, journal={arXiv.org}, author={Garnerone, Silvano and de Oliveira, Thiago R. and Zanardi, Paolo}, year={2009}, month=aug }

@article{Schollwoeck_2011, title={The density-matrix renormalization group in the age of matrix product states}, volume={326}, ISSN={00034916}, DOI={10.1016/j.aop.2010.09.012}, abstractNote={The density-matrix renormalization group method (DMRG) has established itself over the last decade as the leading method for the simulation of the statics and dynamics of one-dimensional strongly correlated quantum lattice systems. In the further development of the method, the realization that DMRG operates on a highly interesting class of quantum states, so-called matrix product states (MPS), has allowed a much deeper understanding of the inner structure of the DMRG method, its further potential and its limitations. In this paper, I want to give a detailed exposition of current DMRG thinking in the MPS language in order to make the advisable implementation of the family of DMRG algorithms in exclusively MPS terms transparent. I then move on to discuss some directions of potentially fruitful further algorithmic development: while DMRG is a very mature method by now, I still see potential for further improvements, as exempliﬁed by a number of recently introduced algorithms.}, note={arXiv:1008.3477 [cond-mat]}, number={1}, journal={Annals of Physics}, author={Schollwoeck, Ulrich}, year={2011}, month=jan, pages={96–192}}

@misc{tirrito2023quantifying,
      title={Quantifying non-stabilizerness through entanglement spectrum flatness}, 
      author={Emanuele Tirrito and Poetri Sonya Tarabunga and Gugliemo Lami and Titas Chanda and Lorenzo Leone and Salvatore F. E. Oliviero and Marcello Dalmonte and Mario Collura and Alioscia Hamma},
      year={2023},
      eprint={2304.01175},
      archivePrefix={arXiv},
      primaryClass={quant-ph}
}

@article{Haug2023stabilizerentropies,
  doi = {10.22331/q-2023-08-28-1092},
  url = {https://doi.org/10.22331/q-2023-08-28-1092},
  title = {Stabilizer entropies and nonstabilizerness monotones},
  author = {Haug, Tobias and Piroli, Lorenzo},
  journal = {{Quantum}},
  issn = {2521-327X},
  publisher = {{Verein zur F{\"{o}}rderung des Open Access Publizierens in den Quantenwissenschaften}},
  volume = {7},
  pages = {1092},
  month = aug,
  year = {2023}
}

@article{Leone_Bittel_2024,
  title = {Stabilizer entropies are monotones for magic-state resource theory},
  author = {Leone, Lorenzo and Bittel, Lennart},
  journal = {Phys. Rev. A},
  volume = {110},
  issue = {4},
  pages = {L040403},
  numpages = {6},
  year = {2024},
  month = {Oct},
  publisher = {American Physical Society},
  doi = {10.1103/PhysRevA.110.L040403},
  url = {https://link.aps.org/doi/10.1103/PhysRevA.110.L040403}
}

@article{Leone_Rizzo_Eisert_Jerbi_2025, title={Entanglement theory with limited computational resources}, volume={21}, rights={2025 The Author(s)}, ISSN={1745-2481}, DOI={10.1038/s41567-025-03048-8}, abstractNote={The precise quantification of the limits to manipulating quantum resources lies at the core of quantum information theory. However, standard information-theoretic analyses do not consider the actual computational complexity involved in performing certain tasks. Here we address this issue within the realm of entanglement theory, finding that accounting for computational efficiency substantially changes what can be achieved using entangled resources. We consider two key figures of merit: the computational distillable entanglement and the computational entanglement cost. These measures quantify the optimal rates of entangled bits that can be extracted from or used to dilute many identical copies of n-qubit bipartite pure states, using computationally efficient local operations and classical communication. We demonstrate that computational entanglement measures diverge considerably from their information-theoretic counterparts. Whereas the information-theoretic distillable entanglement is determined by the von Neumann entropy of the reduced state, we show that the min-entropy governs the computationally efficient setting. On the other hand, computationally efficient entanglement dilution requires maximal consumption of entangled bits, even for nearly unentangled states. Furthermore, in the worst-case scenario, even when an efficient description of the state exists and is fully known, one gains no advantage over state-agnostic protocols. Our findings establish sample-complexity bounds for measuring and testing the von Neumann entropy, fundamental limitations on efficient state compression and efficient local tomography protocols.}, number={11}, journal={Nature Physics}, publisher={Nature Publishing Group}, author={Leone, Lorenzo and Rizzo, Jacopo and Eisert, Jens and Jerbi, Sofiene}, year={2025}, month=nov, pages={1847–1854}}

@article{Nguyen_Devakul_Halbasch_Zaletel_Swingle_2018_ent_pur, title={Entanglement of purification: from spin chains to holography}, volume={2018}, ISSN={1029-8479}, DOI={10.1007/JHEP01(2018)098}, abstractNote={Purification is a powerful technique in quantum physics whereby a mixed quantum state is extended to a pure state on a larger system. This process is not unique, and in systems composed of many degrees of freedom, one natural purification is the one with minimal entanglement. Here we study the entropy of the minimally entangled purification, called the entanglement of purification, in three model systems: an Ising spin chain, conformal field theories holographically dual to Einstein gravity, and random stabilizer tensor networks. We conjecture values for the entanglement of purification in all these models, and we support our conjectures with a variety of numerical and analytical results. We find that such minimally entangled purifications have a number of applications, from enhancing entanglement-based tensor network methods for describing mixed states to elucidating novel aspects of the emergence of geometry from entanglement in the AdS/CFT correspondence.}, number={1}, journal={Journal of High Energy Physics}, author={Nguyen, Phuc and Devakul, Trithep and Halbasch, Matthew G. and Zaletel, Michael P. and Swingle, Brian}, year={2018}, month=jan, pages={98} }

@article{Terhal_Horodecki_Leung_DiVincenzo_2002_ent_pur, title={The entanglement of purification}, volume={43}, ISSN={0022-2488, 1089-7658}, DOI={10.1063/1.1498001}, abstractNote={We introduce a measure of both quantum as well as classical correlations in a quantum state, the entanglement of purification. We show that the (regularized) entanglement of purification is equal to the entanglement cost of creating a state ρ asymptotically from maximally entangled states, with negligible communication. We prove that the classical mutual information and the quantum mutual information divided by two are lower bounds for the regularized entanglement of purification. We present numerical results of the entanglement of purification for Werner states in H2⊗H2.}, number={9}, journal={Journal of Mathematical Physics}, author={Terhal, Barbara M. and Horodecki, Michał and Leung, Debbie W. and DiVincenzo, David P.}, year={2002}, pages={4286–4298}}

@article{Tarabunga_Tirrito_2025, title={Magic transition in measurement-only circuits}, volume={11}, rights={2025 The Author(s)}, ISSN={2056-6387}, DOI={10.1038/s41534-025-01104-y}, number={1}, journal={npj Quantum Information}, publisher={Nature Publishing Group}, author={Tarabunga, Poetri Sonya and Tirrito, Emanuele}, year={2025}, month=oct, pages={166}, language={en} }

@article{PRXQuantum.5.030332,
  title = {Dynamical Magic Transitions in Monitored Clifford+$T$ Circuits},
  author = {Bejan, Mircea and McLauchlan, Campbell and B\'eri, Benjamin},
  journal = {PRX Quantum},
  volume = {5},
  issue = {3},
  pages = {030332},
  numpages = {30},
  year = {2024},
  month = {Aug},
  publisher = {American Physical Society},
  doi = {10.1103/PRXQuantum.5.030332},
  url = {https://link.aps.org/doi/10.1103/PRXQuantum.5.030332}
}

@article{tarabunga2024nonstabilizerness,
  title = {Nonstabilizerness via Matrix Product States in the Pauli Basis},
  author = {Tarabunga, Poetri Sonya and Tirrito, Emanuele and Ba\~nuls, Mari Carmen and Dalmonte, Marcello},
  journal = {Phys. Rev. Lett.},
  volume = {133},
  issue = {1},
  pages = {010601},
  numpages = {7},
  year = {2024},
  month = {Jul},
  publisher = {American Physical Society},
  doi = {10.1103/PhysRevLett.133.010601},
  url = {https://link.aps.org/doi/10.1103/PhysRevLett.133.010601}
}

@misc{iannotti2026nonlocalmagicresourcesfermionic,
      title={Non-Local Magic Resources for Fermionic Gaussian States}, 
      author={Daniele Iannotti and Beatrice Magni and Riccardo Cioli and Alioscia Hamma and Xhek Turkeshi},
      year={2026},
      eprint={2604.27049},
      archivePrefix={arXiv},
      primaryClass={quant-ph},
      url={https://arxiv.org/abs/2604.27049}, 
}

@article{magni2026anticoncentration,
  title = {Anticoncentration and state design of doped real Clifford circuits and tensor networks},
  author = {Magni, Beatrice and Heinrich, Markus and Leone, Lorenzo and Turkeshi, Xhek},
  journal = {Phys. Rev. A},
  volume = {113},
  issue = {6},
  pages = {062446},
  numpages = {12},
  year = {2026},
  month = {Jun},
  publisher = {American Physical Society},
  doi = {10.1103/w6l4-mh6k},
  url = {https://link.aps.org/doi/10.1103/w6l4-mh6k}
}

@misc{turkeshi2026lecturenotesreplicatensor,
      title={Lecture Notes on Replica Tensor Networks for Random Quantum Circuits}, 
      author={Xhek Turkeshi},
      year={2026},
      eprint={2605.11150},
      archivePrefix={arXiv},
      primaryClass={quant-ph},
      url={https://arxiv.org/abs/2605.11150}, 
}

@misc{xiao2026exponentiallyacceleratedsamplingpauli,
      title={Exponentially Accelerated Sampling of Pauli Strings for Nonstabilizerness}, 
      author={Zhenyu Xiao and Shinsei Ryu},
      year={2026},
      eprint={2601.00761},
      archivePrefix={arXiv},
      primaryClass={quant-ph},
      url={https://arxiv.org/abs/2601.00761}, 
}

@misc{sierant2026computingquantummagicstate,
      title={Computing quantum magic of state vectors}, 
      author={Piotr Sierant and Jofre Vallès-Muns and Artur Garcia-Saez},
      year={2026},
      eprint={2601.07824},
      archivePrefix={arXiv},
      primaryClass={quant-ph},
      url={https://arxiv.org/abs/2601.07824}, 
}

@article{perfectPauliSamplingMPS,
  title = {Nonstabilizerness via Perfect {P}auli Sampling of Matrix Product States},
  author = {Lami, Guglielmo and Collura, Mario},
  journal = {Phys. Rev. Lett.},
  volume = {131},
  issue = {18},
  pages = {180401},
  numpages = {6},
  year = {2023},
  month = {Oct},
  publisher = {American Physical Society},
  doi = {10.1103/PhysRevLett.131.180401},
  url = {https://link.aps.org/doi/10.1103/PhysRevLett.131.180401}
}

@article{ITensor,
	title={{The ITensor Software Library for Tensor Network Calculations}},
	author={Matthew Fishman and Steven R. White and E. Miles Stoudenmire},
	journal={SciPost Phys. Codebases},
	pages={4},
	year={2022},
	publisher={SciPost},
	doi={10.21468/SciPostPhysCodeb.4},
	url={https://scipost.org/10.21468/SciPostPhysCodeb.4},
}

@book{Sachdev_2011, 
    place={Cambridge}, 
    edition={2}, 
    title={Quantum Phase Transitions}, 
    publisher={Cambridge University Press}, 
    author={Sachdev, Subir}, 
    year={2011}
}

@article{Lieb1961,
  author  = {Lieb, Elliott and Schultz, Theodore and Mattis, Daniel},
  title   = {Two Soluble Models of an Antiferromagnetic Chain},
  journal = {Annals of Physics},
  volume  = {16},
  pages   = {407--466},
  year    = {1961}
}

@article{Niemeijer1967,
title = {Some exact calculations on a chain of spins 1/2},
journal = {Physica},
volume = {36},
number = {3},
pages = {377-419},
year = {1967},
issn = {0031-8914},
doi = {https://doi.org/10.1016/0031-8914(67)90235-2},
url = {https://www.sciencedirect.com/science/article/pii/0031891467902352},
author = {Th. Niemeijer}
}

@book{Franchini2017,
   title={An Introduction to Integrable Techniques for One-Dimensional Quantum Systems},
   ISBN={9783319484877},
   ISSN={1616-6361},
   url={http://dx.doi.org/10.1007/978-3-319-48487-7},
   DOI={10.1007/978-3-319-48487-7},
   journal={Lecture Notes in Physics},
   publisher={Springer International Publishing},
   author={Franchini, Fabio},
   year={2017} }

@article{Mbeng_2024,
	title = {The quantum {I}sing chain for beginners},
	pages = {82},
	author = {Mbeng, Glen Bigan and Russomanno, Angelo and Santoro, Giuseppe E.},
	journal = {SciPost Phys. Lect. Notes},
	year = {2024},
	publisher = {SciPost},
	doi = {10.21468/SciPostPhysLectNotes.82},
	url = {https://scipost.org/10.21468/SciPostPhysLectNotes.82}
}

@article{DePasquale_Facchi_09,
  title = {$XY$ model on the circle: Diagonalization, spectrum, and forerunners of the quantum phase transition},
  author = {De Pasquale, Antonella and Facchi, Paolo},
  journal = {Phys. Rev. A},
  volume = {80},
  issue = {3},
  pages = {032102},
  numpages = {14},
  year = {2009},
  month = {Sep},
  publisher = {American Physical Society},
  doi = {10.1103/PhysRevA.80.032102},
  url = {https://link.aps.org/doi/10.1103/PhysRevA.80.032102}
}

@misc{appcite:Bittel_Eisert_Leone_Mele_Oliviero_2025, title={A complete theory of the {C}lifford commutant}, url={http://arxiv.org/abs/2504.12263}, DOI={10.48550/arXiv.2504.12263}, abstractNote={The Clifford group plays a central role in quantum information science. It is the building block for many error-correcting schemes and matches the first three moments of the Haar measure over the unitary group -a property that is essential for a broad range of quantum algorithms, with applications in pseudorandomness, learning theory, benchmarking, and entanglement distillation. At the heart of understanding many properties of the Clifford group lies the Clifford commutant: the set of operators that commute with $k$-fold tensor powers of Clifford unitaries. Previous understanding of this commutant has been limited to relatively small values of $k$, constrained by the number of qubits $n$. In this work, we develop a complete theory of the Clifford commutant. Our first result provides an explicit orthogonal basis for the commutant and computes its dimension for arbitrary $n$ and $k$. We also introduce an alternative and easy-to-manipulate basis formed by isotropic sums of Pauli operators. We show that this basis is generated by products of permutations -which generate the unitary group commutant- and at most three other operators. Additionally, we develop a graphical calculus allowing a diagrammatic manipulation of elements of this basis. These results enable a wealth of applications: among others, we characterize all measurable magic measures and identify optimal strategies for stabilizer property testing, whose success probability also offers an operational interpretation to stabilizer entropies. Finally, we show that these results also generalize to multi-qudit systems with prime local dimension.}, note={arXiv:2504.12263 [quant-ph]}, number={arXiv:2504.12263}, publisher={arXiv}, author={Bittel, Lennart and Eisert, Jens and Leone, Lorenzo and Mele, Antonio A. and Oliviero, Salvatore F. E.}, year={2025}, month=apr }

@article{appcite:Chen_Garcia_Bu_Jaffe_2024, title={Magic of Random Matrix Product States}, volume={109}, ISSN={2469-9950, 2469-9969}, DOI={10.1103/PhysRevB.109.174207}, abstractNote={Magic, or nonstabilizerness, characterizes how far away a state is from the stabilizer states, making it an important resource in quantum computing, under the formalism of the Gotteman-Knill theorem. In this paper, we study the magic of the $1$-dimensional Random Matrix Product States (RMPSs) using the $L_{1}$-norm measure. We firstly relate the $L_{1}$-norm to the $L_{4}$-norm. We then employ a unitary $4$-design to map the $L_{4}$-norm to a $24$-component statistical physics model. By evaluating partition functions of the model, we obtain a lower bound on the expectation values of the $L_{1}$-norm. This bound grows exponentially with respect to the qudit number $n$, indicating that the $1$D RMPS is highly magical. Our numerical results confirm that the magic grows exponentially in the qubit case.}, note={arXiv:2211.10350 [quant-ph]}, number={17}, journal={Physical Review B}, author={Chen, Liyuan and Garcia, Roy J. and Bu, Kaifeng and Jaffe, Arthur}, year={2024}, month=may, pages={174207} }

@article{appcite:Garnerone_Oliveira_Haas_Zanardi_2010, title={Statistical properties of random matrix product states}, volume={82}, ISSN={1050-2947, 1094-1622}, DOI={10.1103/PhysRevA.82.052312}, note={arXiv:1003.5253 [quant-ph]}, number={5}, journal={Physical Review A}, author={Garnerone, Silvano and Oliveira, Thiago R. de and Haas, Stephan and Zanardi, Paolo}, year={2010}, month=nov, pages={052312} }

@article{appcite:ITensor,
	title={{The ITensor Software Library for Tensor Network Calculations}},
	author={Matthew Fishman and Steven R. White and E. Miles Stoudenmire},
	journal={SciPost Phys. Codebases},
	pages={4},
	year={2022},
	publisher={SciPost},
	doi={10.21468/SciPostPhysCodeb.4},
	url={https://scipost.org/10.21468/SciPostPhysCodeb.4},
}

@article{appcite:Leone_Bittel_2024,
  title = {Stabilizer entropies are monotones for magic-state resource theory},
  author = {Leone, Lorenzo and Bittel, Lennart},
  journal = {Phys. Rev. A},
  volume = {110},
  issue = {4},
  pages = {L040403},
  numpages = {6},
  year = {2024},
  month = {Oct},
  publisher = {American Physical Society},
  doi = {10.1103/PhysRevA.110.L040403},
  url = {https://link.aps.org/doi/10.1103/PhysRevA.110.L040403}
}

@article{appcite:Schoen_Hammerer_Wolf_Cirac_Solano_2007, title={Sequential Generation of Matrix-Product States in Cavity QED}, volume={75}, ISSN={1050-2947, 1094-1622}, DOI={10.1103/PhysRevA.75.032311}, abstractNote={We study the sequential generation of entangled photonic and atomic multi-qubit states in the realm of cavity QED. We extend the work of C. Schoen et al. [Phys. Rev. Lett. 95, 110503 (2005)], where it was shown that all states generated in a sequential manner can be classified efficiently in terms of matrix-product states. In particular, we consider two scenarios: photonic multi-qubit states sequentially generated at the cavity output of a single-photon source and atomic multi-qubit states generated by their sequential interaction with the same cavity mode.}, note={arXiv:quant-ph/0612101}, number={3}, journal={Physical Review A}, author={Schoen, C. and Hammerer, K. and Wolf, M. M. and Cirac, J. I. and Solano, E.}, year={2007}, pages={032311} }

@misc{appcite:ahmad2025experimentaldemonstrationnonlocalmagic,
      title={Experimental demonstration of non-local magic in a superconducting quantum processor}, 
      author={Halima Giovanna Ahmad and Gianluca Esposito and Viviana Stasino and Jovan Odavic and Carlo Cosenza and Alessandro Sarno and Pasquale Mastrovito and Michele Viscardi and Stefano Cusumano and Francesco Tafuri and Davide Massarotti and Alioscia Hamma},
      year={2025},
      eprint={2511.15576},
      archivePrefix={arXiv},
      primaryClass={quant-ph},
      url={https://arxiv.org/abs/2511.15576}, 
}

@article{appcite:chitambar2019QuantumResourceTheories,
  title = {Quantum Resource Theories},
  author = {Chitambar, Eric and Gour, Gilad},
  year = {2019},
  month = apr,
  journal = {Review of Modern Physics},
  volume = {91},
  pages = {025001--025001},
  doi = {10.1103/RevModPhys.91.025001}
}

@article{appcite:elben2018RenyiEntropiesRandom,
  title = {R\'enyi {{Entropies}} from {{Random Quenches}} in {{Atomic Hubbard}} and {{Spin Models}}},
  author = {Elben, A. and Vermersch, B. and Dalmonte, M. and Cirac, J. I. and Zoller, P.},
  year = {2018},
  month = feb,
  journal = {Physical Review Letters},
  volume = {120},
  pages = {050406--050406},
  doi = {10.1103/PhysRevLett.120.050406}
}

@article{appcite:gu_2024_pseudomagic,
  title = {Pseudomagic Quantum States},
  author = {Gu, Andi and Leone, Lorenzo and Ghosh, Soumik and Eisert, Jens and Yelin, Susanne F. and Quek, Yihui},
  journal = {Phys. Rev. Lett.},
  volume = {132},
  issue = {21},
  pages = {210602},
  numpages = {7},
  year = {2024},
  month = {May},
  publisher = {American Physical Society},
  doi = {10.1103/PhysRevLett.132.210602},
  url = {https://link.aps.org/doi/10.1103/PhysRevLett.132.210602}
}

@article{appcite:hayden2006AspectsGenericEntanglement,
  title = {Aspects of {{Generic Entanglement}}},
  author = {Hayden, Patrick and Leung, Debbie W. and Winter, Andreas},
  year = {2006},
  month = mar,
  journal = {Communications in Mathematical Physics},
  volume = {265},
  number = {1},
  pages = {95--117-95--117},
  doi = {10.1007/s00220-006-1535-6}
}

@article{appcite:islam2015MeasuringEntanglementEntropy,
  title = {Measuring Entanglement Entropy in a Quantum Many-Body System},
  author = {Islam, Rajibul and Ma, Ruichao and Preiss, Philipp M. and Eric Tai, M. and Lukin, Alexander and Rispoli, Matthew and Greiner, Markus},
  year = {2015},
  month = dec,
  journal = {Nature},
  volume = {528},
  number = {7580},
  pages = {77--83},
  publisher = {{Nature Publishing Group}},
  issn = {1476-4687},
  doi = {10.1038/nature15750},
  copyright = {2015 Nature Publishing Group, a division of Macmillan Publishers Limited. All Rights Reserved.},
  langid = {english}
}

@article{appcite:lami2023nonstabilizerness,
  title={Nonstabilizerness via perfect {P}auli sampling of matrix product states},
  author={Lami, Guglielmo and Collura, Mario},
  journal={Physical Review Letters},
  volume={131},
  number={18},
  pages={180401},
  year={2023},
  publisher={APS}
}

@article{appcite:leone2021QuantumChaosQuantum,
  title = {Quantum {{Chaos}} Is {{Quantum}}},
  author = {Leone, Lorenzo and Oliviero, Salvatore F. E. and Zhou, You and Hamma, Alioscia},
  year = {2021},
  month = may,
  journal = {Quantum},
  volume = {5},
  pages = {453--453},
  doi = {10.22331/q-2021-05-04-453}
}

@article{appcite:oliviero2022MeasuringMagicQuantum,
  title = {Measuring Magic on a Quantum Processor},
  author = {Oliviero, Salvatore F. E. and Leone, Lorenzo and Hamma, Alioscia and Lloyd, Seth},
  year = {2022},
  month = dec,
  journal = {npj Quantum Inf},
  volume = {8},
  number = {1},
  pages = {1--8},
  publisher = {{Nature Publishing Group}},
  issn = {2056-6387},
  doi = {10.1038/s41534-022-00666-5},
  copyright = {2022 The Author(s)},
  langid = {english}
}

@article{appcite:popescu2006EntanglementFoundationsStatistical,
  title = {Entanglement and the Foundations of Statistical Mechanics},
  author = {Popescu, Sandu and Short, Anthony J. and Winter, Andreas},
  year = {2006},
  month = nov,
  journal = {Nature Physics},
  volume = {2},
  number = {11},
  pages = {754--758},
  doi = {10.1038/nphys444}
}

@misc{appcite:pz_typicality_mps, title={Typicality in random matrix product states}, url={https://arxiv.org/abs/0908.3877v2}, DOI={10.1103/PhysRevA.81.032336}, journal={arXiv.org}, author={Garnerone, Silvano and de Oliveira, Thiago R. and Zanardi, Paolo}, year={2009}, month=aug }

@misc{appcite:sierant2026computingquantummagicstate,
      title={Computing quantum magic of state vectors}, 
      author={Piotr Sierant and Jofre Vallès-Muns and Artur Garcia-Saez},
      year={2026},
      eprint={2601.07824},
      archivePrefix={arXiv},
      primaryClass={quant-ph},
      url={https://arxiv.org/abs/2601.07824}, 
}

@article{appcite:vairogs2025extracting_randomness,
  title = {Extracting randomness from magic quantum states},
  author = {Vairogs, Christopher and Yan, Bin},
  journal = {Phys. Rev. Res.},
  volume = {7},
  issue = {2},
  pages = {L022069},
  numpages = {6},
  year = {2025},
  month = {Jun},
  publisher = {American Physical Society},
  doi = {10.1103/3ttm-vhdt},
  url = {https://link.aps.org/doi/10.1103/3ttm-vhdt}
}

@book{appcite:watrous_2018,
place={Cambridge},
title={The Theory of Quantum Information},
doi={10.1017/9781316848142},
publisher={Cambridge University Press},
author={Watrous, John},
year={2018}
}

@misc{appcite:webb2016clifford,
      title={The {{Clifford}} group forms a unitary 3-design}, 
      author={Zak Webb},
      year={2016},
      eprint={1510.02769},
      archivePrefix={arXiv},
      primaryClass={quant-ph}
}

@misc{appcite:xiao2026exponentiallyacceleratedsamplingpauli,
      title={Exponentially Accelerated Sampling of Pauli Strings for Nonstabilizerness}, 
      author={Zhenyu Xiao and Shinsei Ryu},
      year={2026},
      eprint={2601.00761},
      archivePrefix={arXiv},
      primaryClass={quant-ph},
      url={https://arxiv.org/abs/2601.00761}, 
}

@misc{appcite:Bittel_Leone_2025_operational, title={Operational interpretation of the Stabilizer Entropy}, url={http://arxiv.org/abs/2507.22883}, DOI={10.48550/arXiv.2507.22883}, abstractNote={Magic-state resource theory is a fundamental framework with far-reaching applications in quantum error correction and the classical simulation of quantum systems. Recent advances have significantly deepened our understanding of magic as a resource across diverse domains, including many-body physics, nuclear and particle physics, and quantum chemistry. Central to this progress is the stabilizer R’enyi entropy, a computable and experimentally accessible magic monotone. Despite its widespread adoption, a rigorous operational interpretation of the stabilizer entropy has remained an open problem. In this work, we provide such an interpretation in the context of quantum property testing. By showing that the stabilizer entropy is the most robust measurable magic monotone, we demonstrate that the Clifford orbit of a quantum state becomes exponentially indistinguishable from Haar-random states, at a rate governed by the stabilizer entropy $M(psi)$ and the number of available copies. This implies that the Clifford orbit forms an approximate state $k$-design, with an approximation error $Theta(exp(-M(psi))$. Conversely, we establish that the optimal probability of distinguishing a given quantum state from the set of stabilizer states is also governed by its stabilizer entropy. These results reveal that the stabilizer entropy quantitatively characterizes the transition from stabilizer states to universal quantum states, thereby offering a comprehensive operational perspective of the stabilizer entropy as a quantum resource.}, note={arXiv:2507.22883 [quant-ph]}, number={arXiv:2507.22883}, publisher={arXiv}, author={Bittel, Lennart and Leone, Lorenzo}, year={2025}, month={July} }

@book{appcite:Gour_2025, place={Cambridge}, title={Quantum {R}esource {T}heories}, publisher={Cambridge University Press}, author={Gour, Gilad}, year={2025}}

@article{appcite:Lami_Haug_Nardis_2024_CAMPS,
  title = {Quantum State Designs with {C}lifford-Enhanced Matrix Product States},
  author = {Lami, Guglielmo and Haug, Tobias and De Nardis, Jacopo},
  journal = {PRX Quantum},
  volume = {6},
  issue = {1},
  pages = {010345},
  numpages = {14},
  year = {2025},
  month = {Mar},
  publisher = {American Physical Society},
  doi = {10.1103/PRXQuantum.6.010345},
  url = {https://link.aps.org/doi/10.1103/PRXQuantum.6.010345}
}

@article{appcite:PhysRevLett.108.110503,
  title = {Measuring $\mathrm{Tr}{\ensuremath{\rho}}^{n}$ on Single Copies of $\ensuremath{\rho}$ Using Random Measurements},
  author = {van Enk, S. J. and Beenakker, C. W. J.},
  journal = {Phys. Rev. Lett.},
  volume = {108},
  issue = {11},
  pages = {110503},
  numpages = {5},
  year = {2012},
  month = {Mar},
  publisher = {American Physical Society},
  doi = {10.1103/PhysRevLett.108.110503},
  url = {https://link.aps.org/doi/10.1103/PhysRevLett.108.110503}
}

@article{appcite:Yashin_2025,
   title={Characterization of non-adaptive {C}lifford channels},
   volume={24},
   ISSN={1573-1332},
   url={http://dx.doi.org/10.1007/s11128-025-04682-0},
   DOI={10.1007/s11128-025-04682-0},
   number={3},
   journal={Quantum Information Processing},
   publisher={Springer Science and Business Media LLC},
   author={Yashin, Vsevolod I. and Elovenkova, Maria A.},
   year={2025},
   month=mar }

@article{appcite:cao2024gravitational,
  title={Gravitational back-reaction is the {H}olographic {D}ual of {M}agic},
  author={Cao, ChunJun and Cheng, Gong and Hamma, Alioscia and Leone, Lorenzo and Munizzi, William and Oliviero, Salvatore FE},
  journal={arXiv preprint arXiv:2403.07056},
  year={2024}
}

@article{appcite:haug2023QuantifyingNonstabilizernessMatrix,
  title = {Quantifying Nonstabilizerness of Matrix Product States},
  author = {Haug, Tobias and Piroli, Lorenzo},
  year = {2023},
  month = jan,
  journal = {Phys. Rev. B},
  volume = {107},
  number = {3},
  pages = {035148},
  publisher = {{American Physical Society}},
  doi = {10.1103/PhysRevB.107.035148}
}

@misc{appcite:huang2026fastexactapproachstabilizer,
	title={A fast and exact approach for stabilizer {R}\'enyi entropy via the {XOR-FWHT} algorithm}, 
	author={Xuyang Huang and Han-Ze Li and Ching Hua Lee and Jian-Xin Zhong},
	year={2026},
	eprint={2512.24685},
	archivePrefix={arXiv},
	primaryClass={quant-ph},
	url={https://arxiv.org/abs/2512.24685}, 
}

@article{appcite:itensor-r0.3,
	title={{Codebase release 0.3 for ITensor}},
	author={Matthew Fishman and Steven R. White and E. Miles Stoudenmire},
	journal={SciPost Phys. Codebases},
	pages={4-r0.3},
	year={2022},
	publisher={SciPost},
	doi={10.21468/SciPostPhysCodeb.4-r0.3},
	url={https://scipost.org/10.21468/SciPostPhysCodeb.4-r0.3}
}

@article{appcite:leone2022stabilizer,
  title={Stabilizer {{R}}{\'e}nyi {{E}}ntropy},
  author={Leone, Lorenzo and Oliviero, Salvatore FE and Hamma, Alioscia},
  journal={Physical Review Letters},
  volume={128},
  number={5},
  pages={050402},
  year={2022},
  publisher={APS}
}

@article{appcite:page_avg_entropy,
  title = {Average entropy of a subsystem},
  author = {Page, Don N.},
  journal = {Phys. Rev. Lett.},
  volume = {71},
  issue = {9},
  pages = {1291--1294},
  numpages = {0},
  year = {1993},
  month = {Aug},
  publisher = {American Physical Society},
  doi = {10.1103/PhysRevLett.71.1291},
  url = {https://link.aps.org/doi/10.1103/PhysRevLett.71.1291}
}

@article{appcite:tarabunga2024nonstabilizerness,
  title = {Nonstabilizerness via Matrix Product States in the Pauli Basis},
  author = {Tarabunga, Poetri Sonya and Tirrito, Emanuele and Ba\~nuls, Mari Carmen and Dalmonte, Marcello},
  journal = {Phys. Rev. Lett.},
  volume = {133},
  issue = {1},
  pages = {010601},
  numpages = {7},
  year = {2024},
  month = {Jul},
  publisher = {American Physical Society},
  doi = {10.1103/PhysRevLett.133.010601},
  url = {https://link.aps.org/doi/10.1103/PhysRevLett.133.010601}
}

@misc{appcite:zhu2016CliffordGroupFails,
  title = {The {{Clifford}} Group Fails Gracefully to Be a Unitary 4-Design},
  author = {Zhu, Huangjun and Kueng, Richard and Grassl, Markus and Gross, David},
  year = {2016},
  month = sep,
  number = {arXiv:1609.08172},
  eprint = {1609.08172},
  eprinttype = {arxiv},
  primaryclass = {quant-ph},
  publisher = {{arXiv}},
  doi = {10.48550/arXiv.1609.08172},
  archiveprefix = {arXiv},
  pages = {1609.08172},
  journaltitle = {}
}

\end{document}

\begin{center}
{\bfseries\large Supplemental Material for ``Stabilizer entropy is trustworthy for mixed states''\par}
\vspace{1em}
Gianluca Esposito,$^{1,2}$ Michele Viscardi,$^{3}$ and Alioscia Hamma$^{1,2,4}$\par
\vspace{0.5em}
{\itshape
$^1$Scuola Superiore Meridionale, Largo S. Marcellino 10, 80138 Napoli, Italy\par
$^2$Istituto Nazionale di Fisica Nucleare (INFN) Sezione di Napoli\par
$^3$Dipartimento di Fisica “E.R. Caianiello”, Università di Salerno, Via Giovanni Paolo II, 132, I-84084 Fisciano (SA), Italy\par
$^4$Universit\`a degli Studi di Napoli Federico II, Dipartimento di Fisica Ettore Pancini\par
}
\vspace{0.5em}
(Dated: \today)
\end{center}

\tableofcontents

\section{Preliminaries}\label{app:preliminaries}
\subsection{Resource theory and resource proxies}
The framework of quantum resource theory (QRT) is a valuable tool for quantifying cost of quantum operations in terms of realistic physical constraints in any field \cite{appcite:chitambar2019QuantumResourceTheories,appcite:Gour_2025}. It involves, at its core, categorizing states and operations as either “free'' or “costly'', namely resourceful, with the motivation either due to operational constraints, or the laws of physics themselves. This approach provides a foundation to rigorously compare the amount of resource held in different quantum states or channels and enable a fine-grained analysis of what fundamental processes and properties drive a certain phenomenon \cite{appcite:chitambar2019QuantumResourceTheories,appcite:Gour_2025}. Below a mathematical definition of  a QRT.
 \begin{definition}[Quantum Resource Theory]\label{app:restheory}
Let $\mathcal{O}$ be a mapping that assigns to any two systems $A$ and $B$ with corresponding Hilbert spaces $\hi_A$ and $\hi_B$ a unique set of quantum channels $\mathcal{O}(A\rightarrow B):=\mathcal{O}(\hi_A\rightarrow \hi_B)$ . Let $\mathcal{F}(\hi):=\mathcal{O}(\mathbb{C}\rightarrow \hi)$ with $\hi$ being an arbitrary Hilbert space. Then the tuple $\mathcal{R}=(\mathcal{F},\mathcal{O})$ is called a \textbf{quantum resource theory} (QRT) if the following two conditions hold \cite{appcite:chitambar2019QuantumResourceTheories}:
    \begin{enumerate}
        \item \textbf{(Doing nothing is free)}: For any physical system $A$ the set $\mathcal{O}(A):=\mathcal{O}(A\rightarrow A)$ contains the identity map $\id_A$. 
        \item \textbf{(Concatenation is free)}: For any three physical systems $A,B$ and $C$, if $\mathcal{E} \in \mathcal{O}(A\rightarrow B)$ and $\Lambda\in \mathcal{O}(B\rightarrow C) $, then $\mathcal{E}\circ\Lambda \in \mathcal{O}(A\rightarrow C)$ 
    \end{enumerate}
    Defining $\mathcal{D}(\hi)$ as the set of state density operators (that is, positive operators of trace one), the set $\mathcal{F}(\hi)\subset\mathcal{D}(\hi)$ defines the set of \textbf{free states} acting on $\hi$, whereas elements belonging to $\mathcal{D}(\hi)/\mathcal{F}(\hi)$ are called \textbf{resource states}.
    In a similar fashion, the quantum channels in $\mathcal{O}(A\rightarrow B)$ are called \textbf{free operations} whereas the quantum channels that do not belong to $\mathcal{O}(A\rightarrow B)$ are called \textbf{resourceful operations}.
\end{definition}
Definition \ref{app:restheory} can be understood from the following physical standpoint: imagine a quantum system under the control of an individual or divided among multiple parties. A QRT defines the feasible actions that can be taken by these parties while abiding by certain limitations or restrictions that stem from technical or experimental boundaries, the rules of a game, or the fundamental laws of physics. The feasible  operations are described by $\mathcal{O}(A\rightarrow B)$, typically a smaller set than the set of all quantum channels. Condition 1 in Def. \ref{app:restheory} states that doing nothing is a free operation, which is a sensible requirement for any QRT, whereas condition 2 implies that $\Lambda\circ\mathcal{E}$ is free whenever $\mathcal{E}$ and $\Lambda$ are both free: this assures that the operations in $\mathcal{O}$ are truly free to perform in any order and any number of times. 

As a consequence one cannot convert a free state into a resource state using only free operations: this concept is usually referred to as \textit{the golden rule of QRTs} \cite{appcite:chitambar2019QuantumResourceTheories}, and it will be of particular importance in the definition of resource proxies. 
\begin{prop}[Golden rule of QRTs]
For any two physical systems $A$ and $B$, if $\mathcal{E}\in \mathcal{O}(A\rightarrow B)$ and $\psi\in \mathcal{F}(A)$, then $\mathcal{E}(\psi)\in \mathcal{F}(B)$.    
\end{prop}
Moreover, another useful definition is QRTs with a tensor product structure:
\begin{definition}
    A QRT $\mathcal{R}=(\mathcal{F},\mathcal{O})$ is said to admit a tensor product structure if the following three conditions hold:
    \begin{enumerate}
        \item For any three physical systems $A,B$ and $C$, if $\mathcal{E}\in \mathcal{O}(A\rightarrow B)$ then $\id_C\otimes \mathcal{E} \in \mathcal{O}(C A\rightarrow C B)$, where $\id_C$ is the identity map on $\mathcal{B}(C)$.
        \item Appending free states is a free operation: for any free state $\sigma\in\mathcal{F}(B)$, the quantum channel $\mathcal{E}_\sigma(\psi):=\psi\otimes\sigma$ belongs to $\mathcal{O}(A\rightarrow AB)$.
        \item Discarding a system is a free operation: for any Hilbert space $\hi$, the set $\mathcal{O}(\hi\rightarrow \mathbb{R})$ is not empty.
Noticing that the set $\mathcal{B}{\hi\rightarrow \mathbb{R}}$ contains only the trace, the last statement is equivalent to state that tracing a system is a free map \cite{appcite:chitambar2019QuantumResourceTheories}.
\end{enumerate}
\end{definition}
To really give significance to a QRT, an operationally meaningful way to quantify the amount of resource in a state or a map is needed. The concept of resource monotones answers this need.
\begin{definition}[Resource monotones]
A function $f:\mathcal{D}(\hi)\rightarrow \mathbb{R}^+$ is called a \textbf{resource monotone} or \textbf{resource measure} for the QRT $\mathcal{R}=(\mathcal{F},\mathcal{O})$ if
\begin{enumerate}
    \item \textbf{(Vanishing on free states)}: $\psi\in \mathcal{F}\implies f(\psi)=0$. If also the converse is true, namely $f(\psi)=0\implies \psi \in \mathcal{F}$, then $f$ is called \textbf{faithful}.
    \item \textbf{(Monotone for free operations)}: For every $ \mathcal{E}\in \mathcal{O}$ and $\psi \in \mathcal{D}(\hi)$, it holds that 
    \begin{equation}
        f(\mathcal{E}(\psi))\leq f(\psi)\,.
    \end{equation}
\end{enumerate}
\end{definition}
The second property is a quantitative enforcement of the golden rule of QRTs. 

It may happen that some function $f$ does not satisfy the requisite for monotonicity for every state, but that one would still want to keep using said function to quantify a resource, either for practical reasons, ease of use, or experimental accessibility. One can then relax the monotonicity condition, upon reassurance that the instances of violations of this condition are rare in the ensemble of states one intends to use such function. To formalize this in the context of qubit systems, we introduce the new concept of \textit{resource proxy}:
\begin{definition}[Resource proxy]\label{app:proxy_def}
    Given a $N-$qubit system Hilbert space $\hi=\mathbb{C}^{2\ot N}$, a QRT $\mathcal{R}=(\mathcal{F},\mathcal{O})$ and an ensemble of states $\mathcal{S}\subseteq \mathcal{D}(\hi)$, a function $f:\mathcal{D}(\hi)\rightarrow \mathbb{R}^+$ is called a $\boldsymbol{\eta}-$\textbf{resource proxy} in the ensemble $\mathcal{S}$ if 
    \begin{enumerate}
        \item It is faithful in the ensemble $\mathcal{S}$;
        \item For every free operation $\mathcal{E}\in \mathcal{O}$ and $\psi \in \mathcal{S}$,
        \begin{equation}
            {\rm Pr}\Big[f[\mathcal{E}(\psi)]-f(\psi)>0\Big]\leq \exp(-\eta N)\;,\eta>0
        \end{equation}
    \end{enumerate}
\end{definition}
 In other words, $f$ is a resource proxy in a state ensemble if it behaves as a monotone with overwhelming probability in that ensemble, i.e. the violations of monotonicity are extremely rare. While not strictly qualifying for the requisites of resource monotones, we believe resource proxies can be of great utility, especially when there are no strict monotones with other practical features like computability or experimental accessibility available. 
\subsection{Non-stabilizerness and stabilizer entropy}
Given a $N-$qubit system with $\hi=\mathbb{C}^{2\otimes N}$ and $d=\dim(\hi)=2^N$, the \textit{Pauli group} $\Tilde{\mathbb{P}}_N$ is defined as
\begin{equation}
    \Tilde{\mathbb{P}}_N=\{\pm 1,\pm i\}\times \{\id,X,Y,Z\}^{\ot N}\,,
\end{equation}
while in this work we refer to the Pauli group modulo phases as $\mathbb{P}_n:=\Tilde{\pauli}_n/\{\pm 1,\pm i\}$ and its elements as \textit{Pauli strings}. We denote the Clifford group as the normalizer of the Pauli group, namely 
\begin{equation}
     \mathcal{C}(d):=\{C\in U(d)\,|\, CPC^\dag=P'\in \pauli(d)\,,\forall P\in \pauli(d)\}\,.
 \end{equation}
 The set of (pure) \textit{stabilizer states} is given by the action of the Clifford group through the computational basis (eigenstates of the $Z$ operator):
\begin{equation}
    \pstab=\left\{C\ket{i},C\in \mathcal{C}(d)\right\}\,.
\end{equation}
Analogously, a pure stabilizer state can be defined as the common eigenstate of $d$ mutually commuting Pauli strings. We define mixed stabilizer states $\stabzero$ as the equal-weight convex combination of orthogonal pure stabilizer states \cite{appcite:cao2024gravitational}. They can alternatively be written as \cite{appcite:leone2022stabilizer}
\begin{equation}
    \stabzero=\left\{\rho=\frac{1}{d}\sum_{P \in G} P\;,\; G\; \text{abelian}\right\}\,.
\end{equation}
Notably, $\pstab\subset\stabzero$ and $\stabzero$ can be obtained starting from $\pstab$ by acting with partial trace and dephasing onto stabilizer bases. These are the non-unitary maps we are going to focus on in this work. We are now ready to define the QRTs of interest of this work, which we will now dub \textit{non-adaptive stabilizer computation} \cite{appcite:Yashin_2025}: the free set $\free$ is previously defined set of mixed stabilizer states $\stabzero$, and free operations $\freeop$ is constituted by: 
\begin{itemize}
    \item Clifford unitaries
    \item Partial trace over a subsystem
    \item Measurements in the computational basis
    \item Appending stabilizer states
\end{itemize}
Notably, these operations are the only CPTP maps that admit a Stinespring dilation consisting of a Clifford unitary, and whose Choi state is a stabilizer state \cite{appcite:Yashin_2025}.

We will use the family of stabilizer entropies as defined in \cite{appcite:leone2022stabilizer}, which are the the unique computable measures of magic for the pure magic-state resource theory. In particular, we focus on $M_2$ and its linear counterpart, ${ M_{\rm lin}}$. 
Linear entropies are of particular importance from the theoretical, computational and experimental point of view. First, one can assess typicality in states ensembles by analyzing statistical moments of such resources \cite{appcite:popescu2006EntanglementFoundationsStatistical,appcite:leone2021QuantumChaosQuantum}; second, the resource inducing properties of quantum maps can be discovered by computing the resource power, even analytically. Most importantly, linear entropies are the only quantities that can be directly accessed experimentally\cite{appcite:PhysRevLett.108.110503,appcite:elben2018RenyiEntropiesRandom,appcite:islam2015MeasuringEntanglementEntropy,appcite:oliviero2022MeasuringMagicQuantum,appcite:ahmad2025experimentaldemonstrationnonlocalmagic} since only they can be expressed by the expectation values of Hermitian operators, without the need for full state tomography.

The linearized SE, $ M_{\rm lin}$ is an example of the desirable properties of linear entropies described so far. In addition, $M_{\rm lin}$ plays a key part in the generation of $\varepsilon$-approximate state $t$-designs \cite{appcite:vairogs2025extracting_randomness}.

 Starting from a pure state $\psi$, one defines the probability distribution $\{\Xi_P(\psi):=\frac{1}{d}\Tr^2(P\psi)\}_{P\in \pauli}$. Then, the $\alpha-$Stabilizer R\'enyi entropy is defined as the $\alpha-$R\'enyi entropy of this distribution \cite{appcite:leone2022stabilizer}
 \begin{equation}
     M_\alpha(\psi)=\frac{1}{1-\alpha}\log d^{-1}\left[\sum_P\Xi_P(\psi)^\alpha\right]\,.
 \end{equation}
 In particular, for $\alpha=2$ one gets the following expression:
 \begin{equation}
     M_2(\psi)=-\log\left[d^{-1}\sum_P \Tr^4(P\psi)\right]=-\log\left[d\Tr(Q\psit)\right]=-\log[\stp(\psi)]\,,
 \end{equation}
 with $\stp(\psi):=d\Tr[Q\psit]$, $Q:=\frac{1}{d^2}\sum_{P\in \pauli} P^{\ot 4}$. Linear 2-SE then reads 
 \begin{equation}
     M_{\lin}^{(2)}=1-\stp(\psi)\,.
 \end{equation}
 2-Stabilizer R\'enyi Entropy has been extended to mixed states in the original paper of \cite{appcite:leone2022stabilizer} in the following way:
 \begin{equation}
     \Tilde{M}_2(\psi)=M_2(\psi)-S_2(\psi)=\log[\pur(\psi)]-\log[\stp(\psi)]
     \label{app:eq:M_2_mixed_states_def}
 \end{equation}
 Even if $\Tilde{M}_2(\psi)$ loses the significance of distillable magic for mixed states, it can be made a good monotone for magic in the mixed state context as well, via convex roof extensions \cite{appcite:Leone_Bittel_2024}, though it loses the ease of computing due to the optimization procedure required to obtain it. 
 Moreover, the null set of $\Tilde{M}_2(\psi)$ is exactly $\stabzero$, since for this class of states it is straightforward to see that $\stp(\psi)=\pur(\psi)=\frac{|G|}{d}$. This is the motivation for choosing that ensemble as our ensemble of free states.
\subsection{Linear stabilizer entropy for mixed states}
Inspired by the extension of $M_2$ to mixed states, we define a new extension of linear stabilizer entropy for mixed states:
\begin{equation}\label{app:mlinmixed}
    \mlin(\psi)=\pur(\psi)-\stp(\psi)
\end{equation}
which retains most of the desirable properties of its logarithmic counterpart, namely:
\begin{itemize}
    \item \textbf{Faithfulness}: $\mlin(\psi)=0 \iff \psi \in {\rm STAB _0}$, otherwise $\mlin(\psi)>0$.
    \item \textbf{Monotonicity for pure-state free operations}, namely Clifford unitaries and appending free states. In particular,
    \begin{equation}
        \begin{split}
            \mlin(C\psi C^\dag)&=\mlin(\psi)\\
            \mlin(\psi\ot \sigma)&=\pur(\psi)\pur(\sigma)-\stp(\psi)\stp(\sigma)\\
            &=\frac{|G_\sigma|}{d}(\pur(\psi)-\stp(\psi))\leq \mlin(\psi)
        \end{split}
    \end{equation}
    since for every $\sigma \in \stabzero$ it holds that $\pur(\sigma)=\stp(\sigma)={|G_\sigma|}/{d}$ with $|G_\sigma|\leq d$.
\end{itemize}
Moreover, $\mlin$ has the following relationship with its logarithmic counterpart, $\Tilde{M}_2$:
\begin{equation}\label{app:m2mlin_rel_Eq}
    \begin{split}
        \Tilde{M}_2(\psi)&\geq \mlin(\psi) \quad\forall \psi \in \mathcal{D}(\hi)\,;\\
       \Tilde{M}_2(\psi)&\geq \log_2[\mlin(\psi)] \quad\forall \psi \in \mathcal{D}(\hi)\,.
    \end{split}
\end{equation}
Proof of these inequalities can be found in Sec.  \ref{app:App_mlinrel}. 
In the rest of this work, we will prove that $\mlin$ is a good resource proxy in different state ensembles.
\subsection{The SE gap for quantum maps}
To quantify the violations of monotonicity of $\mlin$ under a certain free map $\mathcal{E}\in \freeop$, we define the following random variable, called \textit{SE gap}, on the ensemble of states we are interested in:
\begin{equation}
    \mbar^{\mathcal{E}}(\psi):=\mlin(\psi)-\mlin(\mathcal{E}(\psi))
\end{equation}
If $\mlin$ was an actual monotone, then $\mbar^{\mathcal{E}}(\psi)\geq 0\; \forall \psi \in \mathcal{D}(\hi)$ for every free map $\mathcal{E}$, whereas a negative SE gap for some $\psi$ would constitute a violation of monotonicity. In the following sections, we are going to compute the average SE gap, for partial trace and dephasing onto the computational basis, along with an assessment of typicality to confidently state that $\mlin$ is a good magic proxy with respect to these free operations, i.e. that $\mlin$ is a good monotone with overwhelming probability.

\section{\texorpdfstring{$\mlin$}{} and relationship with \texorpdfstring{$\Tilde{M}_2$}{}}\label{app:App_mlinrel}

In this section, we will show the proof of Eq. \eqref{app:m2mlin_rel_Eq}. First, we prove that
\begin{equation}
    \Tilde{M}_2(\psi)\geq \mlin(\psi) \quad\forall \psi \in \mathcal{D}(\hi)
\end{equation}
\textbf{Proof:} First, notice that $\pur(\psi)\geq \stp(\psi)$ for every state, due to the purity being the second moment of the Pauli spectrum and the stabilizer purity being the fourth one. Putting $\pur(\psi)\equiv x$ and $\stp(\psi)\equiv y$, proving Eq. \eqref{app:m2mlin_rel_Eq} amounts to prove the following system of inequalities:
\begin{equation}
    \begin{cases}
    \log_2(\frac xy)\geq x-y\\
    \frac{2}{d+1}\leq y \leq x \leq 1
    \end{cases}\,.
\end{equation}
Define the following function:
\begin{equation}
    f(t)=\log(t)-\log(2)t\,,\quad t\in [0,1]\,.
\end{equation}
Its derivative $f'(t)=\frac1t-\log(2)$ is strictly positive for $0<t<1$, hence $f(t)$ is strictly increasing in such interval. Hence, if $x>y$ it follows that,
\begin{equation}
    \begin{split}
        \log(x)-\log(2)x&>\log(y)-\log(2)y\\
        \implies \log(x)-\log(y)&>\log(2)(x-y)\\
        \implies \log(\frac xy)&>\log(2)(x-y)\\
        \implies \log_2\left(\frac xy\right)&>(x-y)
    \end{split}
\end{equation} \qed

Next, we prove that
\begin{equation}
     \Tilde{M}_2(\psi)\geq \log_2[\mlin(\psi) ]\quad\forall \psi \in \mathcal{D}(\hi)
\end{equation}
\textbf{Proof:} using the same positions as before, it amounts to solving the following inequalities:
\begin{equation}
    \begin{cases}
        \log_2(\frac xy)\geq \log_2(x-y)\\
         \frac{2}{d+1}\leq y \leq x \leq 1\,.
    \end{cases}
\end{equation}
Since the logarithm is strictly increasing, this amounts to solving 
\begin{equation}
    \frac xy \geq x-y \,.
\end{equation}
Multiplying by $y$, we get
\begin{equation}
    x\geq xy-y^2\implies x(1-y)+y^2\geq 0
\end{equation}
which is always true when $x>0$, $y>0$ and $y<1$.\qed

\section{Some useful lemmas}\label{app:lemmas}
In this section we collect a series of useful lemmas used in the calculations throughout the paper.

\begin{lemma}
    Given a random variable $X$ defined on a Hilbert space of dimension $d$, assuming $\exv_\psi X=O(1)$ and $\exv_\psi X >0$ (or larger scaling in $d$) and L\'evy concentration of $X$ around its mean value, namely that for every $\epsilon>0$ then 
        \begin{equation}
            \begin{split}
                {\rm Pr}(X-\exv(X)\geq \epsilon)&\leq 2\exp[-cd\epsilon^2]\\
                {\rm Pr}(X-\exv(X)\leq- \epsilon)&\leq 2\exp[-cd\epsilon^2]\,.
            \end{split}
        \end{equation}
       also the random variable $-\log(X)$ also exhibits L\'evy concentration around $-\log\exv (Y)$ in the following terms:
    \begin{equation}
        \begin{split}
            {\rm Pr}&[-\log(X)+\log(\exv (X))\geq \epsilon]\leq 2\exp[-c'd\epsilon^{ \prime 2}]\\
            {\rm Pr}&[-\log(X)+\log(\exv (X))\leq -\epsilon]\leq 2\exp[-c'd\epsilon^{\prime \prime 2}]
        \end{split}
    \end{equation}
    with $c',\epsilon',\epsilon''$ related to the $c,\epsilon$ typicality constants and error of $X$ by the following relations:
    \begin{equation}
        \begin{split}
            c'&=c\exv(X)^2 \\
            \epsilon'&=1-\exp(-\epsilon)=O(\epsilon)\\
            \epsilon''&=\exp(\epsilon)-1=O(\epsilon)
        \end{split}
    \end{equation}
    
    \begin{proof}
    Starting from the first statement,
    \begin{equation}
        \begin{split}
             {\rm Pr}&[-\log(X)+\log(\exv (X))\geq \epsilon]= {\rm Pr}\left[-\log(\frac{X}{\exv (X)})\geq \epsilon\right]={\rm Pr}\left[\log(\frac{X}{\exv (X)})\leq -\epsilon\right]\\
             {\rm Pr}&\left[X\leq \exv(X)\exp(-\epsilon)\right]=  {\rm Pr}\left[X-\exv(X)\leq -\exv(X)\underbrace{(\exp(-\epsilon)-1)}_{\epsilon'}\right]\overset{(a)}{\leq}2\exp(-cd\,\exv(X)^2\epsilon^{\prime\, 2})\\
             &=2\exp(-c'd\epsilon^{\prime\, 2})
        \end{split}
    \end{equation}
    where in $(a)$ we used the hypothesis statement of L\'evy concentration of $X$. Analogously for the first statement we derive the second one:
    \begin{equation}
        \begin{split}
            {\rm Pr}&\left[-\log(X)+\log[\exv(X)]\leq -\epsilon\right]={\rm Pr}\left[-\log(\frac{X}{\exv(X)})\leq -\epsilon\right]
            {\rm Pr}\left[\log(\frac{X}{\exv(X)})\geq \epsilon\right]\\={\rm Pr}&\left[X\geq\exv(X) \exp(\epsilon)\right]={\rm Pr}\left[X-\exv(X)\geq\exv(X) \underbrace{(\exp(\epsilon)-1)}_{\epsilon''}\right]\overset{(a)}{\leq}2\exp(-cd\,\exv(X)^2\epsilon^{\prime \prime \,2})\\
             &=2\exp(-c'd\epsilon^{\prime \prime\,  2})
        \end{split}
    \end{equation}
    \end{proof}
    \end{lemma}
    Using the former lemma applied to $X=\Tr[\psi_A ^{\alpha}]$ and $X=d^{-1}\sum_P \Tr^{2\alpha}(P\psi)$, using the proof of concentration of the last object for odd $\alpha$ \cite{appcite:gu_2024_pseudomagic} and for $\alpha=2$ \cite{appcite:zhu2016CliffordGroupFails}, one can say that $\alpha-$R\'enyi entropies and $\alpha-$stabilizer R\'enyi entropies both concentrate tightly around their mean value.

\begin{lemma}\label{app:dephasinglemma}
    Given the computational basis $\{\ket i\}$, the dephasing channel over said basis can be written as
    \begin{equation}
        D_B\psi=\frac{1}{d}\sum_{P_z \in \mathbb{Z}_2}P_z \psi P_z\,,
    \end{equation}
    where $\mathbb{Z}_2$ is the (abelian) subgroup of Pauli operator belonging to $\{\id,Z\}^{\ot n}$.
    \proof By induction. Let us begin by recalling the definition of the dephasing channel:
    \begin{equation}
        D_B\psi=\sum_{i\in \{0,1\}^{\times n}}\Pi_i \psi \Pi_i\,.
    \end{equation}
 Let us prove the inductive step for $n=1$ qubit:
    \begin{equation}
       \begin{split}
            D_B ^{(1)}\psi&=\ketbra{0}\psi\ketbra{0}+\ketbra{1}\psi\ketbra{1}\\&=\frac{1}{4}\left[(\id+Z)\psi(\id+Z)+(\id-Z)\psi(\id-Z)\right]\\
            &=\frac{1}{4}(\psi+Z\psi+\psi Z+Z\psi Z+\psi-Z\psi-\psi Z +Z\psi Z)=\frac12 (\psi+Z\psi Z)\,,
       \end{split}
    \end{equation}
    where we used the fact that each single qubit projector onto the computational basis can be written as 
    \begin{equation}
        \Pi_i=\ketbra{i}=\frac{1}{2}(\id_1+(-1)^i Z)\;,i\in \{0,1\};
    \end{equation}
    thus, a $n$-qubit projector onto a computational basis state $\Pi_i$ can then be written as 
    \begin{equation}
        \Pi_i=\bigotimes_{k=1}^n\Pi_{i_k}=\frac{1}{d}\bigotimes_{k=1}^n\left(\id_k+(-1)^{i_k}Z_k\right)\,.
    \end{equation}
   Now, supposing the thesis is true for $n-1$, namely that
   \begin{equation}
       D_B ^{(n-1)}\psi=\frac{1}{2^{n-1}}\sum_{P_z ^{(n-1)} \in\mathbb{Z}_2 ^{(n-1)}}P_z ^{(n-1)}\psi P_z ^{(n-1)}\,,
   \end{equation}
    the dephasing over $n$ can be written as
    \begin{equation}
        \begin{split}
            D_B ^{(n)}\psi&=\sum_{i_n=0}^1(\id_{n-1}\otimes \Pi_{i_n})\sum_{i_1,\dots,i_{n-1}=0}^1(\Pi_{i_1}\otimes\dots\otimes\Pi_{i_{n-1}}\otimes \id_n)\psi(\Pi_{i_1}\otimes\dots\otimes\Pi_{i_{n-1}}\otimes \id_n)(\id_{n-1}\otimes \Pi_{i_n})\\
            &=\sum_{i_n=0}^1(\id_{n-1}\otimes \Pi_{i_n})(D_B ^{(n-1)}\otimes \id_n)(\psi)(\id_{n-1}\otimes \Pi_{i_n})\\
            &=(D_B ^{(n-1)}\otimes D_B ^{(1)})(\psi)=\frac{1}{2^n}\sum_{P_z ^{(n-1)} \in\mathbb{Z}_2 ^{(n-1)}}\sum_{P_z ^{(1)}\in\mathbb{Z}_2 ^{(1)}}(P_z ^{(n-1)}\otimes P_z ^{(1)})\psi(P_z ^{(n-1)}\otimes P_z ^{(1)})\\
            &=\frac1d\sum_{P_z ^{(n)}\in \mathbb{Z}_2 ^{(n)}}P_z ^{(n)}\psi P_z ^{(n)}\,,
           %&=\frac{1}{2^{n-1}}\sum_{i_n=0}^1(\id_{n-1}\otimes \Pi_{i_n})\sum_{P_z ^{(n-1)} \in\mathbb{Z}_2 ^{(n-1)}}(P_z ^{(n-1)}\ot \id_n)\psi (P_z ^{(n-1)}\ot \id_n)
        \end{split}
    \end{equation}
    as $Z_2 ^{(n-1)}\otimes Z_2^{(1)}=Z_2 ^{(n)}$. 
\qed \end{lemma}
\begin{lemma}
   Given the set of $n-$qubit Pauli strings, the following equality holds:
   \begin{equation}
       \sum_{P,P'\in \pauli}\Tr[PP'PP']=d^3
   \end{equation}
   \proof Using the fact that $\Tr[AB]=\Tr[\ttt (A\otimes B)]$, and that the swap can be written as
   \begin{equation}
       \ttt=\frac{1}{d}\sum_P P\ot P\,,
   \end{equation} we can write
   \begin{equation}
       \begin{split}
           \sum_{P,P'\in \pauli}\Tr[PP'PP']&=\sum_{P,P'\in\pauli}\Tr[\ttt(PP'\otimes PP')]=\sum_{P,P'\in\pauli}\Tr\\\
           &=d^2\Tr[\ttt^3]=d^2\Tr[\ttt]=d^2 d=d^3\,,
       \end{split}
   \end{equation}
   where we used the fact that $\ttt^2=\id$ and that $\Tr[\ttt]=d$.
\qed \end{lemma}
\begin{cor}
    Given the set of $n-$qubit Pauli strings and the $\mathbb{Z}_2$ set of strings, the following equality holds:
    \begin{equation}
        \sum_{P\in \mathbb{P}}\sum_{P_z\in \zz}\Tr[PP_z P P_z]=d^3
    \end{equation}
    \proof Again, using the swap trick, we can write 
    \begin{equation}
        \begin{split}
            \sum_{P\in \mathbb{P}}\sum_{P_z\in \zz}\Tr[PP_z P P_z]&=\sum_{P\in \mathbb{P}}\sum_{P_z\in \zz}\Tr\{\ttt[(P\otimes P)(P_z\otimes P_z)]\}\\
            &=d\sum_{P_z\in \zz}\Tr[\ttt\ttt(P_z\otimes P_z)]=d\sum_{P_z\in \zz}\Tr[P_z\otimes P_z]\\
            &=d\sum_{P_z\in \zz}\Tr^2[P_z]=d^3\sum_{P_z\in \zz}\delta_{P_z,\id}=d^3
        \end{split}
    \end{equation}
    \qed
\end{cor}
\begin{lemma}\label{app:twirledswaP_z}
    On an operator level, the following equality holds:
    \begin{equation}
        \sum_{P\in\mathbb{P}}\sum_{P_{z_1},P_{z_2} \in \zz}(P_{z_1}P P_{z_1})\otimes (P_{z_2} \
P P_{z_2})=d^2\sum_{P_z \in \zz} P_z\otimes P_z
    \end{equation}
    \proof We can write a generic $n-$qubit Pauli string as 
    \begin{equation}
        P=\Bar{X}^{x}\Bar{Z}^{z}\,,\Bar{X}:=X^{\ot n}\,,\Bar{Z}:=Z^{\ot n}\,,x,z\in\{0,1\}^{\times n}
    \end{equation}
    a string belonging to $\zz$ can be written as $\Bar{Z}^z$, obviously. Substituting these expressions in the sum (dropping the barred notation for simplicity), we get
    \begin{equation}
       \begin{split}
             \sum_{P\in\mathbb{P}}&\sum_{P_{z_1},P_{z_2} '\in \zz}(P_{z_1}P P_{z_1})\otimes (P_{z_2} P P_{z_2})= \sum_{x,z,z_1,z_2} (Z^{z_1}X^x Z^{z_1} Z^z )\otimes  (Z^{z_2}X^x Z^{z_2} Z^z )\\
             &= \sum_{x,z,z_1,z_2}(-1)^{x\cdot z_1}(-1)^{x\cdot z_2} (Z^{z_1})^2 X^x Z^z \otimes (Z^{z_2})^2 X^x Z^z\\
             &=\sum_{x,z}X^x Z^z \otimes  X^x Z^z\sum_{z_1}(-1)^{x\cdot z_1} \sum_{z_2}(-1)^{x\cdot z_2}=d^2\sum_{x,z}X^x Z^z \otimes  X^x Z^z \delta_{x,0}\\
             &=d^2\sum_{z}Z^z\ot Z^z=d^2\sum_{P_z\in \zz}P_z \ot P_z
       \end{split}
    \end{equation}
    \qed
\end{lemma}
\begin{lemma}\label{app:twirledQz}
    On an operator level, the following equality holds

\begin{equation}
   \begin{split}
       d^2&\sum_{P_{z_1},P_{z_2}P_{z_3},P_{z_4} \in \zz} (P_{z_1}\ot P_{z_2}\ot P_{z_3} \ot P_{z_4})Q(P_{z_1}\ot P_{z_2}\ot P_{z_3} \ot P_{z_4})\\&=\sum_{P\in\mathbb{P}}\sum_{P_{z_1},P_{z_2}P_{z_3},P_{z_4} \in \zz}(P_{z_1}P P_{z_1})\otimes (P_{z_2} \
P P_{z_2})\ot (P_{z_3}P P_{z_3})\otimes (P_{z_4} \
P P_{z_4})=d^4\sum_{P_z \in \zz} P_z ^{\ot 4}  
   \end{split}
\end{equation}
\proof Using the same approach of the previous lemma, we can write
\begin{equation}
    \begin{split}
        \sum_{P\in\mathbb{P}}&\sum_{P_{z_1},P_{z_2}P_{z_3},P_{z_4} \in \zz}(P_{z_1}P P_{z_1})\otimes (P_{z_2} \
P P_{z_2})\ot (P_{z_3}P P_{z_3})\otimes (P_{z_4} \
P P_{z_4})\\
&=\sum_{x,z}X^x Z^z \otimes  X^x Z^z\otimes  X^x Z^z\otimes  X^x Z^z\sum_{z_1}(-1)^{x\cdot z_1} \sum_{z_2}(-1)^{x\cdot z_2}\sum_{z_3}(-1)^{x\cdot z_3} \sum_{z_4}(-1)^{x\cdot z_4}\\
&=d^4\sum_{x,z}X^x Z^z \otimes  X^x Z^z\otimes  X^x Z^z\otimes  X^x Z^z\delta_{x,0}=d^4\sum_{z}Z^z\ot Z^z\ot Z^z\ot Z^z\\
&=d^4\sum_{P_z\in \zz}P_z \ot P_z\ot P_z\ot P_z
    \end{split}
\end{equation}
\qed \end{lemma}
\begin{lemma}
    The following equality holds:
    \begin{equation}
        \sum_{P\in \pauli,P_z\in \zz}\Tr^2[PP_z P P_z]=d^5\,.
    \end{equation}
\proof Since $P$ and $P_z$ either commute or anti-commute, inverting the order of them in the square trace would yield the same result, so we can write
\begin{equation}
    \begin{split}
         &\sum_{P\in \pauli,P_z\in \zz}\Tr^2[PP_z P P_z]= \sum_{P\in \pauli,P_z\in \zz}\Tr^2[PP_z P_z P ]=\sum_{P\in \pauli,P_z\in \zz}\Tr^2[\id]\\
         &=d^2\sum_{P\in \pauli,P_z\in \zz}=d^5
    \end{split}
\end{equation}
\qed \end{lemma}
\begin{lemma}
    The following equality holds:
    \begin{equation}
        \sum_{P\in \pauli,P_z\in \zz}\Tr[PP_z PP_z PP_z PP_z]=d^4\,.
    \end{equation}
    \proof Since $P$ and $P_z$ either commute or anti-commute, inverting the order of two couples of them will not change the result, so we can write
    \begin{equation}
        \begin{split}
              &\sum_{P\in \pauli,P_z\in \zz}\Tr[PP_z PP_z PP_z PP_z]=  \sum_{P\in \pauli,P_z\in \zz}\Tr[PP_z P_z P P_z P  PP_z]=\sum_{P\in \pauli,P_z\in \zz}\Tr[\id]\\
              &=d\sum_{P\in \pauli,P_z\in \zz}=d^4\,.
        \end{split}
    \end{equation}
  
\qed \end{lemma}

Here we also recall the known formula for the Haar average of $k$-copies of a state:

\begin{equation}\label{app:kstateaverage}
    \exv\psi[\psi_U ^{\ot k}]=\binom{d+k-1}{k}^{-1}\Pi_{\sym}=\frac{1}{\prod_{i=0}^{k-1}(d+i)}\sum_{\pi\in S_k}\ttt_\pi\,.
\end{equation}
\begin{lemma}\label{app:UUdag}
    Given $U\in \mathcal{U}(d)$, then
    \begin{equation}\label{app:uudag}
        \exv_U[U^{\ot k}\ot U^{\dag \ot k}]=S_2 \sum_{\pi,\sigma \in S_k}W_{\pi\sigma} ^{(d)}\ttt_\pi\ot\ttt_\sigma
    \end{equation}
    where $S_2$ is the swap operator acting on $\hi^{\ot k}\ot \hi^{\ot k}$
    \proof We will prove that the left hand side is equal to the right hand side of Eq. \eqref{app:uudag} component by component in a basis of $\hi^{\ot k}\ot \hi^{\ot k}$: in particular, we will use Pauli strings $P$ in $\mathcal{B}(\hi^{\ot 2k})$. Due to the structure of Pauli strings, each one of them can be written as $P=P_1 \ot P_2\,,P_1,P_2\in \mathcal{B}(\hi^{\ot k})$. We then have
    \begin{equation}
        \begin{split}
            \exv_U &\Tr[S_2(U^{\ot k}\ot U^{\dag \ot k})P_1\ot P_2 ]=\exv_U \Tr[S_2(U^{\ot k}P_1\ot U^{\dag \ot k}P_2)]\\
            &\overset{(a)}{=}\exv_U \Tr[U^{\ot k}P_1U^{\dag\ot k}P_2]=\sum_{\pi,\sigma \in S_k}W_{\pi\sigma} ^{(d)}\Tr[\ttt_\pi P_1]\Tr[\ttt_\sigma P_2]
        \end{split}
    \end{equation}
    where in passage $(a)$ we used the swap trick. Since the components of r.h.s. and l.h.s. in  are equal in a generic basis, this means the equality holds in an operatorial way: thus we have
    \begin{equation}
         S_2\,\exv_U[U^{\ot k}\ot U^{\dag \ot k}]= \sum_{\pi,\sigma \in S_k}W_{\pi\sigma} ^{(d)}\ttt_\pi\ot\ttt_\sigma\,.
    \end{equation}
    Applying $S_2$ to both sides, exploiting the fact that $S_2 \,^2=\id$ we get the thesis.

\qed \end{lemma}

\section{SE as a magic proxy in the Haar ensemble: techniques}\label{app:haar_app}

In this section, we will prove that linear SE for mixed states is a good magic proxy in the Haar ensemble: to do this, we will compute the SE gap for the partial trace and the computational basis dephasing channel in this example and establish its typicality around positive values for both maps.

\subsection{Average SE gap for partial trace in the Haar ensemble}\label{app:avg_A_haar}
Given $\hi=\hi_A\otimes \hi_B$, with $d_A={\rm dim}\,\hi_A$ and $d_B={\rm dim}\,\hi_B$ and $d={\rm dim}\,\hi=d_A d_B$, let us define the SE gap for the partial trace:
\begin{equation}\label{app:mbarAdef}
    \Bar{M}^{A}(\psi):=\mlin (\psi)-\mlin (\psi^A)\,,
\end{equation}
with $\psi^A=\Tr_B \psi$. Applying a Haar-random unitary to the state $\psi\mapsto \psi_U=U\psi U^\dag$ and averaging over $U$ one is left to compute the following object:
\begin{equation}
    \begin{split}
        \exv_U \mbar^A(\psi_U)&=\exv_U \mlin(\psi_U)-\exv_U \mlin(\psi_U ^A)\\
        &=1-d\exv_U\Tr(Q\psi_U)-\exv_U \pur(\psi_U ^A)+d_A \exv_U \Tr[(Q_A\ot \id_B ^{\ot 4}) \psi_U ^{\ot 4}]
    \end{split}
\end{equation}
using known results from \cite{appcite:page_avg_entropy,appcite:leone2022stabilizer} we know that
\begin{equation}
    \begin{split}
        \exv_U \pur(\psi_U ^A)&=\frac{d_A+d/d_A}{d+1}\\
        \exv_U\Tr(Q\psi_U)&=\frac{4}{d(d+3)}\,.
    \end{split}
\end{equation}
Using standard tools of Weingarten calculus one knows that the Haar average of $k-$copies of a pure state reads
\begin{equation}
   \begin{split}
        \exvpsi \psi_U^{\ot k}&=\frac{\Pi_{{\rm sym}}}{\Tr(\Pi_{{\rm sym}})}=\binom{d+k-1}{k}^{-1}\frac{1}{k!}\sum_{\pi \in S_k}\ttt_\pi\,,
   \end{split}
\end{equation}
where $\ttt_\pi$ are elements of the representation of the group of permutation of $k$ objects $S_k$ on $\hi^{\ot k}$. Using these tools, we compute the partial stabilizer purity:
\begin{equation}
    \begin{split}
         d_A\exv_U \Tr[(Q_A\ot \id_B) \psi_U ^{\ot 4}]&=d_A\binom{d+4}{3}^{-1}\frac{1}{4!}\sum_{\pi \in S_4} \Tr[(Q_A\ot \id_B ^{\ot 4})\ttt_\pi]\\
         &=d_A\binom{d+4}{3}^{-1}\frac{1}{4!}\sum_{\pi \in S_4}\Tr[Q_A \ttt_\pi ^A]\Tr[\ttt_\pi ^B]\,,
    \end{split}
\end{equation}
where we used the fact that the representation operators $\ttt_\pi$ factorize on qubits. This property is fundamental to the direct evaluation of these traces, as this means that the result of those traces will always be some powers of the dimension of the Hilbert space. In formulas, let $\ttt_{\pi} ^{(d)}, Q^{(d)}$ be the operators acting on four copies of a Hilbert space of dimension $d=2^N$, 
\begin{equation}
    \Tr[Q^{(d)}\ttt_\pi ^{(d)}]=\Tr[Q^{(2)}\ttt_\pi ^{(2)}]^N\,.
\end{equation}

Hinging on this fact, one is able to automatically compute these traces for a system with $d=2$ (one qubit) using Wolfram Mathematica and extend the result to a generic number of qubits. We will exploit this property ubiquitously in the following sections, since for averages of higher moments of the Haar distribution the number of permutation operators becomes intractable using other methods. Using this method, we then get
\begin{equation}
     d_A\exvpsi \Tr[(Q_A\ot \id_B^{\ot 4})\psi^{\ot 4}]=\frac{d_A d_B^2+3 d_A+4 d_B}{(d_A d_B+1) (d_A d_B+3)}\,,
\end{equation}
and combining all the pieces one gets
\begin{equation}
        \exv_U [\Bar{M}^{A}(\psi_U)]=\frac{(d_B-1) \left(d_A^2 d_B+1\right)}{(d_A d_B+1) (d_A d_B+3)}=O(1)
    \end{equation}

    \subsection{Variance of SE gap for partial trace in the Haar ensemble}\label{app:variance_A_haar}
    Objective of this section is to calculate the variance of $\Bar{M}^A$. This reads
    \begin{equation}
        \Delta^2 \Bar{M} ^{A}=\Delta^2 \mlin(\psi)+\Delta^2 \mlin(\psi_A)-2{\rm cov}(\mlin(\psi),\mlin(\psi_A))\,.
    \end{equation}
    Let us focus on each term separately.
    \subsubsection{Variance of \texorpdfstring{$\mlin(\psi)$}{}}
$\Delta^2 \mlin(\psi)$ reads
\begin{equation}\label{app:variance_mlin_eq}
    \begin{split}
        \Delta^2 \mlin(\psi)&=\exvpsi[(1-\stp(\psi_U))^2]-[\exvpsi \mlin(\psi_U)]^2\\
        &=1+\exvpsi[(\stp(\psi_U))^2]-2\exvpsi(\stp(\psi_U))-[\exvpsi \mlin(\psi_U)]^2\\
        &=1+d^2\Tr[(Q\otimes Q)\exvpsi(\psi_U ^{\otimes 8})]-2d\Tr[Q\exvpsi(\psi_U ^{\ot 4})]-\left(1-\frac{4}{d+3}\right)^2\\
        &1+d^2\Tr[(Q\otimes Q)\exvpsi(\psi_U ^{\otimes 8})]-2d\Tr[Q\exvpsi(\psi_U ^{\ot 4})]-1-\frac{16}{(d+3)^2}+\frac{8}{d+3}\\
        &=d^2\Tr[(Q\otimes Q)\exvpsi(\psi_U ^{\otimes 8})]-\frac{8}{d+3}-\frac{16}{(d+3)^2}+\frac{8}{d+3}
    \end{split}
\end{equation}
Let us focus on the last term to compute:
\begin{equation}
    \begin{split}
        d^2\Tr[(Q\otimes Q)\exvpsi(\psi_U ^{\otimes 8})]&=\binom{d+7}{8}^{-1} \frac{1}{8!}\sum_{\pi \in S_8}\Tr[(Q\ot Q) \ttt_\pi]
    \end{split}
\end{equation}
again, using the factorization properties of both $Q$ (and hence $Q\ot Q$) and $\ttt_\pi$ we compute these traces for $\hi=\mathbb{C}^{\ot 2}$ and extend it for generic number of qubits, obtaining
\begin{equation}
     d^2\Tr[(Q\otimes Q)\exvpsi(\psi_U ^{\otimes 8})]=\frac{16 \left(d^2+15 d+68\right)}{(d+3) (d+5) (d+6) (d+7)}
\end{equation}
and substituting this expression in Eq. \eqref{app:variance_mlin_eq} we get
\begin{equation}
    \Delta^2 \mlin(\psi)=\frac{96 (d-1)}{(d+3)^2 (d+5) (d+6) (d+7)}= O\left(\frac{1}{d^4}\right)\,.
\end{equation}
\subsubsection{Variance of \texorpdfstring{$\mlin(\psi_A)$}{}}
The variance of $\mlin(\psi_A)$ reads
\begin{equation}
    \begin{split}
        \Delta^2 (\mlin(\psi))&=\exvpsi[(\pur \psi_A-\stp(\psi_A))^2]-\exvpsi[\pur \psi_A-\stp(\psi_A)]^2\\
        &=\exvpsi[\pur^2 \psi_A]+\exvpsi[\stp(\psi_A)^2]-2\exvpsi[\stp(\psi_A)\pur \psi_A]-\left(\frac{(d_A-1) (d_A+1) d_B}{(d_A d_B+1) (d_A d_B+3)}\right)^2\,.
    \end{split}
\end{equation}
\paragraph{Average square purity of reduce density matrix}

The average value of the square of the reduced density matrix purity reads
\begin{equation}
    \begin{split}
        \exvpsi \pur^2(\psi_U ^A)&=\exvpsi \Tr[(\ttt_A^{\ot 2}\ot \id_B ^{\ot 4})\psi_U ^{\ot 4}]\\
        &=\binom{d+3}{4}^{-1}\frac{1}{4!}\sum_{\pi \in S_4}\Tr[\ttt_A ^{\ot 2} \ttt_\pi ^A]\Tr[\ttt_\pi ^B]\,.
    \end{split}
\end{equation}
Direct evaluation of these traces, using the same previously mentioned properties of factorization, yields
\begin{equation}
     \exvpsi \pur^2(\psi_U ^A)=\frac{d^2 (d+4)+(d+4) d_A^4+2 (d (d+5)+1) d_A^2}{(d+1) (d+2) (d+3) d_A^2}
\end{equation}
\paragraph{Average of \texorpdfstring{$\stp(\psi_A)\pur \psi_A$}{}}
We now show computation of this object:
\begin{equation}
    \begin{split}
        \exvpsi[\stp(\psi_A)\pur \psi_U ^A]&=d_A \exvpsi \Tr{[(Q_A\ot \ttt_A)\ot \id_B ^{\ot 6}]\psi_U ^{\ot 6}}\\
        &=\binom{d+5}{6}\frac{1}{6!}\sum_{\pi \in S_6}\Tr{[(Q_A\ot \ttt_A)\ot \id_B ^{\ot 6}]\ttt_\pi}\\
         &=\binom{d+5}{6}\frac{1}{6!}\sum_{\pi \in S_6}\Tr{[(Q_A\ot \ttt_A)\ttt_\pi ^A]}\Tr[\ttt_\pi ^B]\,.
    \end{split}
\end{equation}
Direct evaluation of these sums, using the same trick, yields
\begin{equation}
    \exvpsi[\stp(\psi_A)\pur \psi_U ^A]=\frac{d^2 (d (d+10)+28)+3 (d+6) d_A^4+(d (d (d+13)+58)+12) d_A^2}{(d+1) (d+2) (d+3) (d+5) d_A^2}
\end{equation}
\paragraph{Average of \texorpdfstring{$\stp(\psi_A)^2$}{}}
We now show computation of the following object:
\begin{equation}
    \begin{split}
        \exvpsi\stp(\psi_U ^A)^2&=d_A ^2\exvpsi \Tr{[(Q_A\ot Q_A)\ot \id_B ^{\ot 8}]\psi_U ^{\ot 8}}\\
        &=d_A ^2\binom{d+7}{8}\frac{1}{8!}\sum_{\pi \in S_8} \Tr{[(Q_A\ot Q_A)\ot \id_B ^{\ot 8}]\ttt_\pi}\\
        &=d_A ^2\binom{d+7}{8}\frac{1}{8!}\sum_{\pi \in S_8} \Tr[(Q_A\ot Q_A)\ttt_\pi ^A]\Tr[\ttt_\pi ^B]
    \end{split}
\end{equation}
of which direct evaluation yields 
\begin{equation}
\begin{split}
     \exvpsi\stp(\psi_U ^A)^2&=\frac{1}{(d+1)^2 (d+2) (d+3)^2 (d+5) (d+6) (d+7) d_A^2}\Big\{d^2 (d+8) \big[d \big(d (d+16)+92\big)+200\big]\\&+9 [d (d+16)+76] d_A^4+6 \big\{d \big[d \big(d (d+20)+156\big)+468\big]+96\big\} d_A^2\Big\}
\end{split}
\end{equation}
Putting all elements together, we get to the following result for the variance of $\mlin(\psi_A)$:
\begin{equation}
  \begin{split}
        &\Delta^2(\mlin(\psi_A))\\&=\frac{ 2 (d_A-1) (d_A+1)}{(d+1)^2 (d+2) (d+3)^2 (d+5) (d+6) (d+7) d_A^2}\bigg(d^6+9 d^5-d^4 d_A^2+17 d^4-9 d^3 d_A^2\\&+27 d^3+31 d^2 d_A^2+78 d^2+117 d d_A^2+18 d_A^2\bigg) = O\left(\frac{1}{d^2}\right)
  \end{split}
\end{equation}

\subsubsection{Covariance between \texorpdfstring{$\mlin(\psi)$}{} and \texorpdfstring{$\mlin(\psi_A)$}{}}
The covariance reads
\begin{equation}
   \begin{split}
        {\rm cov}(\mlin(\psi),\mlin(\psi_A))&=\exvpsi[(1-\stp(\psi))(\pur \psi_A-\stp(\psi_A))]-\exvpsi[1-\stp(\psi)]\exvpsi[\pur \psi_A-\stp(\psi_A)]\\
        &=\cancel{\exvpsi[\pur \psi_A-\stp(\psi_A)]}-\exvpsi[\stp(\psi)\pur \psi_A]+\exvpsi[\stp(\psi)\stp(\psi_A)]\\&-\cancel{\exvpsi[\pur \psi_A-\stp(\psi_A)]}+\exvpsi[\stp(\psi)(\pur \psi_A-\stp(\psi_A))]\\
        &=\exvpsi[\stp(\psi)\stp(\psi_A)]+\exvpsi[\stp(\psi)]\exvpsi[\pur \psi_A]\\&-\exvpsi[\stp(\psi)\pur \psi_A]-\exvpsi[\stp(\psi)]\exvpsi[\stp(\psi_A)]      
   \end{split}
\end{equation}
We compute each piece separately.
\paragraph{Average of \texorpdfstring{$\stp(\psi)\stp(\psi_A)$}{}}
We show computation of this object
\begin{equation}
    \begin{split}
        \exvpsi \stp(\psi_U)\stp(\psi_U ^A)&=dd_A\exvpsi\Tr{[(Q_A\ot Q_A)\ot (Q_B\ot \id_B ^{\ot4})]\psi_U ^{\ot 8}}\\
        &=dd_A\binom{d+7}{8}\frac{1}{8!}\sum_{\pi \in S_8} \Tr[(Q_A\ot Q_A)\ttt_\pi ^A]\Tr[(Q_B\ot \id_B^{\ot 4})\ttt_\pi ^B]
    \end{split}
\end{equation}
Direct evaluation of this sum yields
\begin{equation}
    \exvpsi \stp(\psi_U)\stp(\psi_U ^A)=\frac{12 [d (d+15)+70]d_A^2+4 d (d+8) [d (d+11)+34]}{(d+1) (d+3) (d+5) (d+6) (d+7) d_A}\,.
\end{equation}
\paragraph{Average of \texorpdfstring{$\stp(\psi)\pur(\psi_A)$}{}}
We now move to compute the following object:
\begin{equation}
    \begin{split}
        \exvpsi\stp(\psi_U)\pur(\psi_U ^A)&=d\exvpsi \Tr{[(Q_A\ot \ttt_A)\ot (Q_B\ot \id_B ^{\ot 2})]\psi_U ^{\ot 6}}\\
        &=d\binom{d+5}{6}\frac{1}{6!}\sum_{\pi \in S_6}\Tr[(Q_A\ot \ttt_A)\ttt_\pi ^A]\Tr[(Q_B\ot \id_B ^{\ot 2})\ttt_\pi ^B]\,.
    \end{split}
\end{equation}
As usual, using the factorization properties of permutation operators and $Q$, we directly evaluate this sum to be
\begin{equation}
    \exvpsi\stp(\psi_U)\pur(\psi_U ^A)=\frac{4 \left(d+d_A^2\right)}{(d+1) (d+3) d_A}
\end{equation}
Combining all elements together yields to the following result:
\begin{equation}
   {\rm cov}(\mlin(\psi),\mlin(\psi_A))= \frac{96 d (d_A-1) (d_A+1)}{(d+1) (d+3)^2 (d+5) (d+6) (d+7) d_A}
\end{equation}

Finally, summing the terms calculated so fare leaves us with the following final result:
\begin{equation}
    \begin{split}
        \Delta^2(\Bar{M}^A)&=\frac{2 (d-d_A) (d+d_A)}{(d+1)^2 (d+2) (d+3)^2 (d+5) (d+6) (d+7) d_A^2}\Big[d \big[\big(d (d (d+9)-31)-117\big) d_A^2\\&-d (d (d+9)+17)-27\big]-6 \left(3 d_A^2+13\right)\Big]=O\left(\frac{1}{d^2}\right)
    \end{split}
\end{equation}

\subsubsection{Violation probability of SE under partial trace in Haar ensemble}

It is possible to obtain violation probability bounds either Chebyshev and/or L\'evy typicality arguments, and we now show both concentration results for the SE gap for partial trace in the Haar ensemble. As seen before, the variance of the SE gap is exponentially vanishing as $N\rightarrow \infty$ :
    \begin{equation}
        \begin{split}
        \Delta^2 \Bar{M} ^{A}&= O\left(\frac{1}{d^2}\right)
        \end{split}
    \end{equation}
Hence, using the Chebyshev inequality, putting a $O(1)$ discrepancy of the SE gap from its average, we get
\begin{equation}
    \Pr[\mbar^A(\psi)<0]\leq O\left(\frac{1}{d^2}\right)\,.
\end{equation}
A stronger typicality can be proved by means of L\'evy's lemma \cite{appcite:watrous_2018}; it is shown in \cite{appcite:zhu2016CliffordGroupFails} that stabilizer purity is a $\kappa-$Lipschitz function with $\kappa=27/5$, hence, in conjunction with purity being Lipschitz with $\kappa\leq 2$ \cite{appcite:hayden2006AspectsGenericEntanglement} it is straightforward to see the SE gap for partial trace inherits Lipschitzianity with $\kappa\leq 64/5$. Hence, application of the L\'evy to the partial trace SE gap implies the  violation probability to be 

\begin{equation}
     {\rm Pr}(\mbar^A(\psi)\leq 0)\leq \exp \left[-\frac{d \left(\frac{d}{d_A}-1\right)^2 (d d_A+1)^2}{4096 \pi  (d+1)^2 (d+3)^2}\right]\,
\end{equation}
which, recalling Def. \ref{app:proxy_def}, means that $\mlin$ is a $\eta-$proxy for magic, with 
\begin{equation}
\eta=\frac{d (d-d_A)^2 (d d_A+1)^2}{4096 \pi  (d+1)^2 (d+3)^2 d_A^2 \log d}= O\left(\frac{d}{\log d}\right)
\end{equation}
in the Haar ensemble, proving that violations of monotonicity under partial trace are overwhelmingly rare in the Hilbert space.

\subsection{Average SE gap for dephasing in the Haar ensemble}\label{app:avg_db_haar}
In this section we show calculations for the average SE gap for dephasing in the Haar ensemble. Let us write it explicitly: given $\hi\simeq \mathbb{C}^{2\ot N}$, the dephasing channel onto the computational basis $B=\{\ket{i}\}_{i \in \{0,1\}^{\times N}}$ reads
\begin{equation}
    \db:=\sum_{i}\Pi_i \psi \Pi_i\,.
\end{equation}
This channel represents a complete, non post-selective measurement in the computational basis in the system. Alternatively, this channel can also be rewritten in the following form (for proof see Sec.  \ref{app:lemmas}):
\begin{equation}
    D_B\psi=\frac{1}{d}\sum_{P_z \in Z_2}P_z\psi P_z
\end{equation}
where $Z_2$ denotes the (abelian) subgroup of Pauli operators belonging to $\{\id,Z\}^{\otimes N}$. 
As for the partial trace, we define the SE gap as follows:
\begin{equation}
    \Bar{M} ^{D_B}(\psi):=\mlin(\psi)-\mlin(D_B\psi)
\end{equation}
The Haar average of this variable reads\begin{equation}
   \begin{split}
        \exvpsi\mbar^{D_B}(\psi_U)&=1-\exvpsi\stp(\psi_U)-\exvpsi\pur(D_B\psi_U)+\exvpsi\stp(D_B\psi_U)\\
        &=1\frac{4}{d+3}-\frac{2}{d+1}+\exvpsi\stp(D_B\psi_U)\,.
   \end{split}
\end{equation}
Let us focus on the last term: this reads
\begin{equation}
\begin{split}
        \exvpsi\stp(D_B\psi_U)&=d\exvpsi\Tr[Q(D_B\psi_U)^{\ot 4}]=d\exvpsi\Tr[Q D_B ^{\ot 4} \psi_U ^{\ot 4}]\\
        &=d\exvpsi\sum_{i,j,k,l}\Tr[Q\Pi_{ijkl}\psi_U ^{\ot 4}\Pi_{ijkl}]=d\exvpsi\sum_{i,j,k,l}\Tr[\Pi_{ijkl}Q\Pi_{ijkl}\psi_U ^{\ot 4}]\\
        &=d\exvpsi \Tr[(D_B ^{\ot 4}Q)\psi_U ^{\ot 4}]=\binom{d+3}{4}^{-1}\frac{1}{4!}\sum_{\pi \in S_4}\Tr[(D_B ^{\ot 4}Q) \ttt_\pi]\,,
\end{split}
\end{equation}
where $\Pi_{ijkl}=\ketbra{i} \ot \ketbra{j} \ot \ketbra{k} \ot \ketbra{l}$. Since the dephasing channel factorizes on qubits, we are able to automatically evaluate this expression, which turns out to be
\begin{equation}
    \exvpsi\stp(D_B\psi_U)=\frac{d+7}{(d+1) (d+3)}
\end{equation}

\subsection{Variance of SE gap for dephasing in the Haar ensemble}\label{app:variance_db_haar}
The variance of $\Bar{M}^{D_B}$ is given by
\begin{equation}
    \Delta^2 \Bar{M} ^{D_B}=\Delta^2 \mlin(\psi)+\Delta^2 \mlin(D_B\psi)-2{\rm cov}(\mlin(\psi),\mlin(D_B\psi))
\end{equation}
Let us focus on each of these terms separately.
\subsubsection{Variance of \texorpdfstring{$\mlin(D_B\psi)$}{}}
The variance of the SE of the dephased state reads
\begin{equation}
    \begin{split}
        \Delta^2 \mlin(D_B\psi)&=\exvpsi\left[\mlin(D_B\psi_U)^2\right]-\exvpsi\left[\mlin(D_B\psi_U)\right]^2\\
        &=\exvpsi\Tr[(\ttt \ot \ttt)(D_B\psi_U ^{\ot 4})]+d^2\exvpsi\Tr[(Q\ot Q)(D_B\psi_U)^{\ot 8}]-2d\exvpsi\Tr[(\ttt\otimes Q)(D_B\psi_U)^{\ot 6}]\\&-\left(\frac{d-1}{(d+1) (d+3)}\right)^2\,.
    \end{split}
\end{equation}
Again, we split each piece in a separate calculation.
\paragraph{Average of \texorpdfstring{$\Tr[(\ttt \ot \ttt)(D_B\psi_U ^{\ot 4})]$}{}}
We show computation of this object:
\begin{equation}
    \begin{split}
        \exvpsi\Tr[(\ttt \ot \ttt)(D_B\psi_U ^{\ot 4})]&=\exvpsi\Tr[(D_B ^{\ot 4}\ttt \ot \ttt)\psi_U ^{\ot 4}]\\
        &=\binom{d+3}{4}^{-1}\frac{1}{4!}\sum_{\pi \in S_4}\Tr[(D_B ^{\ot 4}\ttt \ot \ttt)\ttt_\pi]
    \end{split}
\end{equation}
and explicit calculation of these traces yield
\begin{equation}
     \exvpsi\Tr[(\ttt \ot \ttt)(D_B\psi_U ^{\ot 4})]=\frac{4 (d+5)}{(d+1) (d+2) (d+3)}
\end{equation}
\paragraph{\texorpdfstring{Average of $\Tr[(\ttt\otimes Q)(D_B\psi)^{\ot 6}]$}{}}
We now show computation of the following:
\begin{equation}
    \begin{split}
        \exvpsi \Tr[(\ttt\otimes Q)(D_B\psi_U)^{\ot 6}]&=\exvpsi \Tr[(D_B ^{\ot 6} \ttt\otimes Q)\psi_U ^{\ot 6}]\\
        &=\binom{d+5}{6}\frac{1}{6!}\sum_{\pi \in S_6}\Tr[(D_B ^{\ot 6}\ttt\otimes Q)\ttt_\pi]
    \end{split}
\end{equation}
and direct evaluation gets us the following result:
\begin{equation}
     \exvpsi \Tr[(\ttt\otimes Q)(D_B\psi_U)^{\ot 6}]=\frac{2 [d (d+13)+58]}{d (d+1) (d+2) (d+3) (d+5)}
\end{equation}
\paragraph{Average of \texorpdfstring{$\Tr[(Q\ot Q)(D_B\psi)^{\ot 8}]$}{}}
We now show computation of this object
\begin{equation}
    \begin{split}
        \exvpsi \Tr[(Q\ot Q)(D_B\psi_U)^{\ot 8}]&= \exvpsi \Tr[(D_B ^{\ot 8}Q\ot Q)\psi_U^{\ot 8}]\\
        &=\binom{d+7}{8}^{-1}\frac{1}{8!}\sum_{\pi \in S_8}\Tr[(D_B ^{\ot 8}Q\ot Q)\ttt_\pi]
    \end{split}
\end{equation}
As usual, direct evaluation of these traces yields
\begin{equation}
   \exvpsi \Tr[(Q\ot Q)(D_B\psi_U)^{\ot 8}]= \frac{d^4+30 d^3+349 d^2+2088 d+5596}{d^2 (d+1) (d+2) (d+3) (d+5) (d+6) (d+7)}
\end{equation}
and finally, combining these pieces we get the following result for the variance of $\mlin(D_B\psi)$:
\begin{equation}
    \begin{split}
        \Delta^2 \mlin(D_B\psi)=\frac{4 (d-1) \left(d^4+9 d^3+11 d^2+9 d+66\right)}{(d+1)^2 (d+2) (d+3)^2 (d+5) (d+6) (d+7)}
    \end{split}
\end{equation}
\subsection{Covariance between SE and dephased SE}
In this section we show complete calculations for the covariance between $\mlin(\psi)$ and $\mlin(D_B\psi)$ over the Haar ensemble. This reads
The covariance reads
\begin{equation}
    \begin{split}
        {\rm cov}(\mlin(\psi),\mlin(D_B\psi))&=\exvpsi\{[1-\stp(\psi_U)][\pur(D_B \psi_U)-\stp(D_B \psi_U)]\}-\exvpsi[\mlin(\psi_U)]\exvpsi[\mlin(D_B \psi_U)]\\
        &=\exvpsi[\pur(D_B \psi_U)-\stp(D_B \psi_U)]-\exvpsi[\stp(\psi)\pur(D_B \psi_U)]+\exvpsi[\stp(\psi)\stp(D_B \psi_U)]\\&-\exvpsi[\mlin(\psi)]\exvpsi[\mlin(D_B \psi_U)]\\
        &=\exvpsi[\mlin(D_B \psi_U)]-\exvpsi[\stp(\psi)\pur(D_B \psi_U)]+\exvpsi[\stp(\psi)\stp(D_B \psi_U)]\\&-\exvpsi[\mlin(\psi_U)]\exvpsi[\mlin(D_B \psi_U)]\\
        &=\frac{d-1}{(d+1) (d+3)}-\exvpsi[\stp(\psi_U)\pur(D_B \psi_U)]+\exvpsi[\stp(\psi_U)\stp(D_B \psi_U)]\\&-\frac{(d-1)^2}{(d+1) (d+3)^2}
    \end{split}
\end{equation}
We now tackle the other two pieces separately.
\paragraph{Average of \texorpdfstring{$\stp(\psi)\pur(D_B\psi)$}{}}
This object reads
\begin{equation}
    \begin{split}
        \exvpsi[\stp(\psi_U)\pur(D_B \psi_U)]&=d\exvpsi\Tr[(Q\ot T)(\psi_U)^{\ot 4}\ot (D_B \psi_U)^{\ot 2}]\\
        &=d\exvpsi\Tr[(Q\ot D_B ^{\ot 2} \ttt)\psi_U ^{\ot 6}]\\
        &=d\binom{d+5}{6}\frac{1}{6!}\sum_{\pi \in S_6}\Tr[(Q\ot D_B^{\ot 2}\ttt)\ttt_\pi]  \,.
    \end{split}
\end{equation}
Explicit evaluation of these traces yields
\begin{equation}
    \exvpsi[\stp(\psi_U)\pur(D_B \psi_U)]=\frac{8}{(d+1) (d+3)}
\end{equation}
\paragraph{Average of \texorpdfstring{$\stp(\psi)\stp(D_B\psi)$}{}}
We now carry out calculations of the following object:
\begin{equation}
    \begin{split}
        \exvpsi[\stp(\psi_U)\stp(D_B \psi_U)]&=d^2\exvpsi\Tr[(Q\ot Q)\psi_U^{\ot 4}\ot (D_B \psi_U)^{\ot 4}]\\
        &=d^2\exvpsi\Tr[(Q\ot D_B^{\ot 4}Q)\psi_U ^{\ot 8}]\\
        &=d^2\binom{d+7}{8}\frac{1}{8!}\sum_{\pi \in S_8}\Tr[(Q\ot D_B^{\ot 4}Q)\ttt_\pi]\,,
    \end{split}
\end{equation}
and again, direct evaluation of these expressions yield
\begin{equation}
    \begin{split}
         \exvpsi[\stp(\psi_U)\stp(D_B \psi_U)]=\frac{4 \left(d^3+22 d^2+167 d+482\right)}{(d+1) (d+3) (d+5) (d+6) (d+7)}\,.
    \end{split}
\end{equation}
After combining all the pieces we get the following expression of the covariance between $\mlin(\psi)$ and $\mlin(D_B \psi)$:
\begin{equation}
    {\rm cov}(\mlin(\psi),\mlin(D_B \psi))=\frac{96 (d-1)}{(d+1) (d+3)^2 (d+5) (d+6) (d+7)}
\end{equation}
\subsubsection{Final expression of the variance of SE gap for dephasing in the Haar ensemble}
Putting the pieces together, we finally get the following expression for $\Delta \mbar^{D_B}$:
\begin{equation}
    \Delta \mbar^{D_B}=\frac{4 (d-1) \left(d^4+33 d^3+59 d^2-15 d+18\right)}{(d+1)^2 (d+2) (d+3)^2 (d+5) (d+6) (d+7)}=O\left(\frac{1}{d^3}\right)
\end{equation}

\subsection{Violation probability of SE under dephasing in the Haar ensemble}

Again, to establish monotonicity violation probability bounds we can use the Chebyshev inequality: putting a $O(1)$ discrepancy of the SE gap from its average, we get
\begin{equation}
    \Pr[\mbar^A(\psi)<0]\leq O\left(\frac{1}{d^3}\right)\,.
\end{equation}

Like the SE gap for partial trace, also the SE dephasing gap is a Lipschitz function with $\kappa=64/5$, hence application of the L\'evy lemma yields the following violation probability:
\begin{equation}
    {\rm Pr}(\mbar^{D_B}(\psi)\leq 0)\leq \exp[-\frac{(d-1)^2 d^3}{4096 \pi  (d+1)^2 (d+3)^2}]
\end{equation}
meaning that $\mlin$ is $\eta-$magic proxy with 

\begin{equation}\eta= \frac{( d-1)^2 d^3}{4096 (d+1)^2 (d+3)^2 \pi \log d}= O\left(\frac{d}{\log d}\right)\,.
\end{equation}

Violations of monotonicity are extremely rare in the Hilbert space for dephasing as well.

\section{SE as a magic proxy in the Clifford orbit: techniques}\label{app:clifford_app}

In this section, we show all the calculations that show that $\mlin$ is a good proxy for magic in the Clifford orbit.
\subsection{Average SE gap for partial trace in the Clifford orbit}\label{app:avg_A_clif}
As in the Haar case, we define the SE gap for partial trace as
\begin{equation}
     \Bar{M}^{A}(\psi):=\mlin (\psi)-\mlin (\psi^A)\,,
\end{equation}

We show calculations to compute the SE gap for partial trace in the Clifford orbit. Starting from a $N-$qubit pure state $\psi$, with $\psi_C:=C\psi C^\dag\,,C\in \mathcal{C}(d)$ and $\psi_C ^A:=\Tr_B \psi_C$, this reads
\begin{equation}
    \begin{split}
        \exvc \mbar^A&=1-\exvc\stp(\psi_C)-\exvc\pur(\psi_C ^A)+\exvc\stp(\psi_C ^A)\\
        &=1-\stp(\psi)-\exvc\Tr[(\ttt_A \ot \id_B) ^{\ot 2}\psi_C ^{\ot 2}]+\exvc\Tr[(Q_A\ot \id_B ^{\ot 4})\psi_C ^{\ot 4}]\\
        &=1-\stp(\psi)-\frac{d_A+d/d_A}{d+1}+d_A\exvc\Tr[(Q_A\ot \id_B ^{\ot 4})\psi_C ^{\ot 4}]
    \end{split}
\end{equation}
where we used the fact that stabilizer purity is invariant under Clifford unitaries and that the Clifford group is a $3-$unitary design \cite{appcite:webb2016clifford}, hence the average reduced density matrix purity is the same as the Haar case.
Focusing on the last term, namely $\exvc\Tr[(Q_A\ot \id_B ^{\ot 4})\psi_C ^{\ot 4}]$, we use the following formula for the Clifford average of 4 copies of a pure state \cite{appcite:leone2021QuantumChaosQuantum}:
\begin{equation}\label{app:psiclifavg4}
    \exvc \psi_C ^{\ot 4}=\alpha(\psi) Q\Pi_{\sym}+\beta(\psi)\Pi_{\sym}
\end{equation}
with
\begin{equation}
    \begin{split}
        \alpha(\psi)&=\frac{6 (d \stp(\psi)+3 \stp(\psi)-4)}{(d-1) (d+1) (d+2) (d+4)}\\
        \beta(\psi)&=\frac{24 [d-\stp(\psi)]}{(d-1) d (d+1) (d+2) (d+4)}\,,
    \end{split}
\end{equation}
and 
\begin{equation}
    \Pi_{\sym}=\frac{1}{4!}\sum_{\pi \in S_4}\ttt_\pi\,.
\end{equation}
Substituting this expression for computing the average reduced density matrix stabilizer purity yields
\begin{equation}
    \begin{split}
        d_A&\exvc\Tr[(Q_A\ot \id_B ^{\ot 4})\psi_C ^{\ot 4}]=\alpha\Tr[(Q_A \ot Q_B)(Q_A \ot \id_B ^{\ot 4})\Pi_{\sym}]+\beta\Tr[(Q_A\ot \id_B ^{\ot 4})\Pi_{\sym}]\\
        &=\frac{\alpha}{24}\sum_{\pi\in S_4}\Tr[Q\ttt_\pi]+\frac{\beta}{24}\sum_{\pi \in S_4}\Tr[Q_A \ttt_\pi ^A]\Tr[\ttt_\pi ^B]\,,
    \end{split}
\end{equation}
where we used both the fact that $Q$ factorizes on tensor products and the fact that it is a projector, namely $Q^2=Q$.
As done in the Haar case, we are able to automatically evaluate these sums using the factorization properties of $Q$ and permutation operators, giving the following result:
\begin{equation}
     d_A\exvc\Tr[(Q_A\ot \id_B ^{\ot 4})\psi_C ^{\ot 4}]=\frac{d^2-d_A^2+d \left(d_A^2-1\right) \stp(\psi)}{\left(d^2-1\right) d_A}\,.
\end{equation}
Summing all the pieces,  and recalling that $\mlin(\psi)=1-\stp(\psi)$ for pure states, we obtain the final expression for the SE gap for partial trace in the Clifford orbit:
\begin{equation}
    \exvc \mbar^A(\psi_C)=\frac{ (d-d_A) (d d_A+1)}{\left(d^2-1\right) d_A}\mlin(\psi)\,.
\end{equation}

Notably, the average SE gap is always positive and governed by the value of magic of the initial state, going to zero for initially pure stabilizer states.
\subsection{Variance of SE gap for partial trace in the Clifford orbit}\label{app:variance_A_clif}
The variance of the SE gap for partial trace over the Clifford orbit reads
\begin{equation}
    \begin{split}
        \Delta_C ^2 \mbar^A&=\Delta_C ^2 [\mlin (\psi_C)]+\Delta_C ^2 [\mlin(\psi_C ^A)]-2 {\rm cov}[\mlin(\psi_C),\mlin(\psi_C ^A)]\\
    &=\Delta_C ^2 [\mlin(\psi_C ^A)]\,,
    \end{split}
\end{equation}
where the first and last term vanish due to the fact that $\mlin(\psi_C)$ is constant along the Clifford orbit.
Let us focus more closely on the variance of reduced density matrix purity over the Clifford orbit: this reads
\begin{equation}
    \begin{split}
       \Delta_C ^2 [\mlin(\psi_C ^A)]&= \exvc[\mlin^2(\psi_C ^A)]-[\exvc \mlin(\psi_C ^A)]^2\\
       &=\exvc[\mlin^2(\psi_C ^A)]-\frac{d^2 (d_A-1)^2 (d_A+1)^2 }{(d-1)^2 (d+1)^2 d_A^2}\mlin(\psi)^2\,.
    \end{split}
\end{equation}
Focusing on the square expectation value of $\mlin(\psi_C ^A)$ leads to
\begin{equation}
    \begin{split}
        \exvc[\mlin^2(\psi_C ^A)]&=\exvc[\pur^2(\psi_C ^A)]+\exvc[\stp^2(\psi_C ^A)]-2\exvc[\stp(\psi_C ^A)\pur(\psi_C ^A)]
    \end{split}
\end{equation}
and we will compute each term separately.
\subsubsection{Average of \texorpdfstring{$\pur^2(\psi_C ^A)$}{}}
We show computation of the following object:
\begin{equation}
    \begin{split}
        \exvc \pur^2(\psi_C ^A)&=\exvc \Tr[(\ttt_A ^{\ot 2} \ot \id_B ^{\ot 4})\psi_C ^{\ot 4}]\\
        &=\alpha\Tr[(\ttt_A ^{\ot 2}\ot \id_B ^{\ot 4})Q\Pi_{\sym}]+\beta\Tr[(\ttt_A\ot \id_B ^{\ot 4})\Pi_{\sym}]\\
        &=\frac{\alpha}{24}\sum_{\pi \in S_4}\Tr[(\ttt_A\ot \ttt_A)Q_A\ttt_\pi ^A]\Tr[Q_B\ttt_\pi ^B]+\frac{\beta}{24}\sum_{\pi \in S_4}\Tr[\ttt_A ^{\ot 2} \ttt_\pi ^A]\Tr[\ttt_\pi ^B]
    \end{split}
\end{equation}
and direct evaluation of these traces yields
\begin{equation}
    \exvc \pur^2(\psi_C ^A)=\frac{(d-1) \left(d^2+3 d d_A^2+d_A^4+d_A^2\right)+(d_A ^2-1) (d_A ^2-d^2)  \mlin(\psi) }{(d-1) (d+1) (d+2) d_A^2}
\end{equation}
\subsubsection{Average of \texorpdfstring{$\pur(\psi_C ^A)\stp(\psi_C ^A)$}{}}
We now show computation of the following object:
\begin{equation}\label{app:sppurAclif}
    \exvc[\pur(\psi_C ^A)\stp(\psi_C ^A)]=d_A\exvc \Tr\{[(Q_A\ot \ttt_A)\ot \id_B ^{\ot 6}]\psi_C ^{\ot 6}\}\,.
\end{equation}
To compute this object, we show the Clifford average of 6 copies of a pure state, shown in \cite{appcite:Bittel_Eisert_Leone_Mele_Oliviero_2025}:
\begin{equation}\label{app:psiclifavg6}
    \exvc \psi_C ^{\ot 6}=\sum_{\Omega \in \mathcal{P}_6}p_{\Omega}(\psi)\frac{\Pi_{\sym}\Omega \Pi_{\sym}}{\Tr[\Omega\Pi_{\sym}]}
\end{equation}
with $\mathcal{P}_6=\{\Omega ^{(6)}_0,\Omega ^{(6)}_1,\Omega ^{(6)}_2,\Omega ^{(6)}_3\}$ being the set of algebraically independent Pauli monomials of order 6, namely the following set of operators acting on $\hi^{\ot 6}$:
\begin{equation}
    \begin{split}
        \Omega^{(6)} _0  &=\id^{\ot 6}\\
        \Omega^{(6)} _1 &=\frac{1}{d}\sum_{P} P^{\ot 4}\ot \id^{\ot 2}\\
        \Omega^{(6)} _2 &=\frac1d \sum_P P^{\ot 6}\\
        \Omega^{(6)} _4 &=\frac{1}{d^2}\sum_{P,Q} P^{\ot 2}\ot (PQ)^{\ot 2}\ot Q^{\ot 2} 
        \end{split}
\end{equation}
and 
\begin{equation}
    p_{\Omega}(\psi)=\sum_{\Omega, \Omega' \in \mathcal{P}_6}(\mathcal{S}^{(6)})^{-1} _{\Omega\Omega'}  \frac{\Tr(\psi^{\ot 6} \Omega')}{\Tr(\Omega' \Pi_{\sym})}
\end{equation}
and finally the $\mathcal{S}^{(6)}$ matrix being defined as
\begin{equation}
    \mathcal{S}^{(6)} _{\Omega\Omega'}:=\frac{\Tr(\Omega\Pi_{\sym}\Omega' \Pi_{\sym})}{\Tr(\Omega \Pi_\sym)\Tr(\Omega' \Pi_\sym)}\,.
\end{equation}
Substituting this expression in Eq. \eqref{app:sppurAclif} we get the following:
\begin{equation}
    \begin{split}
        \exvc[\stp(\psi_C ^A)\pur(\psi_C ^A)]&=d_A \sum_{\Omega \in \mathcal{P}_6}\frac{p_{\Omega}(\psi)}{\Tr(\Omega \Pi_{\sym})}\sum_{\pi,\sigma \in S_6}\Tr{[(Q_A\ot\ttt_A)\ot \id_B ^{\ot 6}]\ttt_\pi \Omega\ttt_\sigma}\\
        &=d_A \sum_{\Omega \in \mathcal{P}_6}\frac{p_{\Omega}(\psi)}{\Tr(\Omega \Pi_{\sym})}\sum_{\pi,\sigma \in S_6}\Tr{[(Q_A\ot\ttt_A)]\ttt_\pi ^A\Omega_A \ttt_\sigma ^A}\Tr[\ttt_\pi ^B\Omega_B \ttt_\sigma ^B]
    \end{split}
\end{equation}

where we used the fact that Pauli monomials, like $Q$ (which is indeed a Pauli monomial) and permutation operators, factorize on qubits. As usual, computing all the traces yields the following result:
\begin{equation}
\begin{split}
\mathbb{E}_{C}\!\left[\mathrm{SP}(\psi_C^A)\,
\mathrm{Pur}(\psi_C^A)\right]
&=
\frac{1}{
17280 (d-2)(d-1)(d+1)(d+2)(d+8)(d+16)d_A^4
}
\\
&\quad \times
\Bigg[
\begin{aligned}
&-\frac{2880 d_A^2}{d(d+23)}
\left(2d^2(7d+38)+(d(d+15)+74)d_A^4
+(d(d(d+43)+240)+76)d_A^2\right) \\
&\quad \times
\left(
\begin{aligned}
&6(d+23)(d(d+8)\M - d(d+4) + 52\M - 36) \\
&+ (d+4)(d+23)^2\oF \\
&+ 432(d+6)\oT
\end{aligned}
\right) \\[4pt]
&+\frac{518400d_A}{(d+23)^2}
\left(12d^2+3(d+6)d_A^4
+(d(d(d+14)+68)+28)d_A^2\right) \\
&\quad \times
\left(
\begin{aligned}
&3(d+23)(3(d+6)\M - 3d - 10) \\
&+ (d+23)^2\oF \\
&+ 12(d+1)(d+14)\oT
\end{aligned}
\right) \\[4pt]
&+144d_A
\left(4d^2+(d+6)d_A^4+(d(d+14)+4)d_A^2\right) \\
&\quad \times
\left(
\begin{aligned}
&24(d+23)(15(d+4)\M - 4(4d+7)) \\
&+ d(d+13)(d+23)^2\oF \\
&+ 17280\oT
\end{aligned}
\right) \\[4pt]
&-\frac{1}{(d+4)(d+23)}
\left(
\begin{aligned}
&d^3(d(d+12)+40)+2(d+8)d_A^6 \\
&+ (d(d(2d+29)+152)+104)d_A^4 \\
&+ d(d(d(3d+43)+196)+120)d_A^2
\end{aligned}
\right) \\
&\quad \times
\left(
\begin{aligned}
&-24(d+23)(d^2+15(d+4)\M-2d-60) \\
&+ (d-32)(d+23)^2\oF \\
&- 17280\oT
\end{aligned}
\right)
\end{aligned}
\Bigg].
\end{split}
\end{equation}

\subsubsection{Average of \texorpdfstring{$\stp^2(\psi_C ^A)$}{}}
We now show computations of the following object:
\begin{equation}\label{app:spA2}
    \begin{split}
        \exvc[\stp^2(\psi_C ^A)]&=d_A ^2\exvc \Tr[(Q_A ^{\ot 2}\ot \id_B ^{\ot 8})\psi_C ^{\ot 8}]\,.
    \end{split}
\end{equation} To compute this object, we show the Clifford average of 8 copies of a pure state, shown in \cite{appcite:Bittel_Eisert_Leone_Mele_Oliviero_2025}:
\begin{equation}\label{app:psiclifavg8}
    \exvc \psi_C ^{\ot 8}=\sum_{\Omega \in \mathcal{P}_8}p_{\Omega}(\psi)\frac{\Pi_{\sym}\Omega \Pi_{\sym}}{\Tr[\Omega\Pi_{\sym}]}
\end{equation}
with $\mathcal{P}_8=\{\Omega_0,\Omega_1,\Omega_2,\Omega_3,\Omega_4,\Omega_5,\Omega_6,\Omega_7,\Omega_8,\Omega_9,\Omega_{10},\Omega_{11},\Omega_{12}\}$ being  set of algebraically independent Pauli monomials of order 8, namely the following set of operators acting on $\hi^{\ot 8}$:
\begin{equation}
    \begin{split}
        \Omega^{(8)} _0  &=\id^{\ot 8}\\
        \Omega^{(8)} _1 &=\frac{1}{d}\sum_{P} P^{\ot 4}\ot \id^{\ot 4}\\
        \Omega^{(8)} _2 &=\frac1d \sum_P P^{\ot 6}\ot \id^{\ot 2}\\
        \Omega^{(8)} _3 &=\frac1d \sum_P P^{\ot 8}\\
        \Omega^{(8)} _4 &=\frac{1}{d^2}\sum_{P,Q} P^{\ot 2}\ot (PQ)^{\ot 2}\ot Q^{\ot 2} \ot \id^{\ot 2}\\
        \Omega^{(8)} _5 &=\frac{1}{d^2}\sum_{P,Q}P^{\ot 3}\ot (PQ)\ot Q^{\ot 3}\\
        \Omega^{(8)} _6 &=\frac{1}{d^2}\sum_{P,Q}P^{\ot 4}\ot Q^{\ot 4}\\
        \Omega^{(8)} _7 &=\frac{1}{d^2}\sum_{P,Q} P^{\ot 2}\ot (PQ)^{\ot 2}\ot Q^{\ot 4}\\
        \Omega^{(8)} _8 &= \frac{1}{d^2}\sum_{P,Q}P^{\ot 2}\ot (PQ)^{\ot 4}\ot Q^{\ot 2}\\
        \Omega^{(8)} _9 &=\frac{1}{d^3}\sum_{P,Q,K}P^{\ot 2}\ot (PK)^{\ot 2}\ot (QK)^{\ot 2}\ot Q^{\ot 2}\\
        \Omega^{(8)} _{10}&=\frac{1}{d^3}\sum_{P,Q,K}P\ot (PK)\ot (PQ)\ot (PQK)\ot (QK)\ot Q \ot K\ot \id \\
        \Omega^{(8)} _{11} &=\frac{1}{d^3}\sum_{P,Q,K} (PK)\ot P\ot (PQK)\ot (PQ)\ot Q^{\ot 2}\ot K^{\ot 2}\\
        \Omega^{(8)} _{12}&=\frac{1}{d^4}\sum_{P,Q,K,L} (PL)\ot P \ot (PQKL)\ot (PQK)\ot (QL)\ot Q \ot (KL)\ot K\,,
    \end{split}
\end{equation}
and 
\begin{equation}
    p_{\Omega}(\psi)=\sum_{\Omega, \Omega' \in \mathcal{P}_8}(\mathcal{S}^{(8)})^{-1} _{\Omega\Omega'}  \frac{\Tr(\psi^{\ot 8} \Omega')}{\Tr(\Omega' \Pi_{\sym})}
\end{equation}
and finally the $\mathcal{S}^{(8)}$ matrix being defined as
\begin{equation}
    \mathcal{S}^{(8)} _{\Omega\Omega'}:=\frac{\Tr(\Omega\Pi_{\sym}\Omega' \Pi_{\sym})}{\Tr(\Omega \Pi_\sym)\Tr(\Omega' \Pi_\sym)}\,.
\end{equation}
Substituting this expression in Eq. \eqref{app:spA2} we get the following:
\begin{equation}
    \begin{split}
        \exvc[\stp^2(\psi_C ^A)]&=d_A ^2 \sum_{\Omega \in \mathcal{P}}\frac{p_{\Omega}(\psi)}{\Tr(\Omega \Pi_{\sym})}\sum_{\pi,\sigma \in S_8}\Tr[(Q_A ^{\ot 2}\ot \id_B ^{\ot 8})\ttt_\pi\Omega \ttt_\sigma]\\
        &=d_A ^2 \sum_{\Omega \in \mathcal{P}}\frac{p_{\Omega}(\psi)}{\Tr(\Omega \Pi_{\sym})}\sum_{\pi,\sigma \in S_8}\Tr[Q_A ^{\ot 2}\ttt_\pi ^A \Omega_A\ttt_\sigma ^A]\Tr[\ttt_\pi ^B\Omega_B \ttt_\sigma ^B]
    \end{split}
\end{equation}

where we used the fact that Pauli monomials, like $Q$ (which is indeed a Pauli monomial) and permutation operators, factorize on qubits. As usual, computing all the traces yields the following result:

\begin{align}
\exvc\!\left[\stp^2(\psi_C^A)\right]
&=
\frac{
N
}{
126(d-8)(d-4)(d-2)(d-1)(d+1)(d+2)(d+4)
(d+8)(d+16)(d+32)(d+64)d_A^4
},
\end{align}
with
\begin{align}
&N\\
&=
C_{11}d^{11}+C_{10}d^{10}+C_9d^9+C_8d^8+C_7d^7
+C_6d^6+C_5d^5+C_4d^4
+C_3d^3+C_2d^2+C_1d+C_0 ,
\end{align}
and \newpage
\begin{equation}
\begin{split}
C_{11}
&=9(d_A^2+13),
\\[4pt]
C_{10}
&=
9\Bigl[
56(3d_A^2+25)
+(d_A^2-1)
\Bigl(
14(-2\M+\TrO{3}+2)d_A^2
\\
&\hspace{5.0cm}
+13(-70\M+28\TrO{2}+\TrO{3}+70)
\Bigr)
\Bigr],
\\[4pt]
C_9
&=
-3\Bigl[
6(42d_A^4-2916d_A^2-21619)
\\
&\hspace{1.0cm}
-7(d_A^2-1)
\Bigl(
6(\M-1)^2d_A^4
\\
&\hspace{2.0cm}
-\bigl(
12\M(\M+110)+180\TrO{2}-672\TrO{3}
+160\TrO{4}
\\
&\hspace{3.0cm}
+261\TrO{5}+238\TrO{8}-1332
\bigr)d_A^2
\\
&\hspace{2.0cm}
-6\bigl(
65\M^2+5440\M-1052\TrO{2}-74\TrO{3}
+1004\TrO{4}
\\
&\hspace{3.0cm}
+2674\TrO{5}+390\TrO{7}-5505
\bigr)
-3482\TrO{8}
\Bigr)
\Bigr],
\\
C_8
&=
-\Bigl[
36(2119d_A^4+23959d_A^2-107446)
\\
&\hspace{1.0cm}
+(d_A^2-1)
\Bigl(
-8976\TrO{10}d_A^2-52451\TrO{11}d_A^2
\\
&\hspace{2.0cm}
+21\Bigl(
6\bigl(
2(111-56\M)\M+\TrO{3}
+21\TrO{5}
\\
&\hspace{3.0cm}
+12\TrO{7}+7\TrO{9}-110
\bigr)d_A^4
\\
&\hspace{2.7cm}
+\bigl(
\M(1083\M+49262)+9744\TrO{2}
-21448\TrO{3}
\\
&\hspace{3.0cm}
+13699\TrO{4}+12028\TrO{5}
-14652\TrO{7}
\\
&\hspace{3.0cm}
+17612\TrO{8}+192\TrO{9}-50345
\bigr)d_A^2
\\
&\hspace{2.7cm}
+15036\M^2
\Bigr)
\\
&\hspace{2.0cm}
+4\Bigl(
3923094\M+445032\TrO{2}
-82428\TrO{3}
\\
&\hspace{2.7cm}
+21\bigl(
85685\TrO{4}+168740\TrO{5}
-79020\TrO{7}
\\
&\hspace{3.2cm}
-5342\TrO{8}-1506\TrO{9}
-3888\TrO{10}
\\
&\hspace{3.2cm}
-190573
\bigr)
-810439\TrO{11}
\Bigr)
\Bigr)
\Bigr],
\\
C_7
&=
3(69d_A^6-831341d_A^4-12451760d_A^2+548128)
\\
&\hspace{1.0cm}
-(d_A^2-1)
\Bigl[
\Bigl(
-21\bigl(
6\M(3787\M-7530)+2052\TrO{2}
\\
&\hspace{2.7cm}
-672\TrO{3}+2126\TrO{4}
-9567\TrO{5}
\\
&\hspace{2.7cm}
-4392\TrO{7}+1718\TrO{8}
-4374\TrO{9}
\\
&\hspace{2.7cm}
+22458
\bigr)
-39788\TrO{11}-1206\TrO{12}
\Bigr)d_A^4
\\
&\hspace{2.0cm}
+4\Bigl(
21\M(10345\M+201751)+541968\TrO{2}
\\
&\hspace{2.7cm}
-895002\TrO{3}+2209137\TrO{4}
+1692243\TrO{5}
\\
&\hspace{2.7cm}
-3991617\TrO{7}+2574684\TrO{8}
+115073\TrO{9}
\\
&\hspace{2.7cm}
-2(76625\TrO{10}+62545\TrO{11}
+555\TrO{12}+2227008)
\Bigr)d_A^2
\\
&\hspace{2.0cm}
+8\Bigl(
42\M(26153\M+416393)+7792512\TrO{2}
\\
&\hspace{2.7cm}
-649932\TrO{3}
+7\bigl(
1965846\TrO{4}+3036168\TrO{5}
\\
&\hspace{3.2cm}
-2812878\TrO{7}+7212\TrO{8}
-70639\TrO{9}
\\
&\hspace{3.2cm}
-166744\TrO{10}-768322\TrO{11}
+5832\TrO{12}
\\
&\hspace{3.2cm}
-2655276
\bigr)
\Bigr)
\Bigr],
\end{split}
\end{equation}
\begin{equation*}
\begin{split}
C_6
&=
-\Bigl[
36(1619d_A^6+1047225d_A^4+5214180d_A^2+2925680)
\\
&\hspace{1.0cm}
+(d_A^2-1)
\Bigl(
-126(21\TrO{5}+12\TrO{7}+7\TrO{9})d_A^6
\\
&\hspace{2.0cm}
+\bigl(
21(556100-286143\M)\M
-2347380\TrO{2}
\\
&\hspace{2.7cm}
+859065\TrO{3}-2917467\TrO{4}
+6067740\TrO{5}
\\
&\hspace{2.7cm}
+2012220\TrO{7}-1315020\TrO{8}
+2685032\TrO{9}
\\
&\hspace{2.7cm}
+215624\TrO{10}-2473933\TrO{11}
-123660\TrO{12}
\\
&\hspace{2.7cm}
-5669097
\bigr)d_A^4
\\
&\hspace{2.0cm}
+4\bigl(
21\M(87711\M+1147310)+4384800\TrO{2}
\\
&\hspace{2.7cm}
+6520704\TrO{3}+24848131\TrO{4}
+40346292\TrO{5}
\\
&\hspace{2.7cm}
-77792652\TrO{7}+29642844\TrO{8}
+1659707\TrO{9}
\\
&\hspace{2.7cm}
-4570232\TrO{10}+4871447\TrO{11}
-157298\TrO{12}
\\
&\hspace{2.7cm}
-25935441
\bigr)d_A^2
\\
&\hspace{2.0cm}
+32\bigl(
42\M(107871\M-251632)+12969432\TrO{2}
\\
&\hspace{2.7cm}
-2012136\TrO{3}+11329192\TrO{4}
-11987472\TrO{5}
\\
&\hspace{2.7cm}
-28065492\TrO{7}+14867622\TrO{8}
-2095667\TrO{9}
\\
&\hspace{2.7cm}
-3444720\TrO{10}+9815176\TrO{11}
+291802\TrO{12}
\\
&\hspace{2.7cm}
+6037962
\bigr)
\Bigr)
\Bigr],
\\
C_5
&=
2\Bigl[
3(90d_A^8-258013d_A^6-16431549d_A^4
+406140672d_A^2-282088448)
\\
&\hspace{1.0cm}
-(d_A^2-1)\Bigl(
\bigl(
21(45\M(19\M-50)+936\TrO{2}
\\
&\hspace{2.7cm}
+983\TrO{4}-4914\TrO{5}
-2196\TrO{7}+740\TrO{8}
\\
&\hspace{2.7cm}
-2187\TrO{9}+1395)
+19894\TrO{11}+603\TrO{12}
\bigr)d_A^6
\\
&\hspace{2.0cm}
+\bigl(
21(337442-510001\M)\M
-26228076\TrO{2}
\\
&\hspace{2.7cm}
+16337418\TrO{3}-27414814\TrO{4}
+51732828\TrO{5}
\\
&\hspace{2.7cm}
+13223952\TrO{7}-16992213\TrO{8}
+13202714\TrO{9}
\\
&\hspace{2.7cm}
+6417492\TrO{10}-22338428\TrO{11}
-1774516\TrO{12}
\\
&\hspace{2.7cm}
+3623739
\bigr)d_A^4
\\
&\hspace{2.0cm}
+4\bigl(
42\M(111469\M-1437661)+16382352\TrO{2}
\\
&\hspace{2.7cm}
+43420476\TrO{3}+24241630\TrO{4}
+97462344\TrO{5}
\\
&\hspace{2.7cm}
-239528142\TrO{7}+69427344\TrO{8}
-9667791\TrO{9}
\\
&\hspace{2.7cm}
-14574232\TrO{10}+83267814\TrO{11}
-736472\TrO{12}
\\
&\hspace{2.7cm}
+55700064
\bigr)d_A^2
\\
&\hspace{2.0cm}
+256\bigl(
21\M(114719\M-1195500)+127344\TrO{2}
\\
&\hspace{2.7cm}
-1534104\TrO{3}-7236726\TrO{4}
-38424792\TrO{5}
\\
&\hspace{2.7cm}
+7346178\TrO{7}+8662668\TrO{8}
-1321523\TrO{9}
\\
&\hspace{2.7cm}
-1071272\TrO{10}+13326082\TrO{11}
+215295\TrO{12}
\\
&\hspace{2.7cm}
+22696401
\bigr)
\Bigr)
\Bigr].
\end{split}
\end{equation*}
\begin{equation*}
\begin{split}
C_4
&=
4\Bigl[
6(5544d_A^8+1946355d_A^6+130634365d_A^4
+547755496d_A^2-437878784)
\\
&\hspace{1.0cm}
-(d_A^2-1)
\Bigl(
\bigl(
21\M(36225\M-81358)+536508\TrO{2}
\\
&\hspace{2.7cm}
-102135\TrO{3}+657447\TrO{4}
-1580082\TrO{5}
\\
&\hspace{2.7cm}
-426132\TrO{7}+236292\TrO{8}
-672266\TrO{9}
\\
&\hspace{2.7cm}
-51662\TrO{10}+631596\TrO{11}
+30915\TrO{12}
\\
&\hspace{2.7cm}
+947793
\bigr)d_A^6
\\
&\hspace{2.0cm}
+\bigl(
-126\M(71117\M+804866)-166976712\TrO{2}
\\
&\hspace{2.7cm}
+78923196\TrO{3}-139225884\TrO{4}
+311595480\TrO{5}
\\
&\hspace{2.7cm}
+191608200\TrO{7}-204402870\TrO{8}
+14483679\TrO{9}
\\
&\hspace{2.7cm}
+46886376\TrO{10}-112578704\TrO{11}
-8484702\TrO{12}
\\
&\hspace{2.7cm}
+110373858
\bigr)d_A^4
\\
&\hspace{2.0cm}
+8\bigl(
42\M(694673\M-7733760)+13438824\TrO{2}
\\
&\hspace{2.7cm}
-32979096\TrO{3}-117500264\TrO{4}
-558984048\TrO{5}
\\
&\hspace{2.7cm}
+348176052\TrO{7}-80056662\TrO{8}
-40885159\TrO{9}
\\
&\hspace{2.7cm}
-10330256\TrO{10}+153404664\TrO{11}
+5072602\TrO{12}
\\
&\hspace{2.7cm}
+295641654
\bigr)d_A^2
\\
&\hspace{2.0cm}
+512\bigl(
84\M(32117\M-388138)+182784\TrO{2}
\\
&\hspace{2.7cm}
-1371072\TrO{3}
-7\bigl(
1540672\TrO{4}+6776064\TrO{5}
\\
&\hspace{3.2cm}
-2889216\TrO{7}-706944\TrO{8}
+217477\TrO{9}
\\
&\hspace{3.2cm}
-1540672\TrO{11}-4272252
\bigr)
+267818\TrO{12}
\bigr)
\Bigr)
\Bigr],\\
C_3
&=
8d_A^2\Bigl[
3(166229d_A^6+29969167d_A^4
+521213596d_A^2-653686528)
\\
&\hspace{1.0cm}
+(d_A^2-1)
\Bigl(
\bigl(
21(638884-347063\M)\M
-6302604\TrO{2}
\\
&\hspace{2.7cm}
+3635688\TrO{3}-5733322\TrO{4}
+13821444\TrO{5}
\\
&\hspace{2.7cm}
+1316322\TrO{7}-2951571\TrO{8}
+3358215\TrO{9}
\\
&\hspace{2.7cm}
+1527748\TrO{10}-5647152\TrO{11}
-444184\TrO{12}
\\
&\hspace{2.7cm}
-6128241
\bigr)d_A^4
\\
&\hspace{2.0cm}
+4\bigl(
189\M(90787\M+190598)+146728260\TrO{2}
\\
&\hspace{2.7cm}
+4114332\TrO{3}
+7\bigl(
21326418\TrO{4}-29100384\TrO{5}
\\
&\hspace{3.2cm}
-53525700\TrO{7}+37205703\TrO{8}
+610142\TrO{9}
\\
&\hspace{3.2cm}
-6861760\TrO{10}+22111114\TrO{11}
-7597395
\bigr)
\\
&\hspace{2.7cm}
+1176360\TrO{12}
\bigr)d_A^2
\\
&\hspace{2.0cm}
-64\bigl(
21\M(1336213\M-7542644)+48357456\TrO{2}
\\
&\hspace{2.7cm}
-25271592\TrO{3}-31110562\TrO{4}
-302135400\TrO{5}
\\
&\hspace{2.7cm}
+282160998\TrO{7}-68843964\TrO{8}
-13055308\TrO{9}
\\
&\hspace{2.7cm}
+4199016\TrO{10}-21595658\TrO{11}
+3461699\TrO{12}
\\
&\hspace{2.7cm}
+130335051
\bigr)
\Bigr)
\Bigr],\\
\end{split}
\end{equation*}
\begin{equation*}
\begin{split}
C_2
&=
-64d_A^2\Bigl[
9(14171d_A^6+2165933d_A^4+34104312d_A^2+2609152)
\\
&\hspace{1.0cm}
-(d_A^2-1)
\Bigl(
\bigl(
126(44905-44148\M)\M
-10134180\TrO{2}
\\
&\hspace{2.7cm}
+5335092\TrO{3}-7036148\TrO{4}
+22217832\TrO{5}
\\
&\hspace{2.7cm}
+7009758\TrO{7}-10929513\TrO{8}
+1006985\TrO{9}
\\
&\hspace{2.7cm}
+2639656\TrO{10}-6782356\TrO{11}
-540125\TrO{12}
\\
&\hspace{2.7cm}
-95382
\bigr)d_A^4
\\
&\hspace{2.0cm}
+\bigl(
1008\M(18413\M+89387)+118143648\TrO{2}
\\
&\hspace{2.7cm}
+72778176\TrO{3}+258192032\TrO{4}
-278647488\TrO{5}
\\
&\hspace{2.7cm}
-720727056\TrO{7}+376712616\TrO{8}
-15898253\TrO{9}
\\
&\hspace{2.7cm}
-64(1357441\TrO{10}-6874147\TrO{11}+1697850)
\\
&\hspace{2.7cm}
-7206886\TrO{12}
\bigr)d_A^2
\\
&\hspace{2.0cm}
-96\bigl(
84\M(43969\M-7714)+10784256\TrO{2}
\\
&\hspace{2.7cm}
-1222848\TrO{3}
+7\bigl(
1233472\TrO{4}-2423040\TrO{5}
\\
&\hspace{3.2cm}
+3385728\TrO{7}-1505664\TrO{8}
-171505\TrO{9}
\\
&\hspace{3.2cm}
-1233472\TrO{11}+30062\TrO{12}
-435060
\bigr)
\bigr)
\Bigr)
\Bigr],
\\[4pt]
C_1
&=
-256d_A^4\Bigl[
288(7917d_A^4+202740d_A^2-235745)
\\
&\hspace{1.0cm}
-(d_A^2-1)
\Bigl(
\bigl(
42\M(9633\M-251926)-17585568\TrO{2}
\\
&\hspace{2.7cm}
+822288\TrO{3}-17473036\TrO{4}
+29172696\TrO{5}
\\
&\hspace{2.7cm}
+38734416\TrO{7}-30383808\TrO{8}
-851445\TrO{9}
\\
&\hspace{2.7cm}
+5512120\TrO{10}-16913176\TrO{11}
-168784\TrO{12}
\\
&\hspace{2.7cm}
+10176306
\bigr)d_A^2
\\
&\hspace{2.0cm}
+2\bigl(
420\M(63945\M-419704)+48484800\TrO{2}
\\
&\hspace{2.7cm}
-26805696\TrO{3}-38347288\TrO{4}
-340560192\TrO{5}
\\
&\hspace{2.7cm}
+289507176\TrO{7}-60181296\TrO{8}
-14376831\TrO{9}
\\
&\hspace{2.7cm}
+3127744\TrO{10}-8269576\TrO{11}
+3676994\TrO{12}
\\
&\hspace{2.7cm}
+149418780
\bigr)
\Bigr)
\Bigr],
\end{split}
\end{equation*}
\begin{equation*}
\begin{split}
C_0
&=
-2048d_A^4\Bigl[
1344(619d_A^4-2399d_A^2-4364)
\\
&\hspace{1.0cm}
-(d_A^2-1)
\Bigl(
\bigl(
84\M(31285\M-159746)-3279360\TrO{2}
\\
&\hspace{2.7cm}
-2821056\TrO{3}-9727424\TrO{4}
-213696\TrO{5}
\\
&\hspace{2.7cm}
+27524448\TrO{7}-12790848\TrO{8}
-174755\TrO{9}
\\
&\hspace{2.7cm}
+2499648\TrO{10}-11197984\TrO{11}
+309034\TrO{12}
\\
&\hspace{2.7cm}
+10790724
\bigr)d_A^2
\\
&\hspace{2.0cm}
+8\bigl(
84\M(8215\M-26834)+4066944\TrO{2}
\\
&\hspace{2.7cm}
-629952\TrO{3}
+7\bigl(
269968\TrO{4}-1755648\TrO{5}
\\
&\hspace{3.2cm}
+1630800\TrO{7}-476256\TrO{8}
-91499\TrO{9}
\\
&\hspace{3.2cm}
-269968\TrO{11}+223428
\bigr)
+112390\TrO{12}
\bigr)
\Bigr)
\Bigr].
\end{split}
\end{equation*}
Putting all these expression together gives the Clifford variance of the SE gap for partial trace.
\subsubsection{Violation probability of SE under partial trace in the Clifford ensemble}
Since the Clifford orbit of a quantum state is a discrete set, we are not able to use L\'evy lemma to establish typicality, hence we will Chebyshev inequality to bound the probability of violations. This is the reason why we computed the variance of the partial trace SE gap on the Clifford orbit in the previous section:
\begin{equation}
    \begin{split}
         \Delta_C ^2 \Bar{M} ^{A}&=O\left(\frac{1}{d_A}\right)+
         O\left(\frac{1}{d}\right)
    \end{split}
\end{equation}
and using Chebyshev inequality, we get that 
\begin{equation}
    {\rm Pr}_C(\mbar^A(\psi)\leq 0)\leq O\left(\frac{1}{d_A}\right)+
    O\left(\frac{1}{d}\right)\,,
\end{equation}
meaning that $\mlin$ is a $\eta-$ magic proxy in the Clifford orbit with $\eta=O(1)$. Although this is a weaker typicality than the one found in the Haar ensemble, violation probability is still exponentially vanishing with the number of qubits.
\subsection{Average SE gap for dephasing in the Clifford orbit}\label{app:avg_db_clif}
In this section, we show complete derivation of the average SE gap for the dephasing map in the Clifford ensemble.
Starting from a pure state $\psi:=\ketbra{\psi}$ and defining $\psi_C:=C\psi C^\dag$, the gap reads
\begin{equation}
    \begin{split}
        \exvc[\mbar^{D_B}(\psi_C)]&=\mlin(\psi)-\exvc[\pur(D_B\psi_C)]+\exvc[\stp(D_B \psi_C)]\,.\\
        &=\mlin(\psi)-\frac{2}{d+1}+\exvc[\stp(D_B \psi_C)],.
    \end{split}
\end{equation}
Focusing now on the last term, which reads
\begin{equation}
    \begin{split}
        \exvc[\stp(D_B \psi_C)]&=d\exvc \Tr[Q (D_B \psi_C)^{\ot 4}]=d\exvc\Tr[(D_B ^{\ot 4}Q)\psi_C]\,,
    \end{split}
\end{equation}
we substitute the expression in Eq. \eqref{app:psiclifavg4} for the Clifford averages of 4 tensor copies of a pure state and we get the following:
\begin{equation}
    \begin{split}
        \exvc[\stp(D_B \psi_C)]&=d\bigg(\frac{\alpha(\psi)}{24}\sum_{\pi \in S_4} \Tr[(D_B ^{\ot 4}Q)Q\ttt_\pi]+\frac{\beta(\psi)}{24}\sum_{\pi \in S_4}\Tr[(D_B ^{\ot 4}Q)\ttt_\pi]\bigg)
    \end{split}
\end{equation}
and explicitly evaluating these traces we get the following result:
\begin{equation}
   \exvc[\stp(D_B \psi_C)]= \frac{2-\mlin(\psi)}{d+1}
\end{equation}
and summing the terms together we get the following result for the SE gap for dephasing in the Clifford orbit:
\begin{equation}
\exvc[\mbar^{D_B}(\psi_C)]=\frac{d }{d+1}\mlin(\psi)\,.
\end{equation}
\subsection{Variance of SE gap for dephasing in the Clifford orbit}\label{app:variance_db_clif}
In this section we focus closely on the computation of the following object:
\begin{equation}
\begin{split}
    \Delta^2 \mbar^{D_B}&=\Delta^2[\mlin(\psi_C)]+\Delta^2[\mlin(D_B\psi_C)]-2{\rm cov}[\mlin(\psi_C),\mlin(D_B\psi_C)]\\&=\Delta^2[\mlin(D_B\psi_C)]
\end{split}
\end{equation}
where again the first and last terms vanish due to $\mlin$ being constant on the Clifford orbit. Inspecting the variance of dephased states, this reads
\begin{equation}
\begin{split}
    \Delta^2[\mlin(D_B\psi_C)]&=\exvc [\mlin ^2(D_B\psi_C)]-[\exvc \mlin(D_B\psi_C)]^2\\&=
    \exvc [\mlin ^2(D_B\psi_C)]-\left(\frac{\mlin(\psi)}{d+1}\right)^2
\end{split}
\end{equation}
Focusing on the expectation value of the square of dephased $\mlin$ yields
\begin{equation}
    \begin{split}
        \exvc [\mlin ^2(D_B\psi_C)]&=\exvc[\pur^2(D_B\psi_C)]+\exvc[\stp^2(D_B\psi_C)]-2\exvc[\stp(D_B \psi_C)\pur(D_B\psi_C)]
    \end{split}
\end{equation}
We will focus on each piece in a separate subsubsection.
\subsubsection{Average of \texorpdfstring{$\pur^2(D_B\psi_C)$}{}}
We show computation of the following object:
\begin{equation}
    \begin{split}
        \exvc[\pur^2 (D_B\psi_C)]&=\exvc \Tr[\ttt\ot \ttt (D_B\psi_C)^{\ot 4}]=\exvc \Tr[(D_B ^{\ot 4}\ttt\ot \ttt) \psi_C^{\ot 4}]\\
        &=\alpha\Tr[(D_B ^{\ot 4}\ttt\ot \ttt)Q\Pi_{\rm sym}]+\beta\Tr[(D_B ^{\ot 4}\ttt\ot \ttt)\Pi_{\rm sym}]\\
        &=\frac{\alpha}{24}\sum_{\pi \in S_4}\Tr[(D_B ^{\ot 4}\ttt\ot \ttt)Q\ttt_\pi]+\frac{\beta}{24}\sum_{\pi \in S_4}\Tr[(D_B ^{\ot 4}\ttt\ot \ttt)\ttt_\pi]\,.
    \end{split}
\end{equation}
exploiting the usual properties of factorization, direct evaluation of these traces yield the result:
\begin{equation}
    \exvc[\pur^2 (D_B\psi_C)]=\frac{2 (3-\mlin)}{(d+1) (d+2)}
\end{equation}
\subsubsection{Average of \texorpdfstring{$\pur(D_B\psi_C)\stp(D_B\psi_C)$}{}}
In this section, we show complete computation of the following object:
\begin{equation}
    \begin{split}
        \exvc &\pur(D_B\psi_C)\stp(D_B\psi_C)=d\exvc \Tr[\ttt\ot Q (D_B\psi_C)^{\ot 6}]\\
        &=\exvc\Tr[(D_B ^{\ot 6}\ttt\ot Q)\psi_C ^{\ot 6}]
    \end{split}
\end{equation}
Substituting the expression of the six-copy Clifford average of a state shown in Eq. \eqref{app:psiclifavg6} \cite{appcite:Bittel_Eisert_Leone_Mele_Oliviero_2025} we get the following:
\begin{equation}
    \begin{split}
        \exvc &\pur(D_B\psi_C)\stp(D_B\psi_C)=d\sum_{\Omega \in \mathcal{P}_6}\frac{p_{\Omega}(\psi)}{\Tr(\Omega \Pi_{\sym})}\sum_{\pi,\sigma \in S_6}\Tr{(D_B ^{\ot 6}\ttt\ot Q)\ttt_\pi \Omega\ttt_\sigma}
    \end{split}
\end{equation}
Again, due to the factorization properties of the operators involved, we are able to completely carry out these sums of traces, yielding
\begin{equation}
      \exvc \pur(D_B\psi_C)\stp(D_B\psi_C)=4\frac{d (1-\mlin)+\stp_3(\psi)+1}{(d+1) (d+2)}=O\left(\frac{1}{d}\right)
\end{equation}
\subsubsection{Average of \texorpdfstring{$\stp^2(D_B\psi_C)$}{}}
In this section we compute the following object:
\begin{equation}
    \begin{split}
        \exvc & \stp^2(D_B \psi_C)=d^2\exvc \Tr[Q\ot Q(D_B\psi_C)^{\ot 8}]\\
        &=d^2\exvc \Tr[(D_B ^{\ot 8} Q\ot Q)\psi_C ^{\ot 8}]\,.
    \end{split}
\end{equation}
Substituting the expression of the 8-copy Clifford average of a pure state shown in Eq. \eqref{app:psiclifavg8} \cite{appcite:Bittel_Eisert_Leone_Mele_Oliviero_2025}, we get
\begin{equation}
    \begin{split}
         \exvc & \stp^2(D_B \psi_C)=d^2\sum_{\Omega \in \mathcal{P}_8}\frac{p_{\Omega}(\psi)}{\Tr[\Omega\Pi_{\sym}]}\Tr[(D_B ^{\ot 8}Q\ot Q)\Pi_{\sym}\Omega \Pi_{\sym}]\\
         &=\frac{d^2}{8!^2}\sum_{\Omega \in \mathcal{P}_8}\frac{p_{\Omega}(\psi)}{\Tr[\Omega\Pi_{\sym}]}\sum_{\pi,\sigma \in S_6}\Tr[(D_B ^{\ot 8}Q\ot Q)\ttt_\pi\Omega \ttt_\sigma]\,.
    \end{split}
\end{equation}
As usual, we are able to directly compute these traces using the factorization properties of the involved operators, yielding the following result:
\begin{equation}
\begin{split}
        \exvc  \stp^2(D_B \psi_C)&=\frac{\mlin(\psi)^2-4 \mlin(\psi)+\stp_4(\psi)+\Tr[\Omega_8\psi^{\ot 8}]+4}{(d+1) (d+2)}=O\left(\frac{1}{d^2}\right)
\end{split}
\end{equation}
\subsubsection{Final expression of \texorpdfstring{$\Delta^2 \mbar^{D_B}$}{}}
Summing all the elements together yields the final expression of $\Delta^2 \mbar^{D_B}$, which reads
\begin{equation}
    \begin{split}
        \Delta^2 \mbar^{D_B}&=\frac{(6+2d-4d^2)(1-\mlin(\psi))-\mlin(\psi)^2+(-4 d-4) \stp_3(\psi)+(d+1) \stp_4(\psi)+(d+1) \Tr[\Omega_8 \psi^{\ot 8}]}{(d+1)^2(d+2)}\\
        &\leq \frac{(-4 d^2-2 d+2 (d+1)+2)(1-\mlin(\psi))-\mlin(\psi)^2}{(d+1)^2(d+2)}=O\left(\frac{\mlin(\psi)}{d}\right)
    \end{split}
\end{equation}
with $\stp_3(\psi)$ and $\stp_4(\psi)$ being the stabilizer purity of order $3$ and $4$ respectively \cite{appcite:Bittel_Leone_2025_operational}.
\subsubsection{Violation probability of SE under dephasing in the Clifford orbit}
We have seen that
\begin{equation}
    \begin{split}
         \Delta_C ^2 \Bar{M} ^{D_B}&=O\left(\frac{\mlin(\psi)}{d}\right)
    \end{split}    
\end{equation}
Hence, using Chebyshev's inequality, we obtain an upper bound to the probability of violations establishing $\mlin$ as a $\eta-$magic proxy for dephasing with $\eta=O(1)$ in the Clifford orbit.

\section{SE as a proxy in random MPS: techniques}

\subsection{Random MPS}\label{app:kcopiesmps}
Given a (not normalized) state belonging in $\hi=\hi_F ^{\ot N}=\mathbb{C}^{2\ot N}$ , it can be expressed in the following way:
\begin{equation}\label{app:mpsdef}
    \ket\psi=\sum_{i_1,\dots,i_N}\Tr[A^{i_1}[1]\dots A^{i_N}[N] ]\ket{i_1,\dots,i_N}
\end{equation}
where the $A^{i_k} [k]$ are $\chi\times\chi$ matrices: $\chi$ is referred to as the \textit{bond dimension} of the state, and this representation of states is called Matrix Product State representation with periodic boundary conditions (PBC-MPS). If the set $\{A^0[k],A^1 [k]\}$ is the same for all qubits indices $k$, the MPS is called \textit{homogeneous} or \textit{translationally invariant}. An extensive treatment of SE in translationally invariant MPS can be found in \cite{appcite:haug2023QuantifyingNonstabilizernessMatrix}, where they show that it can be efficiently computed in such states.

We will draw and develop the techniques of MPS sequential generation, found in \cite{appcite:pz_typicality_mps,appcite:Garnerone_Oliveira_Haas_Zanardi_2010,appcite:Chen_Garcia_Bu_Jaffe_2024,appcite:Lami_Haug_Nardis_2024_CAMPS,appcite:Schoen_Hammerer_Wolf_Cirac_Solano_2007}, to perform analytical averages over the ensembles of MPS with given bond dimension $\chi$ and prove that $\mlin$ is typically non increasing under dephasing and partial trace over such ensembles.

In this section we will show the full derivation of $k$-tensor copies the average $n-$qubit PBC-MPS with bond dimension $\chi$. 
Consider the initial qubit state to be $\ket{0}^{\ot N}$ and an ancillary system $\hi_A:=\mathbb{C}^{\chi}$. Let $U_k$ be a unitary operation on $\hi_A\ot \hi_F\simeq \mathbb{C}^{2\chi}$ acting on the ancillary system and the $k$-th qubit of the state. The $A[k]$ matrices are then defined by
\begin{equation}\label{app:akdef}
    A^{i} _{\alpha,\beta}[k]:=\bra{i,\alpha}U_k\ket{0,\beta}=\Tr_F \{(\ketbra{i_k}{0}\ot \id_A)U_k\}\,.
\end{equation}
We can write the density matrix of the state in Eq. \eqref{app:mpsdef} as
\begin{equation}\label{app:mpsdensitymatrixdef}
    \begin{split}
        \psi&=\ketbra{\psi}=\sum_{\mathbf{i},\mathbf{m}}\Tr_A\left[\prod_{s=1}^N A^{i_s}[s]\right]\Tr_A\left[\prod_{s=1}^N A^{m_s}[s]\right]^* \ketbra{\mathbf{i}}{\mathbf{m}}
    \end{split}
\end{equation}
where $\mathbf{i}=i_1,\dots, i_N$, and we now exploit the following property:
\begin{equation}\label{app:cyclicalA}
    \Tr_A\left[\prod_{s=1} ^N A^{i_s}[s]\right]=\Tr_{A^{\ot N}}\left[S_A \bigotimes_{s=1}^N A^{i_s}[s]\right]
\end{equation}
where $S_A$ is the operator that cyclically permutes the states in $A^{\ot N}$:
\begin{equation}
    S_A \ket{a}_1\dots\ket{a}_N=\ket{a}_N\ket{a}_1\dots\ket{a}_{N-1}\,.
\end{equation}
Substituting the expression in Eq. \eqref{app:akdef} in Eq. \eqref{app:cyclicalA} we get
\begin{equation}
    \begin{split}
        \Tr_A\left[\prod_{s=1} ^N A^{i_s}[s]\right]&=\Tr_{A^{\ot N}}\left\{S_A \bigotimes_{s=1}^N \Tr_F[(\ketbra{i_s}{0}\ot \id_A)U_s]\right\}\\
        &=\Tr_{A^{\ot N}}\left\{S_A \Tr_{F^{\ot N}}[(\ketbra{\mathbf{i}}{\mathbf{0}}\ot \id_A ^{\ot N})\bigotimes_{s=1} ^N U_s]\right\}\\
        &=\Tr_{A^{\ot N}}\left\{\Tr_{F^{\ot N}}[(\ketbra{\mathbf{i}}{\mathbf{0}}\ot S_A)\bigotimes_{s=1} ^N U_s]\right\}\\
        &=\Tr_{A^{\ot N}\ot F^{\ot N}}\left[ (\ketbra{\mathbf{i}}{\mathbf{0}}\ot S_A)\bigotimes_{s=1} ^N U_s\right]
    \end{split}
\end{equation}
and then the density operator in \eqref{app:mpsdensitymatrixdef} reads
\begin{equation}\label{app:mpsdensitymatrix}
    \begin{split}
        \psi&=\sum_{\mathbf{i},\mathbf{m}}\Tr\left[ (\ketbra{\mathbf{i}}{\mathbf{0}}\ot S_A)\bigotimes_{s=1} ^N U_s\right]\Tr\left[ (\ketbra{\mathbf{m}}{\mathbf{0}}\ot S_A)\bigotimes_{s=1} ^N U_s\right]^* \ketbra{\mathbf{i}}{\mathbf{m}}\\
        &=\sum_{\mathbf{i},\mathbf{m}}\Tr\Big\{[(\ketbra{\mathbf{i}}{\mathbf{0}}\ot S_A)\ot \ketbra{\mathbf{0}}{\mathbf{m}}\ot S_A ^{\dag}]\bigotimes_{s=1}^N U_s \ot U_s ^\dag\Big\} \ketbra{\mathbf{i}}{\mathbf{m}}
    \end{split}
\end{equation}
extracting the unitaries $U_s$ according to the (2$\chi$-dimensional) Haar unitary measure, one obtains the ensemble of Random Matrix Product States (RMPS) of bond dimension $\chi$.

For simplicity, we will carry out the calculation for $k=4$ copies, but the derivation is the same for $k=2$.
Recalling the density matrix expression of Eq. \eqref{app:mpsdensitymatrix}:
\begin{equation}
    \psi=\sum_{\mathbf{i},\mathbf{m}}\Tr\Big\{[(\ketbra{\mathbf{i}}{\mathbf{0}}\ot S_A)\ot \ketbra{\mathbf{0}}{\mathbf{m}}\ot S_A ^{\dag}]\bigotimes_{s=1} U_s \ot U_s ^\dag\Big\} \ketbra{\mathbf{i}}{\mathbf{m}}
\end{equation}
From now on, we drop the bold notation for the multi-indices, to lighten the notation. Four copies of $\psi$ now read
\begin{equation}\label{app:psi4mps}
    \begin{split}
        \psi^{\ot 4}&=\sum_{i,j,k,l,m,n,o,p}\Tr[\ketbra{i}{0}\ot S_A)\bigotimes_{s=1}^N U_s ]\Tr[\ketbra{j}{0}\ot S_A)\bigotimes_{s=1}^N U_s ]\Tr[\ketbra{k}{0}\ot S_A)\bigotimes_{s=1}^N U_s ]\\&\Tr[\ketbra{l}{0}\ot S_A)\bigotimes_{s=1}^N U_s ]\Tr[\ketbra{0}{m}\ot S_A ^\dag)\bigotimes_{s=1}^N U_s ^\dag ]\Tr[\ketbra{0}{n}\ot S_A ^\dag)\bigotimes_{s=1}^N U_s ^\dag ]\\& \Tr[\ketbra{0}{o}\ot S_A ^\dag)\bigotimes_{s=1}^N U_s ^\dag ]\Tr[\ketbra{0}{p}\ot S_A ^\dag)\bigotimes_{s=1}^N U_s ^\dag ]\ketbra{ijkl}{mnop}\\
    &=\!\!\sum_{i,j,k,l,m,n,o,p}\!\Tr[(\ketbra{ijkl}{0000}\ot S_A ^{\ot 4})\ot(\ketbra{0000}{mnop}\ot S_A ^{\dag \ot 4})\bigotimes_{s=1}^N U_s ^{\ot 4}\ot U_s ^{\dag \ot 4}]\ketbra{ijkl}{mnop}
    \end{split}
\end{equation}
Performing the Haar average of this object amounts to perform the $N$ independent averages of $U_s^{\ot 4}\ot U_s^{\dag\ot 4}$. To perform these, we make use of Lemma \ref{app:UUdag}, which reads
\begin{equation}
    \exv_{U\in\mathcal{U}(d)} [U^{\ot 4}\ot U^{\dag \ot 4}]=S_2\sum_{\pi,\sigma \in S_4}W_{\pi,\sigma}(d)\ttt_\pi(d)\ot \ttt_\sigma(d)
\end{equation}
where $S_2$ swaps the first four copies of $\hi$ with the second four copies. Applying this formula, we get
\begin{equation}
    \begin{split}
    \bigotimes_{s=1}^N \exv_{U_s}[U_s ^{\ot 4}\ot U_s ^{\dag \ot 4}]&=\bigotimes_{s=1}^N S_2 ^{(s)}\sum_{\pi_s,\sigma_s }W_{\pi_s,\sigma_s}^{(2\chi)}\ttt_{\pi_s}^{(2\chi)}\ot\ttt_{\sigma_s}^{(2\chi)}\\
    &=S_2\bigotimes_{s=1}^N\sum_{\pi_s,\sigma_s }W_{\pi_s,\sigma_s}^{(2\chi)}\ttt_{\pi_s}^{(2\chi)}\ot\ttt_{\sigma_s}^{(2\chi)}
    \end{split}
\end{equation}
Substituting this expression in Eq. \eqref{app:psi4mps}, we get
\begin{equation}
    \begin{split}
        \exvrmps{\chi}(\psi^{\ot 4})&=\sum_{i,j,k,l,m,n,o,p}\sum_{\pi_1,\sigma_1,\dots \pi_N,\sigma_N}W_{\pi_1,\sigma_1} ^{(2\chi)}\dots W_{\pi_N,\sigma_N} ^{(2\chi)}{\rm Tr}\Bigg[S_2\Big(\bigotimes_{s=1}^{N}\ttt_{\pi_s} ^{(2\chi)}\ot \ttt_{\sigma_s}^{(2\chi)}\Big)(\ketbra{ijkl}{0000}\ot S_A^{\ot 4})\\&\ot(\ketbra{0000}{mnop}\ot S_A ^{\dag \ot 4})\Bigg]\ketbra{ijkl}{mnop}
    \end{split}
\end{equation}
we now use the fact that given $\hi=\hi_A\ot \hi_B$, $\ttt_\pi=\ttt_\pi ^A \ot \ttt_\pi ^B$ and write
\begin{equation}
   \begin{split}
        \exvrmps{\chi}(\psi^{\ot 4})&=\sum_{i,j,k,l,m,n,o,p}\sum_{\pi_1,\sigma_1,\dots \pi_N,\sigma_N}W_{\pi_1,\sigma_1} ^{(2\chi)}\dots W_{\pi_N,\sigma_N} ^{(2\chi)}{\rm Tr}\bigg[S_2\bigotimes_{s=1}^{N} (\ttt_{\pi_s} ^{(2)}\ketbra{ijkl}{0000}\ot \ttt_{\pi_s} ^{(\chi)}S_A^{\ot 4})\\&\ot(\ttt_{\sigma_s}^{(2)}\ketbra{0000}{mnop}\ot \ttt_{\sigma_s}^{(\chi)}S_A ^{\dag \ot 4})\bigg]\ketbra{ijkl}{mnop}\\
        &=\sum_{i,j,k,l,m,n,o,p}\sum_{\pi_1,\sigma_1,\dots \pi_N,\sigma_N}W_{\pi_1,\sigma_1} ^{(2\chi)}\dots W_{\pi_N,\sigma_N} ^{(2\chi)}{\rm Tr}\bigg[S_2\bigg(\bigotimes_{s=1}^{N} (\ttt_{\pi_s} ^{(2)}\ketbra{ijkl}{0000}\ot \ttt_{\pi_s} ^{(\chi)}S_A^{\ot 4})\\&\ot(\ketbra{0000}{mnop}\ot \ttt_{\sigma_s}^{(\chi)}S_A ^{\dag \ot 4})\bigg)\bigg]\ketbra{ijkl}{mnop}\\
        &\overset{(a)}{=}\bigotimes_{s=1}^N\sum_{i_s,j_s,k_s,l_s,m_s,n_s,o_s,p_s}\sum_{\pi_s,\sigma_s}W_{\pi_s,\sigma_s} ^{(2\chi)}\\&\Tr[\ttt_{\pi_s}^{(2)}\ketbra{i_s j_s k_s l_s}{m_s n_s o_s p_s}\ot \ttt_{\pi_s}^{(\chi)}S_A^{\ot 4}\ttt_{\sigma_s}^{(\chi)}S_A ^{\dag\ot 4}]\ketbra{i_s j_s k_s l_s}{m_s n_s o_s p_s}\\
        &\overset{(b)}{=}\bigotimes_{s=1}^N \sum_{\pi_s,\sigma_s}W_{\pi_s,\sigma_s} ^{(2\chi)}\Tr[\ttt_{\pi_s}^{(\chi)}S_A^{\ot 4}\ttt_{\sigma_s}^{(\chi)}S_A ^{\dag\ot 4}]\ttt_{\pi_s}^{(2)}
   \end{split}
\end{equation}
where in $(a)$ we used the swap trick $\Tr[S(A\ot B)]=\Tr[AB]$ and in $(b)$ we recognized the decomposition of $\ttt_{\pi_s} ^{(2)}$ in the computational basis:
\begin{equation}
    \ttt_{\pi_s} ^{(2)}=\sum_{i_s,j_s,k_s,l_s,m_s,n_s,o_s,p_s}\Tr[\ttt_{\pi_s}^{(2)}\ketbra{i_s j_s k_s l_s}{m_s n_s o_s p_s}]\ketbra{i_s j_s k_s l_s}{m_s n_s o_s p_s}\,.
\end{equation}
Now, the adjoint action of $S_A^{\ot 4}$ over $\ttt_{\sigma_s}^{(\chi)}$ is to cyclically permute the qubit upon $\ttt_{\sigma_s}^{(\chi)}$ will act, hence
\begin{equation}
    \Tr[\ttt_{\pi_s}^{(\chi)}S_A^{\ot 4}\ttt_{\sigma_s}^{(\chi)}S_A ^{\dag\ot 4}]=\Tr[\ttt_{\pi_s}^{(\chi)}\ttt_{\sigma_{s-1}}^{(\chi)}]\,.
\end{equation}
Substituting this last expression, we get to the final expression for the average $4-$copy RMPS:
\begin{equation}\label{app:avgmpsnotnorm}
    \exvrmps{\chi}(\psi^{\ot 4})=\bigotimes_{s=1}^N \sum_{\pi_s,\sigma_s}W_{\pi_s,\sigma_s} ^{(2\chi)}\Tr[\ttt_{\pi_s}^{(\chi)}\ttt_{\sigma_{s-1}}^{(\chi)}]\ttt_{\pi_s}^{(2)}
\end{equation}
\subsection{Normalization constants of random MPS}\label{app:norm_MPS}
The expression obtained in Eq. \eqref{app:avgmpsnotnorm} is not normalized, as we started from an initial state which is not normalized. In order to get a normalized expression for the average $4-$copy of a RMPS, we will compute the trace of Eq. \eqref{app:avgmpsnotnorm}:
\begin{equation}\label{app:norm4def}
    \mathcal{N}_4=\Tr[\exvrmps{\chi}(\psi^{\ot 4})]
\end{equation}
we will show computation of this normalization for $N=2$ qubits, for simplicity, but the derivation is easily extendable for a generic number of qubits: in this case, Eq. \eqref{app:norm4def} reads
\begin{equation}
  \begin{split}
        \nnf &=\sum_{\pi_1,\sigma_1}\sum_{\pi_2,\sigma_2}W_{\pi_1,\sigma_1}^{(2\chi)}W_{\pi_2,\sigma_2}^{(2\chi)}\Tr[\ttt_{\pi_1}^{(\chi)}\ttt_{\sigma_2}^{(\chi)}]\Tr[\ttt_{\pi_2}^{(\chi)}\ttt_{\sigma_1}^{(\chi)}]\Tr[\ttt_{\pi_1}^{(2)}]\Tr[\ttt_{\pi_2}^{(2)}]\\
        &=\sum_{\sigma_1,\sigma_2}\left(\sum_{\pi_1}\Tr[\ttt_{\pi_1}^{(2)}]\Tr[\ttt_{\pi_1}^{(\chi)}\ttt_{\sigma_2}^{(\chi)}]W_{\pi_1,\sigma_1}^{(2\chi)}\right)\left(\sum_{\pi_2}\Tr[\ttt_{\pi_2}^{(2)}]\Tr[\ttt_{\pi_2}^{(\chi)}\ttt_{\sigma_1}^{(\chi)}]W_{\pi_2,\sigma_2}^{(2\chi)}\right)
  \end{split}
\end{equation}
Denoting the $24\times 24$ matrix $\Tilde{W}^{(4)}$ as 
\begin{equation}
    \begin{split}
        \Tilde{W}^{(4)}_{\pi,\sigma}=\sum_{\gamma\in S_4}\Tr[\ttt_{\gamma}^{(2)}]\Tr[\ttt_{\gamma}^{(\chi)}\ttt_{\pi}^{(\chi)}]W_{\gamma,\sigma}^{(2\chi)}
    \end{split}
\end{equation}
we have
\begin{equation}
    \nnf=\sum_{\sigma_1,\sigma_2} \Tilde{W}^{(4)}_{\sigma_1,\sigma_2}\Tilde{W}^{(4)}_{\sigma_2,\sigma_1}=\Tr[(\Tilde{W}^{(4)})^2]
\end{equation}
hence, for N qubits we have
\begin{equation}
\begin{split}
    \nnf&=\Tr[(\Tilde{W}^{(4)})^N]\\
    &=1+\left(3\ 2^{-N}+4\right) \left(\frac{\chi ^2-4}{4 \chi ^2-1}\right)^N+\left(6\ 2^{N}+4\right) \left(\frac{\chi ^2-1}{4 \chi ^2-1}\right)^N+2(f_+ ^N +f_- ^N)+l_+ ^N + l_- ^N\sim O(1)
    \end{split}
\end{equation}
with \begin{equation}
    \begin{split}
        f_{\pm}&=\frac{24 \chi ^5-114 \chi ^3+135 \chi\pm\sqrt{\chi ^2 \left(9-4 \chi ^2\right)^2 \left(4 \left(\chi ^2-5\right) \chi ^2+97\right)} }{8(16 \chi ^5-40 \chi ^3+9 \chi )}\\
        l_{\pm}&=\frac{6 \chi  \left(4 \chi ^4-25 \chi ^2+51\right)\pm2 \sqrt{\chi ^2 \left(16 \chi ^8+280 \chi ^6-1007 \chi ^4-4230 \chi ^2+13041\right)}}{8(16 \chi ^5-40 \chi ^3+9 \chi) }
    \end{split}
\end{equation}
Similar procedure can be carried out for $k=2$ to see that these normalization constants quickly converge to 1.

\subsection{Average SE gap for partial trace in random MPS}\label{app:avg_A_MPS}
In this section we show explicit derivations for obtaining the SE gap for partial trace in the random MPS ensemble.
We define the $\mbar ^A$ random variable as
\begin{equation}
    \mbar ^A:=1-\stp(\psi)-\pur(\psi_A)+\stp(\psi_A)
\end{equation}
and compute each of these pieces separately.
Instrumental to such averages is the computation of the average over two and four copies of a RMPS, which read (see Sec.  \ref{app:kcopiesmps} for details)
\begin{equation}\label{app:avgmpsstate}
    \begin{split}
        \exvrmps{\chi}(\psi^{\ot 4})&=\frac{1}{\mathcal{N}_4}\bigotimes_{s=1}^N\sum_{\pi_s\,\sigma_s\in S_4} W_{\pi_s,\sigma_s}^{(4)}(2\chi) \Tr[\ttt_{ \pi_s} ^{(4)}(\chi)\ttt_{\sigma_{s-1}} ^{(4)}(\chi)]\ttt_{\pi_s} ^{(4)}(2)\\
          \exvrmps{\chi}(\psi^{\ot 2})&=\frac{1}{\mathcal{N}_2}\bigotimes_{s=1}^N\sum_{\pi_s\,\sigma_s\in S_2} W_{ 2 
          \pi_s,\sigma_s} ^{(2)}(2\chi) \Tr[\ttt_{ \pi_s} ^{(2)}(\chi)\ttt_{\sigma_{s-1}} ^{(2)}(\chi)]\ttt_{\pi_s} ^{(2)}(2)
    \end{split}
\end{equation}
with $\mathcal{N}_2$ and $\mathcal{N}_4$ being normalization constants (see Sec.  \ref{app:norm_MPS}), $W_{\pi_s,\sigma_s}^{(4)}(2\chi)$ and $W_{\pi_s,\sigma_s} ^{(2)}(2\chi)$ being the Weingarten functions related to the commutant of $U^{\ot 4}$ and $U^{\ot 2}$ respectively, with $ U\in \mathcal{U}(2\chi)$, $T_{\pi_s} ^{(4)}(x)$ and $T_{\pi_s} ^{(2)}(x)$ with $\,x\in \{\chi,2\}$ being permutation operators over four and two copies of an $x$-dimensional Hilbert space respectively. We will omit the number of copies of the Hilbert spaces when it can be inferred by the context, for simplicity of notation.

\subsubsection{Average stabilizer purity of random MPS}\label{app:app_SP_MPS}

Objective of this section is to show the result of the following computation:
\begin{equation}
\exvrmps{\chi}\stp(\psi):=2^N\exvrmps{\chi}\Tr[Q\psi^{\ot 4}]
\end{equation}
We will exploit an important fact about the expression Eq. \eqref{app:avgmpsstate}: the operatorial part of this expression is factorized on each qubit, paired with the fact that $Q=Q_1 ^{(1)}\otimes\dots \otimes Q_N ^{(1)}$, with $Q_i ^{(1)}=\frac14\sum_{P\in\{\id,X,Y,Z\}}P^{\ot 4}$.
Using these properties, and substituting the expression in Eq. \eqref{app:avgmpsstate}, the average stabilizer purity then reads
\begin{equation}
    \begin{split}
        \exvrmps{\chi}\stp(\psi)&=2^N\Tr[Q\exvrmps{\chi}(\psi^{\ot 4})]\\
        &=\frac{2^N}{4^N\nnf}\sum_{P_1,\dots,P_N}\sum_{\pi_1,\sigma_1,\dots \pi_N,\sigma_N}W_{\pi_1,\sigma_1} ^{(2\chi)}\dots W_{\pi_N,\sigma_N} ^{(2\chi)}\Tr[\ttt_{\pi_1} ^{(\chi)}\ttt_{\sigma_N}^{(\chi)}]\dots\Tr[\ttt_{\pi_N} ^{(\chi)}\ttt_{\sigma_{N-1}}^{(\chi)}]\\&\Tr[(P_1\ot\dots P_N)^{\ot 4}(\ttt_{\pi_1}^{(2)}\ot\dots \ttt_{\pi_N}^{(2)})]\\
        &=\frac{2^N}{\nnf}\sum_{\pi_1,\sigma_1,\dots \pi_N,\sigma_N}W_{\pi_1,\sigma_1} ^{(2\chi)}\dots W_{\pi_N,\sigma_N} ^{(2\chi)}\Tr[\ttt_{\pi_1} ^{(\chi)}\ttt_{\sigma_N}^{(\chi)}]\dots\Tr[\ttt_{\pi_N} ^{(\chi)}\ttt_{\sigma_{N-1}}^{(\chi)}]\\
        &\frac{1}{4^N}\sum_{P_1,\dots,P_N}\Tr[P_1 ^{\ot 4} \ttt_{\pi_1} ^{(2)}]\dots \Tr[P_N ^{\ot 4}\ttt_{\pi_N} ^{(2)}]\\
        &=\frac{2^N}{\nnf}\sum_{\pi_1,\sigma_1,\dots \pi_N,\sigma_N}W_{\pi_1,\sigma_1} ^{(2\chi)}\dots W_{\pi_N,\sigma_N} ^{(2\chi)}\Tr[\ttt_{\pi_1} ^{(\chi)}\ttt_{\sigma_N}^{(\chi)}]\dots\Tr[\ttt_{\pi_N} ^{(\chi)}\ttt_{\sigma_{N-1}}^{(\chi)}]\\&\left(\frac14\sum_{P_1}\Tr[P_1 ^{\ot 4}\ttt_{\pi_1}^{(2)}]\right)\dots\left(\frac14\sum_{P_N} \Tr[P_N ^{\ot 4}\ttt_{\pi_N} ^{(2)}]\right)\\
        &=\frac{2^N}{\nnf} \prod_{s=1}^N \sum_{\pi_s,\sigma_s}W_{\pi_s,\sigma_s} ^{(2\chi)}\Tr[\ttt_{\pi_s} ^{(\chi)}\ttt_{\sigma_{s-1}}^{(\chi)}]^{(2\chi)}\Tr[Q^{(1)}\ttt_{\pi_s}^{(2)}]\\
        &=2^N\prod_{s=1}^N\sum_{\sigma_s}\underbrace{\left(\sum_{\pi_s}\Tr[Q^{(1)}\ttt_{\pi_s}]\Tr[\ttt_{\pi_s} ^{(\chi)}\ttt_{\sigma_{s-1}}^{(\chi)}]W_{\pi_s,\sigma_s}^{(2\chi)}\right)}_{\Tilde{W}^Q _{\sigma_{s-1},\sigma_s}}\\
        &=\frac{2^N}{\nnf}\sum_{\sigma_1,\dots,\sigma_N}\Tilde{W}^Q _{\sigma_{N},\sigma_1}\Tilde{W}^Q _{\sigma_{1},\sigma_2}\dots \Tilde{W}^Q _{\sigma_{N-1},\sigma_N}=\frac{2^N}{\nnf}\sum_{\sigma_N}(\Tilde{W}^Q)^N _{\sigma_N,\sigma_N}\\
        &=\frac{2^N}{\nnf}\Tr[(\Tilde{W}^Q)^N]
    \end{split}
\end{equation}
with

\begin{equation}
    \Tilde{W}^Q _{\pi,\sigma}=\sum_{\gamma\in S_4}\Tr[\ttt_\gamma {(2)}Q^{(1)}]\Tr[\ttt_\gamma {(\chi)}\ttt_\pi {(\chi)}]W_{\gamma,\sigma} {(2\chi)}\,.
\end{equation}

\subsubsection{Average purity of traced out Random MPS}
The partial trace of the first $N_B$ of the average $2-$copy average RMPS reads
\begin{equation}
  \begin{split}
         &\exvrmps{\chi}\psi_A ^{\ot 2}=\Tr_B\exvrmps{\chi}\psi^{\ot 2}=\frac{1}{\nnt}\sum_{\sigma_1,\dots,\sigma_N}\sum_{\pi_1,\dots,\pi_{N_B}}\underbrace{\Tr[\ttt_{\pi_1}^{(2)}]\Tr[\ttt_{\pi_1}^{(\chi)}\ttt_{\sigma_N} ^{(\chi)}]W_{\pi_1,\sigma_1} ^{(2\chi)}}_{\Tilde{W}^{(2)} _{\sigma_N,\sigma_1}}\dots\\&\dots\underbrace{\Tr[\ttt_{\pi_{N_B}}^{(2)}]\Tr[\ttt_{\pi_{N_B}}^{(\chi)}\ttt_{\sigma_{N_B-1}} ^{(\chi)}]W_{\pi_{N_B},\sigma_{N_B}}^{(2\chi)}}_{\Tilde{W}^{(2)}_{\sigma_{N_B-1},\sigma_{N_B}}}\times\\&\sum_{\pi_{N_B+1},\dots,\pi_{N}}W_{\pi_{N_B+1},\sigma_{N_B+1}} ^{(2\chi)}\dots W_{\pi_{N},\sigma_{N}} ^{(2\chi)}(\ttt_{\pi_{N_B+1}} ^{(2)}\otimes\dots\otimes \ttt_{\pi_N})\\
         &=\frac{1}{\nnt}\sum_{\pi_{N_B+1},\dots,\pi_N,\sigma_{N_B},\dots,\sigma_N}(\Tilde{W}^{(2)})^{N_B} _{\sigma_N,\sigma_{N_B}}W_{\pi_{N_B+1},\sigma_{N_B+1}} ^{(2\chi)}\dots W_{\pi_{N},\sigma_{N}} ^{(2\chi)}\Tr[\ttt_{\pi_{N_B+1}}^{(\chi)}\ttt_{\sigma_{N_B}}^{(\chi)}]\dots\Tr[\ttt_{\pi_N} ^{(\chi)}\ttt_{\sigma_1} ^{(\chi)}]\\&(\ttt_{\pi_{N_B+1}} ^{(2)}\otimes\dots\otimes \ttt_{\pi_N}^{(2)})
  \end{split}
\end{equation}
Hence the purity then reads
\begin{equation}
    \begin{split}
       &\exvrmps{\chi}\pur(\psi_A)=\frac{1}{\nnt}\sum_{\sigma_{N_B},\sigma_N}(\Tilde{W}^{(2)})^{N_B} _{\sigma_N,\sigma_{N_B}}\prod_{s={N_B}}^N \sum_{\sigma_s}\underbrace{\sum_{\pi_s}\Tr[\ttt\ttt_{\pi_s}^{(2)}]\Tr[\ttt_{\pi_s} ^{(\chi)}\ttt_{\sigma_{s-1}}]W_{\pi_s,\sigma_s}^{(2\chi)}}_{\Tilde{W}^{(\pur)} _{\sigma_{s-1},\sigma_s}}\\
       &=\frac{1}{\nnt}\sum_{\sigma_N,\sigma_{N_B}}(\Tilde{W}^{(2)})^{N_B} _{\sigma_N,\sigma_{N_B}}(\Tilde{W}^{(\pur)})^{N-N_B} _{\sigma_{N_B},\sigma_{N}}=\frac{1}{\nnt}\Tr[(\Tilde{W}^{(2)})^{N_B}\cdot(\Tilde{W}^{(\pur)})^{N-N_B}]
    \end{split}
\end{equation}

\subsubsection{Average stabilizer purity of traced out RMPS}\label{app:app:ptracempsstabpurity}
By the same procedure utilized in the partial trace of the $2-$copy average RMPS, the $4-$copy partial trace of the average RMPS reads
\begin{equation}
 \begin{split}
        \Tr_B \exvrmps{\chi}(\psi^{\ot 4})&=\frac{1}{\nnf}\sum_{\pi_{N_B+1},\dots,\pi_N,\sigma_{N_B},\dots,\sigma_N}(\Tilde{W}^{(4)})^{N_B} _{\sigma_N,\sigma_{N_B}}W_{\pi_{N_B+1},\sigma_{N_B+1}} ^{(2\chi)}\dots W_{\pi_{N},\sigma_{N}} ^{(2\chi)}\\&\Tr[\ttt_{\pi_{N_B+1}}^{(\chi)}\ttt_{\sigma_{N_B}}^{(\chi)}]\dots\Tr[\ttt_{\pi_N} ^{(\chi)}\ttt_{\sigma_1} ^{(\chi)}](\ttt_{\pi_{N_B+1}} ^{(2)}\otimes\dots\otimes \ttt_{\pi_N} ^{(2)})
 \end{split}
\end{equation}
hence the stabilizer purity reads
\begin{equation}
    \begin{split}
       &\exvrmps{\chi}\stp(\psi_A)=\frac{1}{\nnf}\sum_{\sigma_{N_B},\sigma_N}(\Tilde{W}^{(4)})^{N_B} _{\sigma_N,\sigma_{N_B}}\prod_{s={N_B}}^N \sum_{\sigma_s}\underbrace{\sum_{\pi_s}\Tr[Q\ttt_{\pi_s}^{(2)}]\Tr[\ttt_{\pi_s} ^{(\chi)}\ttt_{\sigma_{s-1}}]W_{\pi_s,\sigma_s}^{(2\chi)}}_{\Tilde{W}^{(Q)} _{\sigma_{s-1},\sigma_s}}\\
       &=\frac{1}{\nnf}\sum_{\sigma_N,\sigma_{N_B}}(\Tilde{W}^{(f)})^{N_B} _{\sigma_N,\sigma_{N_B}}(\Tilde{W}^{(Q)})^{N-N_B} _{\sigma_{N_B},\sigma_{N}}=\frac{1}{\nnf}\Tr[(\Tilde{W}^{(4)})^{N_B}\cdot(\Tilde{W}^{(Q)})^{N-N_B}]
    \end{split}
\end{equation}

\subsection{Average SE gap for dephasing in random MPS}\label{app:avg_db_MPS}
In this section we show explicit derivations for obtaining the SE gap for dephasing in the random MPS ensemble.
As for the general Haar case, we define the random variable $\mbar^{D_B}(\psi)$ as
\begin{equation}
   \mbar^{D_B}(\psi)=\mlin(\psi)-\mlin(D_B\psi) 
\end{equation}
and perform the average over the RMPS ensemble.

To calculate the $\mbar^{D_B}$ of a RMPS, we will need to compute the average stabilizer purity (already computed in Sec.  \ref{app:app_SP_MPS}), the purity after dephasing and the stabilizer purity after dephasing.

\subsubsection{Average purity of dephased RMPS}\label{app:mpspuritydb}
In this section we show how to compute the following object:
\begin{equation}\label{app:mps_pur_db_def}
    \exvrmps{\chi}[\pur(\db)]=\exvrmps{\chi}\Tr[\ttt(\db)^{\ot 2}]
\end{equation}
Let us first recall the definition of the dephasing channel over $N$ qubits:
\begin{equation}
    \db=\sum_{i_1,\dots i_N}(\Pi_{i_1}\ot\dots\Pi_{i_N})\psi(\Pi_{i_1}\ot\dots\Pi_{i_N})=(D_B)^{(1)}\ot\dots\ot (D_B)^{(N)}\psi
\end{equation}
with
\begin{equation}
    (D_B(\cdot))^{(k)}=\sum_{i_k=0}^1 \Pi_{i_k}(\cdot)\Pi_{i_k}
\end{equation}
as we see, the dephasing channel neatly factorizes on each qubit.
Applying this fact, we then substitute the expression of the average $2-$copy RMPS:
\begin{equation}
   \begin{split}
        \exvrmps{\chi}[(D_B\psi)^{\ot 2}]&=\frac{1}{\mathcal{N}_2}\bigotimes_{s=1}^N\sum_{\pi_s\,\sigma_s\in S_2} W_{ 
          \pi_s,\sigma_s} ^{(2\chi)} \Tr[\ttt_{ \pi_s} ^{(\chi)}\ttt_{\sigma_{s-1}} ^{(\chi)}](D_B ^{\ot 2}\ttt_{\pi_s} ^{(2)})
   \end{split}
\end{equation}
with
\begin{equation}
    (D_B ^{\ot 2}\ttt_{\pi_s} ^{(2)}):=\sum_{i=0}^1\sum_{j=0}^1(\Pi_i \ot \Pi_j)\ttt_{\pi_s} ^{(2)}(\Pi_i \ot \Pi_j)
\end{equation}
Substituting this expression in Eq. \eqref{app:mps_pur_db_def} yields
\begin{equation}
    \begin{split}
        \exvrmps{\chi}[\pur(\db)]&=\frac{1}{\mathcal{N}_2}\prod_{s=1}^N\sum_{\pi_s\,\sigma_s\in S_2} W_{ 
          \pi_s,\sigma_s} ^{(2\chi)} \Tr[\ttt_{ \pi_s} ^{(\chi)}\ttt_{\sigma_{s-1}} ^{(\chi)}]\Tr[\ttt(D_B ^{\ot 2}\ttt_{\pi_s} ^{(2)})]\\
          &=\frac{1}{\mathcal{N}_2}\prod_{s=1}^N\sum_{\sigma_s}\underbrace{\left(\sum_{\pi_s}\Tr[\ttt(D_B ^{\ot 2}\ttt_{\pi_s} ^{(2)})]\Tr[\ttt_{ \pi_s} ^{(\chi)}\ttt_{\sigma_{s-1}} ^{(\chi)}]W_{ 
          \pi_s,\sigma_s} ^{(2\chi)} \right)}_{\Tilde{W}^{(\pur)(D_B) }_{\sigma_{s-1},\sigma_s}}\\
          &=\frac{1}{\nnt}\sum_{\sigma_1,\dots,\sigma_N}\Tilde{W}^{(\pur)(D_B) }_{\sigma_{N},\sigma_1}\Tilde{W}^{(\pur)(D_B) }_{\sigma_{1},\sigma_2}\dots \Tilde{W}^{(\pur)(D_B) }_{\sigma_{N-1},\sigma_N}\\
          &=\frac{1}{\nnt}\Tr[(\Tilde{W}^{(\pur)(D_B) })^N]
    \end{split}
\end{equation}
direct evaluation of this trace yields
\begin{equation}
    \exvrmps{\chi}[\pur(\db)]=\frac{(2\chi^2+\chi-1)^N+(2\chi^2-\chi-1)^N}{(2\chi^2-2)^N+(4\chi^2-1)^N}
\end{equation}

\subsubsection{Average stabilizer purity of dephased RMPS}\label{app:mpsstpuritydb}
In this section we will show the derivation of the following object:
\begin{equation}
    \exvrmps{\chi}[\stp(\db)]=2^N\exvrmps{\chi}\Tr[Q(\db)^{\ot 4}]
\end{equation}
Again, using the factorization of the dephasing channel and of the $Q$ operator, we get
\begin{equation}
    \begin{split}
        \exvrmps{\chi}[\stp(\db)]&=\frac{2^N}{\mathcal{N}_4}\prod_{s=1}^N\sum_{\pi_s\,\sigma_s\in S_4} W_{ 
          \pi_s,\sigma_s} ^{(2\chi)} \Tr[\ttt_{ \pi_s} ^{(\chi)}\ttt_{\sigma_{s-1}} ^{(\chi)}]\Tr[Q(D_B ^{\ot 4}\ttt_{\pi_s} ^{(2)})]\\
          &=\frac{2^N}{\mathcal{N}_4}\prod_{s=1}^N\sum_{\sigma_s}\underbrace{\left(\sum_{\pi_s}\Tr[Q(D_B ^{\ot 4}\ttt_{\pi_s} ^{(2)})]\Tr[\ttt_{ \pi_s} ^{(\chi)}\ttt_{\sigma_{s-1}} ^{(\chi)}]W_{ 
          \pi_s,\sigma_s} ^{(2\chi)} \right)}_{\Tilde{W}^{(Q)(D_B) }_{\sigma_{s-1},\sigma_s}}\\
          &=\frac{1}{\nnf}\sum_{\sigma_1,\dots,\sigma_N}\Tilde{W}^{(Q)(D_B) }_{\sigma_{N},\sigma_1}\Tilde{W}^{(Q)(D_B) }_{\sigma_{1},\sigma_2}\dots \Tilde{W}^{(Q)(D_B) }_{\sigma_{N-1},\sigma_N}\\
          &=\frac{1}{\nnf}\Tr[(\Tilde{W}^{(Q)(D_B) })^N]
    \end{split}
\end{equation}

\subsection{Typicality of the results}
Assessing the typicality of this result in the ensemble of Random MPS makes use of the L\'evy lemma. As before, both $\mbar^A$ and $\mbar ^{D_B}$ are Lipschitz functions with $\eta\leq{64}/{5}$ in this ensemble. When acting on the ensemble of ${\rm MPS_{PBC}}(\chi)$ (which, equipped with the any Schatten norm satisfies the definition of a metric space), we can consider $\mbar$ as functions $\mbar: {\rm MPS_{PBC}}(\chi) \mapsto \mathbb{R}$. The set of $\{\bigotimes_{s=1}^N U_s\}$ used for the sequential generation of the $A_i[k]$ matrices provides and invariant measure on said set, hence the L\'evy lemma implies
\begin{equation}
    {\rm Pr}(\mbar(\psi)-\exvrmps{\chi}(\mbar)\leq -\epsilon)\leq 2 \exp (-\frac{\epsilon^2  {\rm dim}({\rm MPS_{PBC}}(\chi))}{4096\pi})
\end{equation}
The only thing left to establish is the dimension (number of independent parameters) of the ${\rm MPS_{PBC}}(\chi) $ set. For $N$ qubits, one needs two $\chi\times\chi$ matrices for each qubit for a MPS to be specified: hence, the number of independent parameters seems to be $2N\chi^2$. However, one notices that invertible transformations $G\in {\rm GL}(\chi)$ applied to each matrix leave the MPS unchanged: it is a gauge degree of freedom that reduces the number of free parameters to $(2N-1)\chi^2$. Finally, taking into account the invariance of quantum states under global complex phases, yields the dimension of the ${\rm MPS_{PBC}}(\chi)$ ensemble to be $(2N-1)\chi^2-1$. So, the L\'evy lemma now reads
\begin{equation}\label{app:levy_MPS}
     {\rm Pr}(\mbar(\psi)-\exvrmps{\chi}(\mbar)\leq -\epsilon)\leq 2 \exp [-\frac{\big((2N-1)\chi^2-1\big)\epsilon^2 }{4096\pi}]\,.
\end{equation}

Applying Eq. \eqref{app:levy_MPS} for $\mbar$ either for dephasing or partial trace by setting $\epsilon_\mathcal{E}=\exvrmps{\chi}\mbar^{\mathcal{E}}(\psi)\;,\mathcal{E}\in \{D_B,\Tr_B\}$, as we have done in previous cases, would not yield a useful formula for establishing the scaling of violation probability, as the final result is written in terms of the dressed $\Tilde{W}$ matrices. However, one can check that the gap quickly converges to the Haar value, which is $O(1)$, so we can set $\epsilon=1$ to get an upper bound to violation probability:
\begin{equation}\label{app:levy_mbar_A}
     {\rm Pr}(\mbar(\psi)\leq 0)\leq {\rm Pr}(\mbar(\psi)-\exvrmps{\chi}(\mbar)\leq -1)\leq 2 \exp [-\frac{(2N-1)\chi^2-1 }{4096\pi}]\,,
\end{equation}
which implies that $\mbar$ is a $\eta-$magic proxy for both partial trace and dephasing, with
\begin{equation}
    \begin{split}
        \eta&=\frac{(2 N-1) \chi ^2-1-4096 \pi  \log 2}{4096 \pi  N}=O(\chi^2)+O\left(\frac{1}{N}\right)
    \end{split}
\end{equation}
which guarantees typicality even with fixed values of $\chi$, though if $\chi$ scales monotonically with the number of qubits, which is coherent with any reasonable physical setting in which the MPS framework is generally used in,  the stronger $\mlin$ as a proxy becomes. It is interesting to notice that if $\chi = O(\sqrt{d})$, we recover the general case of L\'evy typicality: this is consistent with the fact that a generic state in the Hilbert space can be represented as a MPS with $\chi=O(\sqrt{d})$.

\subsection{SE gaps for MPS in OBC}
\label{app:app:averages_and_variances_MPS_OBC}
Having established the behavior of SE gaps for MPS in PBC, we now turn to the Open Boundary Condition (OBC) setting. This case is of independent interest: OBC are the natural choice in tensor-network algorithms such as DMRG, and the different structure at the boundaries may, in principle, affect the positivity and scaling of the SE gap. Our goal is to verify that both $\mlin$ and the 2-Stabilizer R\'enyi Entropy $\mtwo$ retain positive gaps under OBC across a range of system sizes, bond dimensions, and physically motivated quantum channels.

We consider families of random MPS in OBC with fixed bond dimension $\chi$ and varying system size $N$~\footnote{Random MPS at fixed bond dimension are generated using the ITensor Julia package~\cite{appcite:ITensor,appcite:itensor-r0.3}.}. For each pair $(N,\chi)$ with $\chi\in[2,14]$ and $N\in[2,16]$, we sample $3.5\times 10^4$ states and evaluate the magic gap of $\mlin$ with respect to three channels: complete dephasing in the computational basis, partial trace of the first qubit, and partial trace of the first half of the system.

\begin{figure*}[t]
    \centering
    \includegraphics[width=\textwidth]{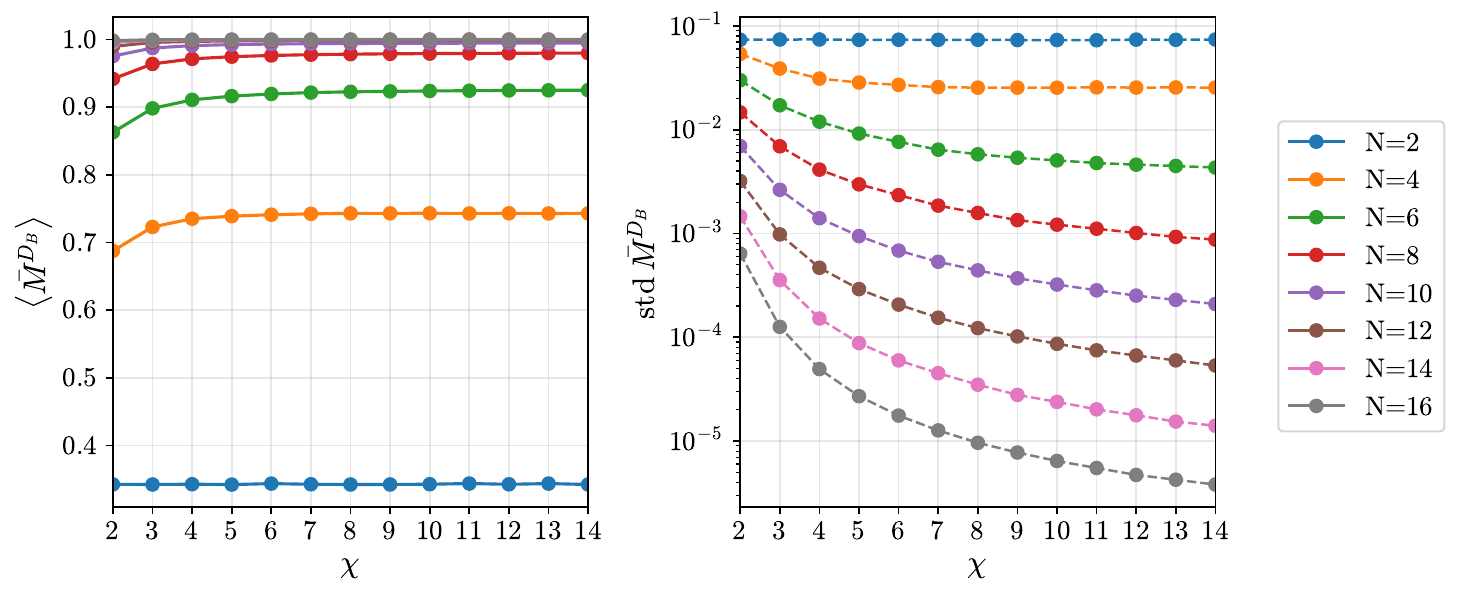}  
    \caption{Average magic gap of $\mlin$ (left) and its standard deviation (right) under complete dephasing in the computational basis for MPS in OBC, plotted as a function of the bond dimension $\chi\in[2,14]$ for system sizes $N\in[2,16]$. For each pair $(N,\chi)$, $3.5\times 10^4$ random MPS are sampled.}
    \label{app:fig:magic_gap_D_B_MPS_OBC} 
\end{figure*}

\begin{figure*}[t]
    \centering
    \includegraphics[width=\textwidth]{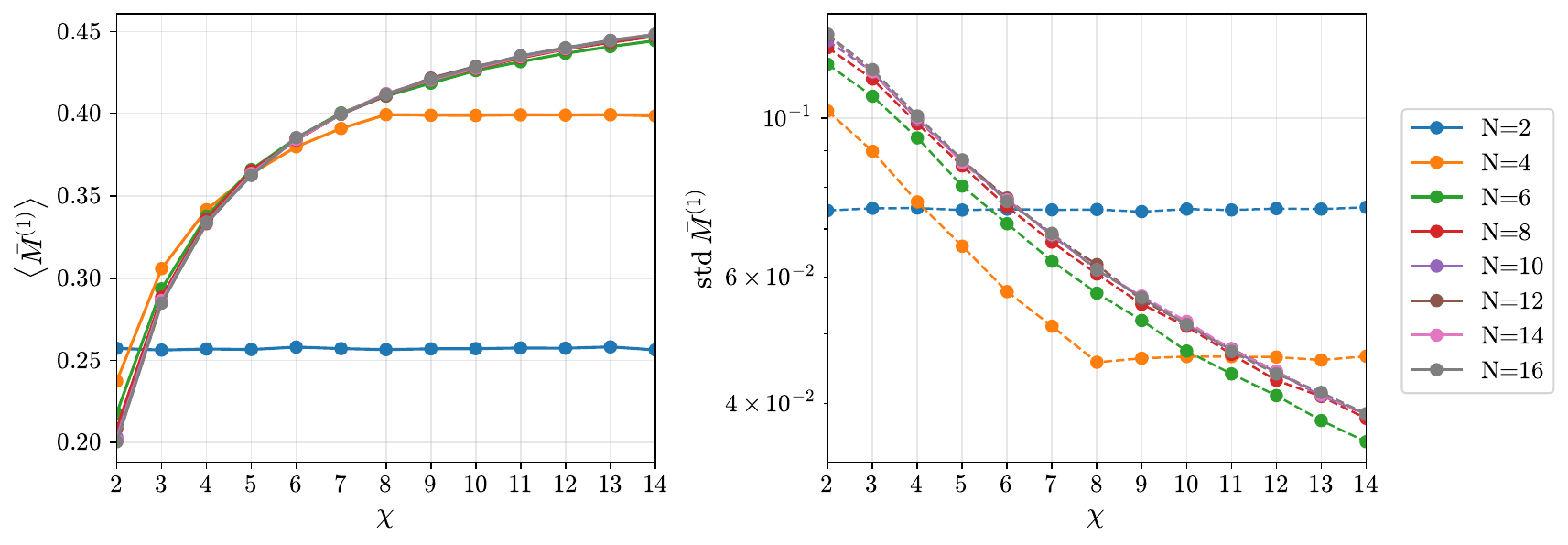}  
    \caption{Average magic gap of $\mlin$ (left) and its standard deviation (right) under the partial trace of the first qubit for MPS in OBC, plotted as a function of the bond dimension $\chi\in[2,14]$ for system sizes $N\in[2,16]$. For each pair $(N,\chi)$, $3.5\times 10^4$ random MPS are sampled.}
    \label{app:fig:magic_gap_Tr1_MPS_OBC} 
\end{figure*}

\begin{figure*}[t]
    \centering
    \includegraphics[width=\textwidth]{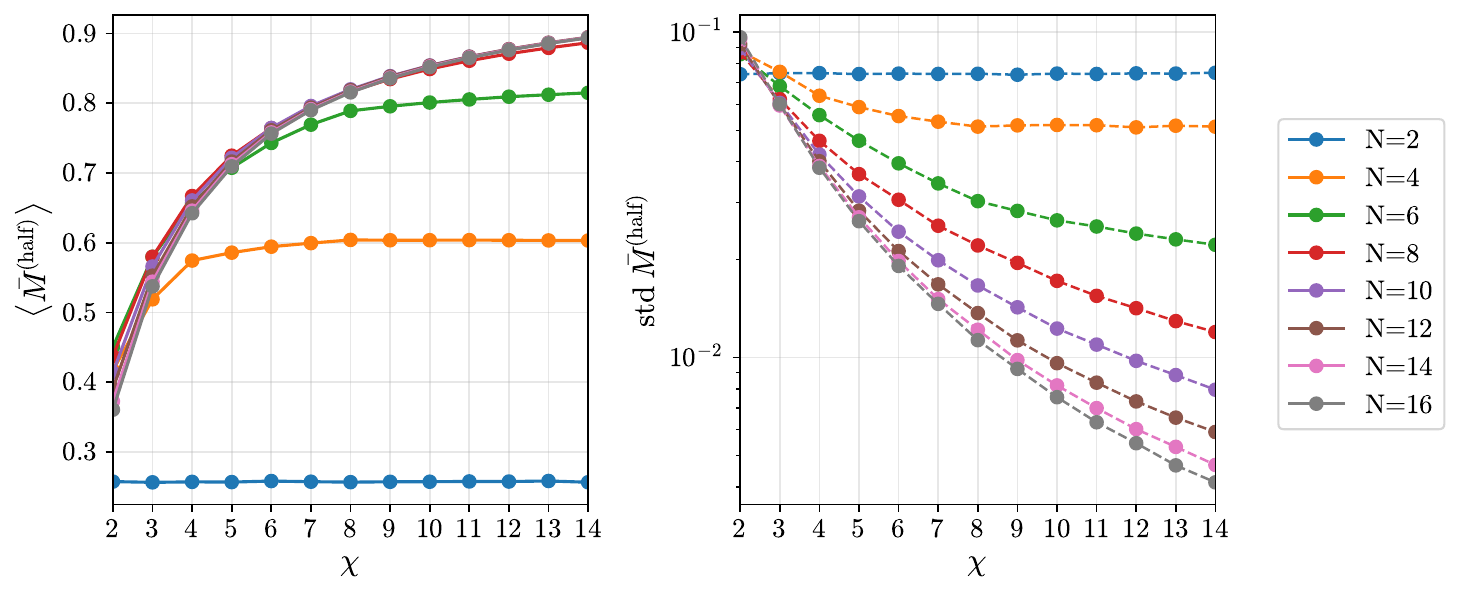}  
    \caption{Average magic gap of $\mlin$ (left) and its standard deviation (right) under the partial trace of the first half of the system for MPS in OBC, plotted as a function of the bond dimension $\chi\in[2,14]$ for system sizes $N\in[2,16]$. For each pair $(N,\chi)$, $3.5\times 10^4$ random MPS are sampled.}
    \label{app:fig:magic_gap_Trhalf_MPS_OBC} 
\end{figure*}

Figures~\ref{app:fig:magic_gap_D_B_MPS_OBC}--\ref{app:fig:magic_gap_Trhalf_MPS_OBC} report the average magic gap and its standard deviation for all three channels. In every case the average gap is strictly positive---even accounting for the standard deviation---and increases monotonically with both $N$ and $\chi$. 

A comparison across channels reveals a clear hierarchy: the gap is larger for the partial trace of half the system with respect to the single-qubit partial trace. This behavior is expected, since tracing out $N/2$ qubits discards a larger fraction of the system's degrees of freedom than tracing out a single qubit, therefore erasing more magic from the initial state and  yielding to a wider gap. 

\begin{figure*}[t]
    \centering
    \includegraphics[width=\textwidth]{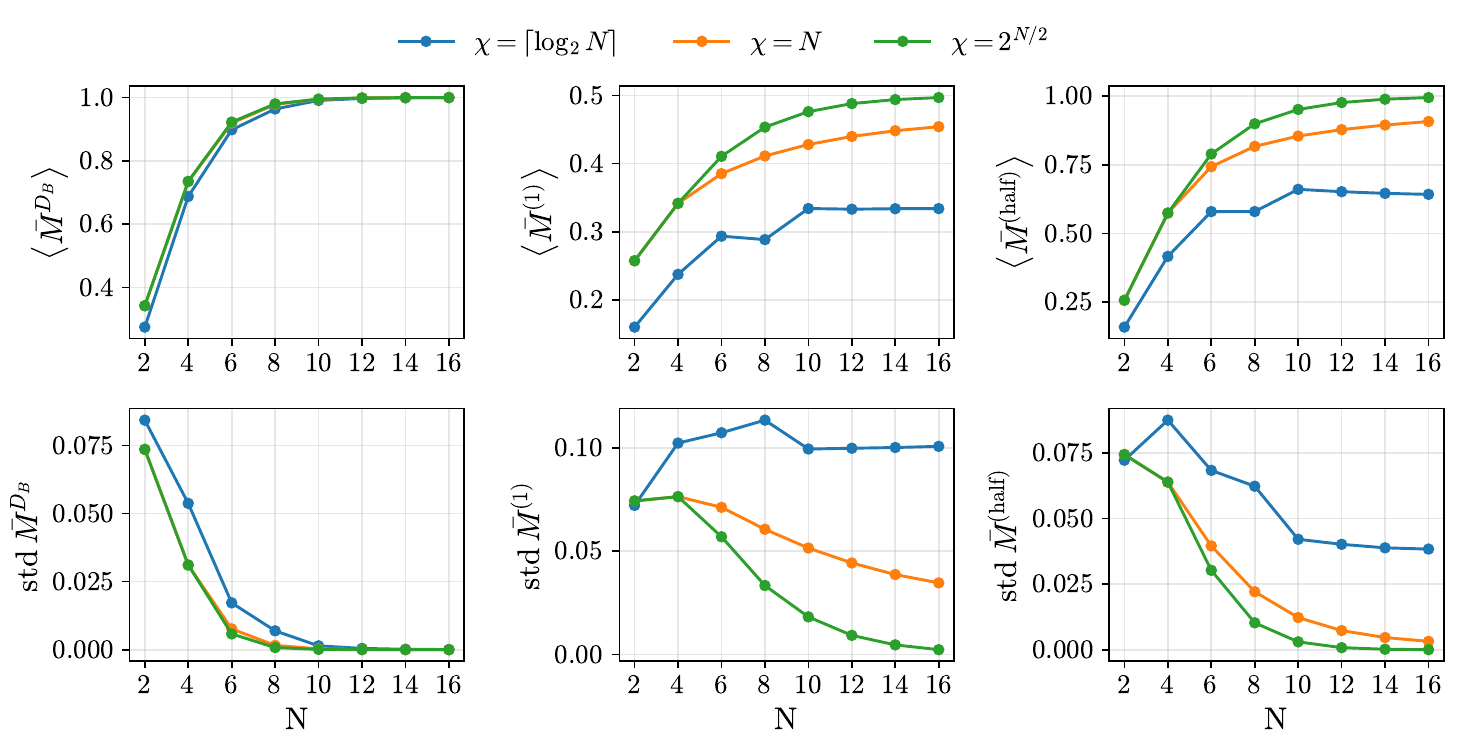}  
    \caption{Average magic gaps (top row) and their standard deviation (bottom row) versus the number of qubits $N$ for MPS in OBC with maximum bond dimensions $\chi(N) = \lceil \log_2 N \rceil$, $\chi(N) = N$, and $\chi(N) =  2^{N/2} $. Columns, left to right: (1) magic gap of the dephasing channel in the computational basis, (2) magic gap after tracing out all but the first qubit, and (3) magic gap after tracing out half of the system.}
    \label{app:fig:scalings_in_chi_MPS_OBC} 
\end{figure*}

To probe the interplay between bond dimension and system size more systematically, we study three scaling regimes: $\chi(N)=\lceil\log_2 N\rceil$, $\chi(N)=N$, and $\chi(N)=2^{N/2}$. The results, reported in Fig.~\ref{app:fig:scalings_in_chi_MPS_OBC}, show that the average magic gap is positive already at $N=2$ and grows with $N$ for all three scalings and all three channels.

\begin{figure*}[t]
    \centering
    \includegraphics[width=\textwidth]{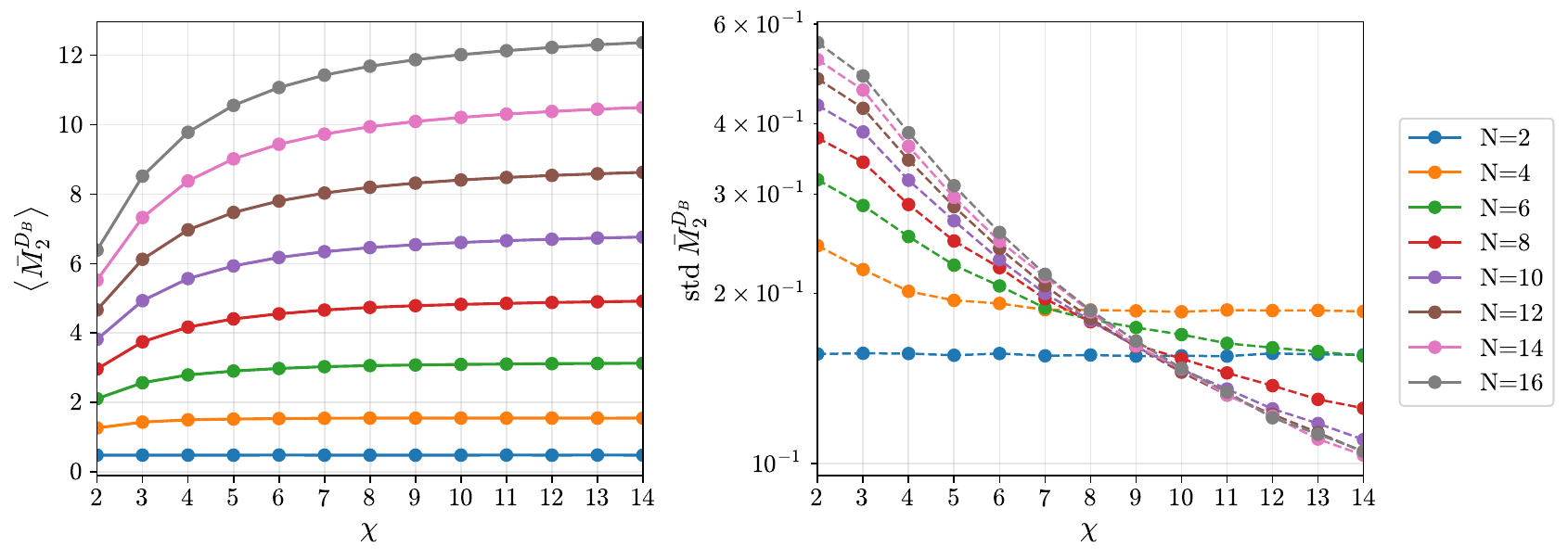}  
    \caption{Average 2-SE gap $\mbar_2$ (left) and its standard deviation (right) under complete dephasing in the computational basis for MPS in OBC, with the same parameters as Fig.~\ref{app:fig:magic_gap_D_B_MPS_OBC}. The gap is positive for all $(N,\chi)$ and increases with both the system size and the bond dimension.}
    \label{app:fig:M2_gap_D_B_MPS_OBC} 
\end{figure*}

\begin{figure*}[t]
    \centering
    \includegraphics[width=\textwidth]{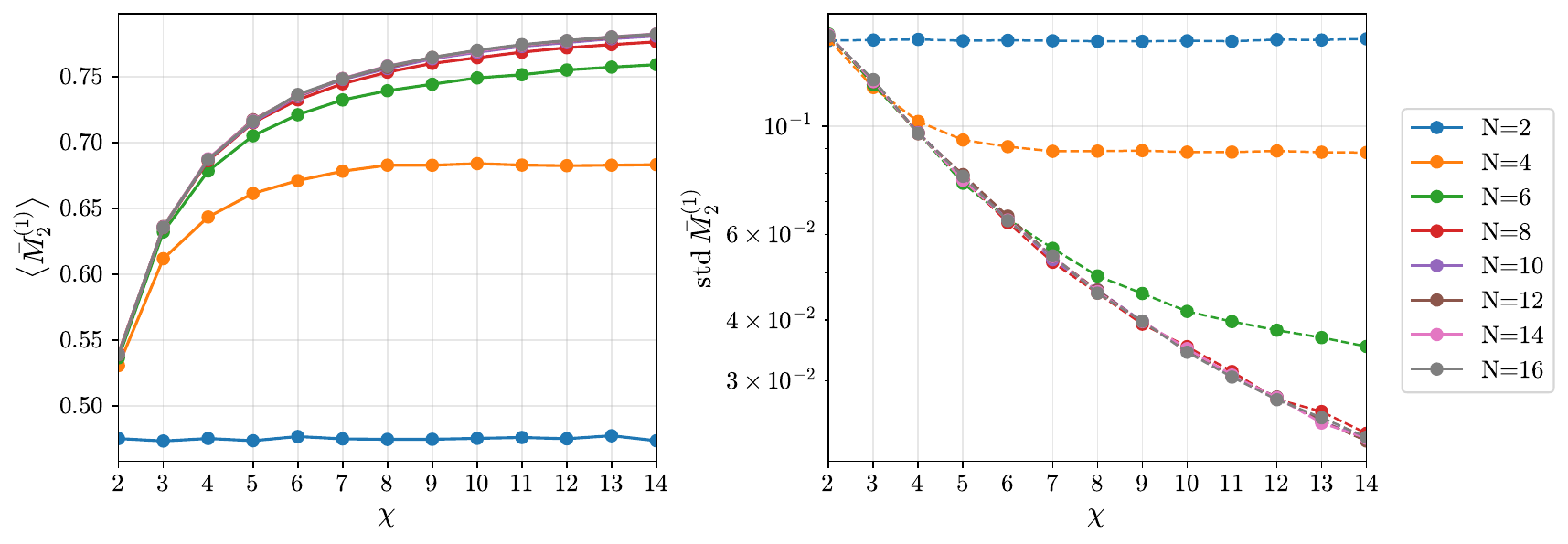}  
    \caption{Average 2-SE gap $\mbar_2$ (left) and its standard deviation (right) under partial trace of the first qubit for MPS in OBC, with the same parameters as Fig.~\ref{app:fig:magic_gap_Tr1_MPS_OBC}. The gap is positive for all $(N,\chi)$ and increases with both the system size and the bond dimension.}
    \label{app:fig:M2_gap_Tr1_MPS_OBC} 
\end{figure*}

\begin{figure*}[t]
    \centering
    \includegraphics[width=\textwidth]{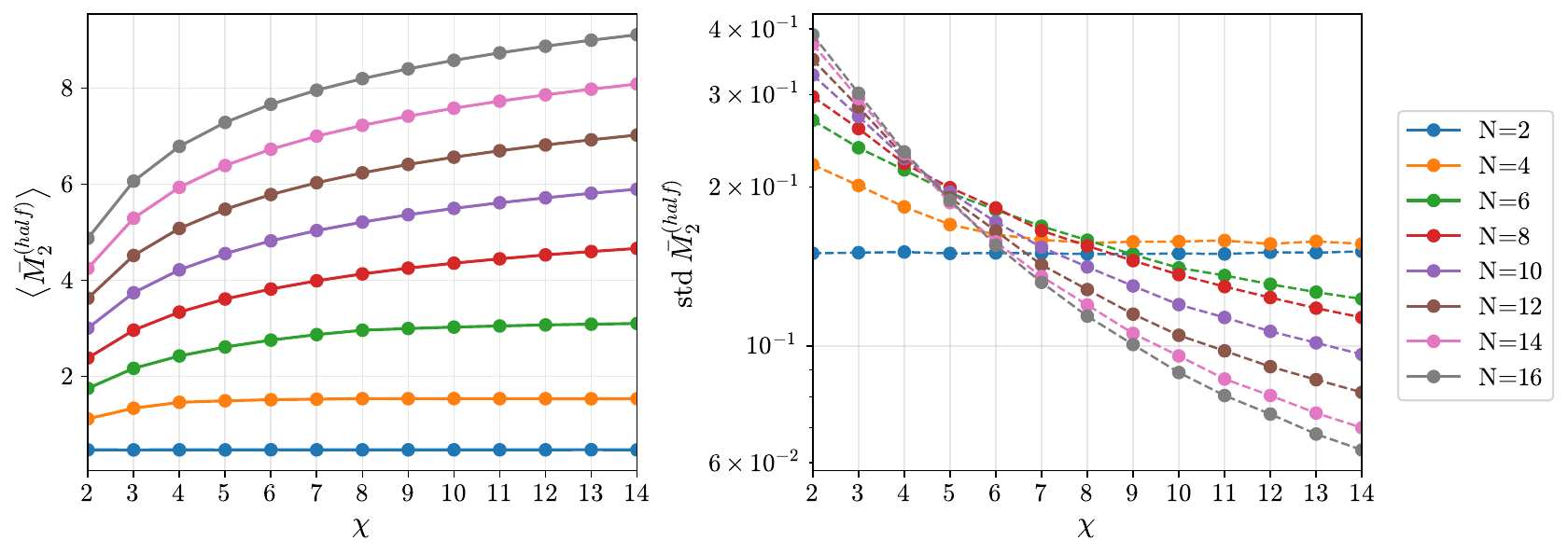}  
    \caption{Average 2-SE gap $\mbar_2$ (left) and its standard deviation (right) under partial trace of the first half of the system for MPS in OBC, with the same parameters as Fig.~\ref{app:fig:magic_gap_Trhalf_MPS_OBC}. The gap is positive for all $(N,\chi)$ and increases with both the system size and the bond dimension.}
    \label{app:fig:M2_gap_Trhalf_MPS_OBC} 
\end{figure*}

Because of its central importance in the resource theory of magic and quantum many-body systems, we also investigate the magic gap of the 2-SE in MPS with OBC. Recalling the definition of the 2-SE for mixed states in Eq.~\ref{app:eq:M_2_mixed_states_def}, we define the 2-SE gap for a quantum channel $\mathcal{E}$ acting on a state $\psi$ as:
\begin{equation}
    \mbar_2^{\mathcal{E}}(\psi):=\mtwo(\psi)-\mtwo(\mathcal{E}(\psi))
\end{equation}
Figures~\ref{app:fig:M2_gap_D_B_MPS_OBC}--\ref{app:fig:M2_gap_Trhalf_MPS_OBC} report the average 2-SE gap and its standard deviation for the same three channels studied above. The qualitative picture mirrors that of $\mlin$: the gap is positive for every sampled $(N,\chi)$---also when accounting for the standard deviation---and increases with the system size.

Overall, these simulations provide strong numerical evidence that both $\mlin$ and $\mtwo$ exhibit robustly positive magic gaps for MPS in OBC across all tested system sizes, bond dimensions, and channels. This confirms their reliability as magic proxies in MPS ensembles that are most relevant to quantum many-body systems' simulations.

\section{Computing purity and stabilizer purity of mixed states using Walsh-Hadamard transform}

In this appendix, we start by reviewing the numerical methods introduced in Ref.~\cite{appcite:huang2026fastexactapproachstabilizer} for computing the stabilizer entropy of an arbitrary pure statevector of an $N$-qubit system. We then extend these techniques to evaluate the purity and the stabilizer purity of both reduced density operators and states dephased in the computational basis.

Section~\ref{app:sec:SP_of_pure_states} presents a concise introduction of the approach developed in Ref.~\cite{appcite:huang2026fastexactapproachstabilizer}, closely following its notation and conceptual framework in order to ensure consistency with the existing literature and clarity. In Section~\ref{app:sec:SP_of_mixed_states}, we analyze the action of computational-basis dephasing and partial-trace channels in the Pauli operator basis for density matrices. We exploit the characteristics of such states and some properties of the Walshâ€“Hadamard transform to generalize the methods of Ref.~\cite{appcite:huang2026fastexactapproachstabilizer}, enabling the efficient computation of the purity and the stabilizer purity for both dephased and  reduced states.

\subsection{Computing the stabilizer purity of pure states of qubits systems using the Walsh-Hadamard Transform}
\label{app:sec:SP_of_pure_states}

Let us start by considering a quantum system of $N$ qubits with Hilbert space $\mathcal{H} \simeq \mathbb{C}^{2^N}$, and an arbitrary quantum state $\ket{\psi} \in \mathcal{H}$:
\begin{equation}
    \ket{\psi} = \sum_{i \in \mathbb{Z}_2^N} \psi_i \ket{i}.
\end{equation}

The object we will be computing is the stabilizer purity. For a pure state, this requires computing $d^2$ Pauli strings' expectation values, which rapidly becomes prohibitive already for modest system sizes\footnote{The cost of computing all the $d^2 = 2^{2N}$ Pauli strings' expectation values for a pure state of a system of $N$ qubits scales as $\mathcal{O}(2^{3N})$.}.

For MPS there exists many numerical methods  \cite{appcite:lami2023nonstabilizerness,appcite:tarabunga2024nonstabilizerness} that reduces this scaling to $\mathcal{O}\big(d^2 N\text{poly}(\chi)\big)$, where $\chi$ is the bond dimension of the MPS.
Recently, many new methods were proposed to reduce this scaling by exploiting the properties of the Walsh-Hadamard Transform in the expression of the second moment of the Pauli spectrum \cite{appcite:huang2026fastexactapproachstabilizer,appcite:xiao2026exponentiallyacceleratedsamplingpauli,appcite:sierant2026computingquantummagicstate}. Such new methods were able to reduce the cost of computing the stabilizer purity and the 2-SE of a pure state of $N$ qubits to $\mathcal{O}(d^2 N)$.

Before delving in the computation of the stabilizer purity, let us remind that, up to an unimportant phase factor, any Pauli string can be written as the product of two Pauli strings: one that is only composed of $I$ and $X$, which we denote as $X^x$, and one made of $I$ and $Z$, which we denote as $Z^z$, with $x,z \in \mathbb{Z}_2^N$ being the binary strings selecting on which qubits a Pauli $X$ or $Z$ operator is applied. 
The action of Pauli strings on any state $\ket{\psi}$ can, therefore, be understood from the action of $X^x$ and $Z^z$ Pauli strings on computational basis states:
\begin{equation}
    X^x \ket{i} = \ket{i \oplus x} \; , \;  Z^z \ket{i} = (-1)^{z \cdot i} \ket{i} \; , \; i \in \mathbb{Z}_2^N
\end{equation}
such that, for an arbitrary state $\ket{\psi}$, it holds that:
\begin{equation}
    X^x Z^z \ket{\psi} = \sum_{i \in \mathbb{Z}_2^N}  (-1)^{z \cdot i}  \psi_i \ket{i \oplus x} 
\end{equation}
and the expectation value of a Pauli string can be written as:
\begin{equation}
    \Tr ( X^x Z^z \psi ) = \bra{\psi}X^x Z^z \ket{\psi} = \sum_{i \in \mathbb{Z}_2^N}  (-1)^{z \cdot i} \psi_{i \oplus x}^* \psi_i
\end{equation}
Following the same notation of Ref.~\cite{appcite:huang2026fastexactapproachstabilizer}, we define:
\begin{equation}
    g_x(i) :=  \psi_{i \oplus x}^* \psi_i
\end{equation}
such that:
\begin{equation}
     \Tr( X^x Z^z \psi)  = \sum_{i \in \mathbb{Z}_2^N} (-1)^{z \cdot i} g_x(i).
     \label{app:eq:exp_value_X_x_Z_z_psi}
\end{equation}
At this point, we can rewrite each term which is involved in the computation of the stabilizer purity as:
\begin{equation}
     \Tr ( X^x Z^z \psi)^4 =  \sum_{i_1, i_2, i_3, i_4  \in \mathbb{Z}_2^N} \prod_{n=1}^4 (-1)^{z \cdot i_n} g_x(i_n) =  \sum_{i_1, i_2, i_3, i_4  \in \mathbb{Z}_2^N}  (-1)^{z \cdot \big( \bigoplus_{n=1}^4 i_n \big) } g_x(i_1) g_x(i_2)^* g_x(i_3) g_x(i_4)^*
\end{equation}
It is important to notice that:
\begin{equation}
    \sum_{z\in \mathbb{Z}_2^N} (-1)^{z \cdot s} = d \delta_{s,0}
\label{app:eq:delta_sum_over_z}
\end{equation}
where $s \in \mathbb{Z}_2^N $ and $\delta_{s,0}$ takes value one only when $s$ is the binary string of all zeros. This simple result can be immediately exploited in the following summation:
\begin{equation}
    \sum_{z  \in \mathbb{Z}_2^N } \Tr (X^x Z^z \psi)^4 = d \sum_{\substack{i_1, i_2, i_3, i_4  \in \mathbb{Z}_2^N , \\ i_1 \oplus i_2 \oplus i_3 \oplus i_4 = 0}} g_x(i_1) g_x(i_2)^* g_x(i_3) g_x(i_4)^* .
\end{equation}
Now, we use $i_4 = i_1 \oplus i_2 \oplus i_3$ and rewrite: 
\begin{equation}
    \sum_{z  \in \mathbb{Z}_2^N} \Tr ( X^x Z^z \psi)^4 = d \sum_{i_1, i_2, i_3 \in \mathbb{Z}_2^N} g_x(i_1) g_x(i_2)^* g_x(i_3) g_x(i_1 \oplus i_2 \oplus i_3)^* .
\end{equation}
Similarly, we define $i := i_1 \oplus i_2$, such that $ i_1 = i \oplus i_2 $: 
\begin{equation}
    \sum_{z \in \mathbb{Z}_2^N} \Tr ( X^x Z^z \psi)^4= d \sum_{i, i_2, i_3 \in \mathbb{Z}_2^N} g_x(i_2 \oplus i) g_x(i_2)^* g_x(i_3) g_x(i_3\oplus i)^* .
    \label{app:eq:before_xor_convolution} 
\end{equation}
The latter summation can further be rewritten in terms of xor-convolutions of $g_x$ functions. We define:
\begin{equation}
    C_x(s) := \sum_{i \in \mathbb{Z}_2^N} g_x(i) g_x(i \oplus s)^*
\end{equation}
and then we plug it in Eq.~\ref{app:eq:before_xor_convolution}:
\begin{equation}
    \sum_{z \in \mathbb{Z}_2^N} \Tr ( X^x Z^z \psi)^4 = d \sum_{i \in \mathbb{Z}_2^N}|C_x(i)|^2
\end{equation}
At this point, we can rewrite the latter summation exploiting the nice properties of the Walsh-Hadamard transform. We will not delve into the properties or details of such a transform, but the interested reader can refer to Ref.~\cite{appcite:huang2026fastexactapproachstabilizer} for a comprehensive exposition. First of all, let us define the Walsh-Hadamard transform of $g_x$ functions as:
\begin{equation}
    G_x := WHT(g_x)
    \label{app:eq:def_WHT_G_x}
\end{equation}
such that, exploiting the xor-convolution theorem for the Walsh-Hadamard transform we can rewrite:
\begin{equation}
    WHT(C_x) = d^{1/2} G_x G_x^* = d^{1/2} |G_x|^2
\end{equation}
and further exploiting Parseval's identity we write:
\begin{equation}
    \sum_{i \in \mathbb{Z}_2^N} |C_x(i)|^2 = \sum_{k \in \mathbb{Z}_2^N} |WHT(C_x)(k)|^2 = d \sum_{k \in \mathbb{Z}_2^N} |G_x(k)|^4
\end{equation}
At this point, we have all the ingredients for computing the stabilizer purity as:
\begin{equation}
    \stp(\ket{\psi}) = d^{-1} \sum_{x,z \in \mathbb{Z}_2^N}  \Tr (X^x Z^z \psi)^4 = d  \sum_{x, k \in \mathbb{Z}_2^N}  |G_x(k)|^4
    \label{app:eq:stab_purity_WHT}
\end{equation}

\subsection{Computing the Stabilizer Purity of dephased and reduced states of qubits systems using the Walsh-Hadamard Transform}
\label{app:sec:SP_of_mixed_states}

We start by considering the expression of an arbitrary pure state density matrix in the Pauli basis:
\begin{equation}
    \psi = \ketbra{\psi}{\psi} = \frac{1}{d} \sum_{P \in \mathcal{P}_N} \Tr(P\psi) P
\end{equation}
When dephasing $\psi$ in computational basis, the only elements of the Pauli spectrum of $\psi$ that survive will be Pauli strings of the type $Z^z$, since:
\begin{equation}
    D_B(\psi) = \frac{1}{d} \sum_{z \in  \in \mathbb{Z}_2^{N} } \Tr(Z^z\psi) Z^z
\end{equation}
Therefore, its stabilizer purity will simply be:
\begin{equation}
    \stp(D_B(\psi)) = d \sum_{z \in \mathbb{Z}_2^{N}} \Tr (Z^z \psi)^4
\end{equation}
and considering the expression of the stabilizer purity in Eq.~\ref{app:eq:stab_purity_WHT}, we can write:
\begin{equation}
     \stp(D_B(\psi)) = d \sum_{k \in \mathbb{Z}_2^N}  |G_0(k)|^4
\end{equation}

Computing the stabilizer purity of reduced states is slightly more challenging. Let us start by considering an arbitrary bipartition of our quantum system of interest $\mathcal{H} = \mathcal{H}_A \otimes \mathcal{H}_B$, where $A$ and $B$ are subsystems of, respectively, $N_A$ and $N_B$ qubits and with Hilbert spaces of dimensions $d_A=2^{N_A}$ and $d_B=2^{N_B}$.
We rewrite and arbitrary quantum state $\ket{\psi} \in \mathcal{H}$ as:
\begin{equation}
    \ket{\psi} = \sum_{\substack{i_A \in \mathbb{Z}_2^{N_A}, \\i_B \in \mathbb{Z}_2^{N_B}}} \psi_{i_A,i_B} \ket{i_A} \ket{i_B}
\end{equation}
The corresponding density matrix can be, therefore, written as:
\begin{equation}
    \psi = \sum_{\substack{i_A, j_A \in \mathbb{Z}_2^{N_A}, \\i_B, j_B \in \mathbb{Z}_2^{N_B}}} \psi_{i_A,i_B} \psi_{j_A,j_B}^* \ketbra{i_A}{j_A} \ketbra{i_B}{j_B}
\end{equation}
and tracing out the subsystem $A$ we obtain:
\begin{equation}
     \psi_B = \Tr_A \psi = \sum_{\substack{i_A \in \mathbb{Z}_2^{N_A}, \\i_B, j_B \in \mathbb{Z}_2^{N_B}}} \psi_{i_A,i_B} \psi_{i_A,j_B}^* \ketbra{i_B}{j_B}
\end{equation}

As for in Sec.~\ref{app:sec:SP_of_pure_states}, we are interested in computing the expectation value of Pauli strings $P_B$ of the type $X^x Z^z$, where in this setting $x,z \in \mathbb{Z}_2^{N_B}$. Then, we can write:
\begin{equation}
    \Tr ( X^x Z^z {\psi_B} ) = \sum_{\substack{i_A \in \mathbb{Z}_2^{N_A}, \\i_B, j_B \in \mathbb{Z}_2^{N_B}}} \psi_{i_A,i_B} \psi_{i_A,j_B}^* \bra{j_B} X^x Z^z \ket{i_B} =  \sum_{\substack{i_A \in \mathbb{Z}_2^{N_A}, \\i_B, j_B \in \mathbb{Z}_2^{N_B}}} \psi_{i_A,i_B} \psi_{i_A,j_B}^* (-1)^{z \cdot i_B} \delta_{j_B, i_B \oplus x}
\end{equation}
and by enforcing $ \delta_{j_B, i_B \oplus x}$ and defining:
\begin{equation}
    \Tilde{g_x}(i_B) := \sum_{i_A \in \mathbb{Z}_2^{N_A}} \psi_{i_A,i_B} \psi_{i_A, i_B \oplus x}^*
\end{equation}
we can rewrite the expectation value as:
\begin{equation}
    \Tr ( X^x Z^z {\psi_B}) =  \sum_{i_B, j_B \in \mathbb{Z}_2^{N_B}} (-1)^{z \cdot i_B} \Tilde{g_x}(i_B) 
    \label{app:eq:exp_value_X_x_Z_z_psi_B}
\end{equation}

Closely following Sec.~\ref{app:sec:SP_of_pure_states}, we now shift our attention to the following quantity:
\begin{equation}
    \sum_{z \in \mathbb{Z}_2^{N_B}} \Tr ( X^x Z^z {\psi_B} )^4  = \sum_{i_1, i_2, i_3, i_4  \in \mathbb{Z}_2^{N_B}}  (-1)^{z \cdot \big( \bigoplus_{n=1}^4 i_n \big) } \Tilde{g_x}(i_1) \Tilde{g_x}(i_2)^* \Tilde{g_x}(i_3) \Tilde{g_x}(i_4)^*
\end{equation}
which, by the same arguments of Sec.~\ref{app:sec:SP_of_pure_states}, reduces to:
\begin{equation}
     \sum_{z \in \mathbb{Z}_2^{N_B}} \Tr ( X^x Z^z {\psi_B} )^4   = d_B \sum_{i, i_2, i_3  \in \mathbb{Z}_2^{N_B}} \Tilde{g_x}(i_2 \oplus i) \Tilde{g_x}(i_2)^* \Tilde{g_x}(i_3) \Tilde{g_x}(i_3 \oplus i)^*
\end{equation}
In the latter expression, again, we recognize two xor-convolutions. To simplify the notation we define:
\begin{equation}
    \Tilde{C}_x(s) := \sum_{i \in \mathbb{Z}_2^{N_B}} \Tilde{g_x}(i) \Tilde{g_x}(i \oplus s)^*
\end{equation}
and so we can write:
\begin{equation}
     \sum_{z \in \mathbb{Z}_2^{N_B}} \Tr ( X^x Z^z {\psi_B} )^4 = d_B \sum_{i  \in \mathbb{Z}_2^{N_B}} | \Tilde{C}_x(i)|^2
\end{equation}
Now, using Parseval's identity for the WHT we can write:
\begin{equation}
    \sum_{i  \in \mathbb{Z}_2^{N_B}} | \Tilde{C}_x(i)|^2 = \sum_{k  \in \mathbb{Z}_2^{N_B}} | WHT(\Tilde{C}_x)(k)|^2
\end{equation}
where :
\begin{equation}
    WHT(\Tilde{C}_x) = d_B^{1/2} |\Tilde{G}_x|^2
\end{equation}
with $\Tilde{G}_x := WHT(\Tilde{g}_x)$. Finally, we can write the stabilizer purity of the reduced state $\psi_B$ as:
\begin{equation}
    \stp(\psi_B) = d_B \sum_{x,k \in \mathbb{Z}_2^{N_B}} |\Tilde{G}_x(k)|^4
\end{equation}

\subsection{Computing the Purity of dephased and reduced states of qubits systems using the Walsh-Hadamard Transform}
\label{app:sec:Pur_of_mixed_states}
In this section, we start by briefly deriving the expression of the purity of a pure state in terms of the WHT of the $g_x$ coefficients, and later we extend this method for computing the purity of dephased and reduced states. 
It is important to highlight that, unlike stabilizer purity, computing the purity using the WHT has no computational advantage with respect to other methods. We nonetheless adopt this formulation because, in practice, we are interested in computing $\mlin(\cdot)$, and since the WHT is already performed to compute the stabilizer purity, both quantities can be extracted from the same transforms at no additional cost.

The purity of a quantum state can be written as:
\begin{equation}
\pur(\psi) = d^{-1} \sum_{P}  \Tr ( P \psi)^2
\end{equation}
and an arbitrary Pauli string can always be written as:
\begin{equation}
    P = i^{-x \cdot z} X^x Z^z
\end{equation}
This means that, in the expression of the purity, we can substitute:
\begin{equation}
    \sum_ P \Tr (P \psi )^2 = \sum_{x,z} (-1)^{x \cdot z}  \Tr( X^x Z^z \psi)^2
\end{equation}
and so:
\begin{equation}
    \sum_ P \Tr (P \psi )^2 = \sum_ P |\Tr (P \psi )|^2 = \sum_{x,z} |\Tr( X^x Z^z \psi)|^2
\end{equation}
Now, we consider the argument of the latter summation and use Eq.~\ref{app:eq:exp_value_X_x_Z_z_psi}, which holds for pure states, to obtain: 
\begin{equation}
     |\Tr ( X^x Z^z \psi)|^2 = \sum_{i_1, i_2\in \mathbb{Z}_2^N}  (-1)^{z \cdot (i_1 \oplus i_2) } g_x(i_1) g_x(i_2)^*
\end{equation}
We are interested in the following summation:
\begin{equation}
    \sum_{z  \in \mathbb{Z}_2^N } |\Tr (X^x Z^z \psi)|^2 = d \sum_{\substack{i_1, i_2 \in \mathbb{Z}_2^N , \\ i_1 \oplus i_2  = 0}} g_x(i_1) g_x(i_2)^*  = d \sum_{\substack{i \in \mathbb{Z}_2^N}} |g_x(i)|^2
\end{equation}
where in the first passage we exploited, again, the result of Eq.~\ref{app:eq:delta_sum_over_z}. So we can write the purity of a pure state $\psi$ as:
\begin{equation}
\pur(\psi) =  \sum_{\substack{x,i \in \mathbb{Z}_2^N}} |g_x(i)|^2 =  \sum_{\substack{x,i \in \mathbb{Z}_2^N}} |G_x(i)|^2 
\end{equation}
where $G_x$, as introduced in Eq.~\ref{app:eq:def_WHT_G_x} is the WHT of the $g_x$ and in the last step we exploited Parseval's identity.

At this point, it is straightforward to derive the expression for the purity of the respective dephased state in computational basis, which will simply be:
\begin{equation}
\pur(D_B(\psi)) = \sum_{\substack{i \in \mathbb{Z}_2^N}} |G_0(i)|^2 
\end{equation}

We can proceed very similarly for reduced states, for which the purity reads:
\begin{equation}
    \pur (\psi_B) = d_B^{-1} \sum_{x,z \in \mathbb{Z}_2^{N_B}} |\Tr(X^x Z^z \psi_B)|^2
\end{equation}
Starting from Eq.~\ref{app:eq:exp_value_X_x_Z_z_psi_B}, we can write:
\begin{equation}
    \sum_{z \in \mathbb{Z}_2^{N_B}} |\Tr ( X^x Z^z {\psi_B} )|^2  = \sum_{z, i_1, i_2  \in \mathbb{Z}_2^{N_B}}  (-1)^{z \cdot (i_1 \oplus i_2) } \Tilde{g_x}(i_1) \Tilde{g_x}(i_2)^* = d_B  \sum_{i  \in \mathbb{Z}_2^{N_B}} |\Tilde{g_x}(i)|^2
    \end{equation}
where in the last passage we used Eq.~\ref{app:eq:delta_sum_over_z}.
We  can, again, use Parseval's identity to rewrite the purity of $\psi_B$ as:
\begin{equation}
    \pur (\psi_B) = \sum_{x,i \in \mathbb{Z}_2^{N_B}}  |\Tilde{g_x}(i)|^2 =  \sum_{x,k \in \mathbb{Z}_2^{N_B}}  |\Tilde{G_x}(k)|^2
\end{equation}

\bibliographystyle{apsrev4-2}
\bibliography{prl_sub}